\documentclass[10pt]{article}
\usepackage{epsf}
\usepackage{axodraw}
\usepackage{epsfig}                            
\usepackage{graphicx}
\usepackage{rotate}
\usepackage{latexsym}
\renewcommand{\cal}{\mathcal}
\begin{document}
\marginparwidth 3cm
%
%
%
%
\newcommand{\nl}{\nonumber\\}
\newcommand{\nn}{\nonumber}
\newcommand{\ds}{\displaystyle}
\newcommand{\mpar}[1]{{\marginpar{\hbadness10000%
                      \sloppy\hfuzz10pt\boldmath\bf#1}}%
                      \typeout{marginpar: #1}\ignorespaces}
\def\mnew{\mpar{\hfil NEW \hfil}\ignorespaces}
\newcommand{\lpar}{\left(}                            
\newcommand{\rpar}{\right)} 
\newcommand{\lrbr}{\left[}
\newcommand{\rrbr}{\right]}
\newcommand{\lcbr}{\left\{}
\newcommand{\rcbr}{\right\}} 
\newcommand{\rbrak}[1]{\lrbr#1\rrbr}
\newcommand{\bq}{\begin{equation}}                    
\newcommand{\eq}{\end{equation}}
\newcommand{\bqa}{\begin{eqnarray}}
\newcommand{\eqa}{\end{eqnarray}}
\newcommand{\ba}[1]{\begin{array}{#1}}
\newcommand{\ea}{\end{array}}
\newcommand{\ben}{\begin{enumerate}}
\newcommand{\een}{\end{enumerate}}
\newcommand{\bei}{\begin{itemize}}
\newcommand{\eei}{\end{itemize}}
\newcommand{\eqn}[1]{Eq.(\ref{#1})}
\newcommand{\eqns}[2]{Eqs.(\ref{#1}--\ref{#2})}
\newcommand{\eqnss}[1]{Eqs.(\ref{#1})}
\newcommand{\eqnsc}[2]{Eqs.(\ref{#1},~\ref{#2})}
\newcommand{\tbn}[1]{Tab.(\ref{#1})}
\newcommand{\tbns}[2]{Tabs.(\ref{#1}--\ref{#2})}
\newcommand{\tbnsc}[2]{Tabs.(\ref{#1},~\ref{#2})}
\newcommand{\fig}[1]{Fig.~\ref{#1}}
\newcommand{\figs}[2]{Figs.~\ref{#1}--\ref{#2}}
\newcommand{\sect}[1]{Sect.~\ref{#1}}
%
%
\newcommand{\TeV}{\;\mathrm{TeV}}                     
\newcommand{\GeV}{\;\mathrm{GeV}}
\newcommand{\MeV}{\;\mathrm{MeV}}
\newcommand{\nb}{\;\mathrm{nb}}
\newcommand{\pb}{\;\mathrm{pb}}
\newcommand{\fb}{\;\mathrm{fb}}
\def\Re{\mathop{\operator@font Re}\nolimits}
\def\Im{\mathop{\operator@font Im}\nolimits}
\newcommand{\ord}[1]{{\cal O}\lpar#1\rpar}
\newcommand{\group}{SU(2)\otimes U(1)}
\newcommand{\ib}{i}
\newcommand{\asums}[1]{\sum_{#1}}
\newcommand{\asumt}[2]{\sum_{#1}^{#2}}
\newcommand{\asum}[3]{\sum_{#1=#2}^{#3}}
%
%
\newcommand{\tmi}{\times 10^{-1}}
\newcommand{\tmii}{\times 10^{-2}}
\newcommand{\tmiii}{\times 10^{-3}}
\newcommand{\tmiv}{\times 10^{-4}}
\newcommand{\tmfv}{\times 10^{-5}}
\newcommand{\tmfvi}{\times 10^{-6}}
\newcommand{\tmfvii}{\times 10^{-7}}
\newcommand{\tmfviii}{\times 10^{-8}}
\newcommand{\tmfix}{\times 10^{-9}}
\newcommand{\tmfx}{\times 10^{-10}}
%
%
\newcommand{\fer}{{\rm{fer}}}
\newcommand{\bos}{{\rm{bos}}}
\newcommand{\gen}{\rm{g}}
\newcommand{\dbl}{\rm{d}}
\newcommand{\philone}{\phi}
\newcommand{\philoneb}{\phi_{0}}
\newcommand{\phiind}[1]{\phi_{#1}}
\newcommand{\gBi}[2]{B_{#1}^{#2}}
\newcommand{\gBn}[1]{B_{#1}}
%
%
\newcommand{\ph}{\gamma}
\newcommand{\ab}{A}
\newcommand{\abr}{A^r}
\newcommand{\abb}{A^{0}}
\newcommand{\abi}[1]{A_{#1}}
\newcommand{\abri}[1]{A^r_{#1}}
\newcommand{\abbi}[1]{A^{0}_{#1}}
\newcommand{\wb}{W}            
\newcommand{\wbi}[1]{W_{#1}}           
\newcommand{\wbp}{W^{+}}
\newcommand{\wbm}{W^{-}}
\newcommand{\wbpm}{W^{\pm}}
\newcommand{\wbpi}[1]{W^{+}_{#1}}
\newcommand{\wbmi}[1]{W^{-}_{#1}}
\newcommand{\wbpmi}[1]{W^{\pm}_{#1}}
\newcommand{\wbli}[1]{W^{[+}_{#1}}
\newcommand{\wbri}[1]{W^{-]}_{#1}}
\newcommand{\zb}{Z}
\newcommand{\zbi}[1]{Z_{#1}}
\newcommand{\vb}{V}
\newcommand{\vbi}[1]{V_{#1}}      
\newcommand{\vbiv}[1]{V^{*}_{#1}}      
\newcommand{\Pb}{P}
\newcommand{\Sb}{S}
\newcommand{\Bb}{B}
%
%
\newcommand{\hk}{K}
\newcommand{\hkg}{\phi}
\newcommand{\hkn}{\phi^{0}}                 
\newcommand{\hkp}{\phi^{+}}
\newcommand{\hkm}{\phi^{-}}
\newcommand{\hkpm}{\phi^{\pm}}
\newcommand{\hkmp}{\phi^{\mp}}
\newcommand{\hki}[1]{\phi^{#1}}
\newcommand{\hb}{H}
\newcommand{\hbi}[1]{H_{#1}}
\newcommand{\hkl}{\phi^{[+\cgfi\cgfi}}
\newcommand{\hkr}{\phi^{-]}}
%
%
\newcommand{\fpx}{X}
\newcommand{\fpy}{Y}
\newcommand{\fpxp}{X^+}
\newcommand{\fpxm}{X^-}
\newcommand{\fpxpm}{X^{\pm}}
\newcommand{\fpxi}[1]{X^{#1}}
\newcommand{\fpyZ}{Y^{\ssZ}}
\newcommand{\fpyA}{Y^{\ssA}}
\newcommand{\fpyZA}{Y_{\ssZ,\ssA}}
\newcommand{\fpbxi}[1]{{\overline{X}}^{#1}}
\newcommand{\fpbyZ}{{\overline{Y}}^{\ssZ}}
\newcommand{\fpbyA}{{\overline{Y}}^{\ssA}}
\newcommand{\fpbyZA}{{\overline{Y}}^{\ssZ,\ssA}}
%
%
\newcommand{\Flone}{F}
\newcommand{\fpsi}{\psi}
\newcommand{\fpsii}[1]{\psi^{#1}}
\newcommand{\fpsib}{\psi^{0}}
\newcommand{\fpsir}{\psi^r}
\newcommand{\fpsiL}{\psi_{_L}}
\newcommand{\fpsiR}{\psi_{_R}}
\newcommand{\fpsiLi}[1]{\psi_{_L}^{#1}}
\newcommand{\fpsiRi}[1]{\psi_{_R}^{#1}}
\newcommand{\fpsiLbi}[1]{\psi_{_{0L}}^{#1}}
\newcommand{\fpsiRbi}[1]{\psi_{_{0R}}^{#1}}
\newcommand{\fpsiLR}{\psi_{_{L,R}}}
\newcommand{\fbpsi}{{\overline{\psi}}}
\newcommand{\fbpsii}[1]{{\overline{\psi}}^{#1}}
\newcommand{\fbpsir}{{\overline{\psi}}^r}
\newcommand{\fbpsiL}{{\overline{\psi}}_{_L}}
\newcommand{\fbpsiR}{{\overline{\psi}}_{_R}}
\newcommand{\fbpsiLi}[1]{\overline{\psi_{_L}}^{#1}}
\newcommand{\fbpsiRi}[1]{\overline{\psi_{_R}}^{#1}}
\newcommand{\fe}{e}
\newcommand{\ff}{f}
\newcommand{\fep}{e^{+}}
\newcommand{\fem}{e^{-}}
\newcommand{\fepm}{e^{\pm}}
\newcommand{\fp}{f^{+}}
\newcommand{\fm}{f^{-}}
\newcommand{\fhp}{h^{+}}
\newcommand{\fhm}{h^{-}}
\newcommand{\fh}{h}
\newcommand{\flm}{\mu}
\newcommand{\flmp}{\mu^{+}}
\newcommand{\flmm}{\mu^{-}}
\newcommand{\fll}{l}
\newcommand{\fllp}{l^{+}}
\newcommand{\fllm}{l^{-}}
\newcommand{\flt}{\tau}
\newcommand{\fltp}{\tau^{+}}
\newcommand{\fltm}{\tau^{-}}
\newcommand{\fq}{q}
\newcommand{\ffQ}{Q}
\newcommand{\fu}{u}
\newcommand{\fd}{d}
\newcommand{\fc}{c}
\newcommand{\fs}{s}
\newcommand{\fqp}{q'}
\newcommand{\fup}{u'}
\newcommand{\fdp}{d'}
\newcommand{\fcp}{c'}
\newcommand{\fsp}{s'}
\newcommand{\fdpp}{d''}
\newcommand{\ffi}[1]{f_{#1}}
\newcommand{\bffi}[1]{{\overline{f}}_{#1}}
\newcommand{\ffpi}[1]{f'_{#1}}
\newcommand{\bffpi}[1]{{\overline{f}}'_{#1}}
\newcommand{\ft}{t}
\newcommand{\ffb}{b}
\newcommand{\ffp}{f'}
\newcommand{\fft}{{\tilde{f}}}
\newcommand{\fl}{l}
\newcommand{\fnu}{\nu}
\newcommand{\fU}{U}
\newcommand{\fD}{D}
\newcommand{\fUc}{\overline{U}}
\newcommand{\fDc}{\overline{D}}
\newcommand{\fnul}{\nu_l}
\newcommand{\fnue}{\nu_e}
\newcommand{\fnum}{\nu_{\mu}}
\newcommand{\fnut}{\nu_{\tau}}
\newcommand{\fbe}{{\overline{e}}}
\newcommand{\fbu}{{\overline{u}}}
\newcommand{\fbd}{{\overline{d}}}
\newcommand{\fbf}{{\overline{f}}}
\newcommand{\fbfp}{{\overline{f}}'}
\newcommand{\fbl}{{\overline{l}}}
\newcommand{\fbnu}{{\overline{\nu}}}
\newcommand{\fbnul}{{\overline{\nu}}_{\fl}}
\newcommand{\fbnue}{{\overline{\nu}}_{\fe}}
\newcommand{\fbnum}{{\overline{\nu}}_{\flm}}
\newcommand{\fbnut}{{\overline{\nu}}_{\flt}}
\newcommand{\fuL}{u_{_L}}
\newcommand{\fdL}{d_{_L}}
\newcommand{\ffL}{f_{_L}}
\newcommand{\fbuL}{{\overline{u}}_{_L}}
\newcommand{\fbdL}{{\overline{d}}_{_L}}
\newcommand{\fbfL}{{\overline{f}}_{_L}}
\newcommand{\fuR}{u_{_R}}
\newcommand{\fdR}{d_{_R}}
\newcommand{\ffR}{f_{_R}}
\newcommand{\fbuR}{{\overline{u}}_{_R}}
\newcommand{\fbdR}{{\overline{d}}_{_R}}
\newcommand{\fbfR}{{\overline{f}}_{_R}}
%
%
\newcommand{\barf}{\overline f}                
\newcommand{\barl}{\overline l}
\newcommand{\barq}{\overline q}
\newcommand{\barqp}{\overline{q}'}
\newcommand{\barb}{\overline b}
\newcommand{\bart}{\overline t}
\newcommand{\barc}{\overline c}
\newcommand{\baru}{\overline u}
\newcommand{\bard}{\overline d}
\newcommand{\bars}{\overline s}
\newcommand{\barv}{\overline v}
\newcommand{\barnu}{\overline{\nu}}
\newcommand{\barne}{\overline{\nu}_{\fe}}
\newcommand{\barnm}{\overline{\nu}_{\flm}}
\newcommand{\barnt}{\overline{\nu}_{\flt}}
%
%
\newcommand{\glu}{g}
%
%
\newcommand{\prot}{p}
\newcommand{\aprot}{{\bar{p}}}
\newcommand{\Nucln}{N}
%
%
\newcommand{\tM}{{\tilde M}}
\newcommand{\tMs}{{\tilde M}^2}
\newcommand{\tW}{{\tilde \Gamma}}
\newcommand{\tWs}{{\tilde\Gamma}^2}
\newcommand{\fphi}{\phi}
\newcommand{\fJpsi}{J/\psi}
\newcommand{\fgpsi}{\psi}
\newcommand{\Glone}{\Gamma}
\newcommand{\Gloni}[1]{\Gamma_{#1}}
\newcommand{\Glones}{\Gamma^2}
\newcommand{\Glonec}{\Gamma^3}
\newcommand{\glone}{\gamma}
\newcommand{\glones}{\gamma^2}
\newcommand{\gloneq}{\gamma^4}
\newcommand{\gloni}[1]{\gamma_{#1}}
\newcommand{\glonis}[1]{\gamma^2_{#1}}
\newcommand{\Grest}[2]{\Gamma_{#1}^{#2}}
\newcommand{\grest}[2]{\gamma_{#1}^{#2}}
\newcommand{\resampl}{A_{_R}}
\newcommand{\resasyi}[1]{{\cal{A}}_{#1}}
\newcommand{\sSrest}[1]{\sigma_{#1}}
\newcommand{\Srest}[2]{\sigma_{#1}\lpar{#2}\rpar}
\newcommand{\Gdist}[1]{{\cal{G}}\lpar{#1}\rpar}
\newcommand{\sGdist}{{\cal{G}}}
\newcommand{\Aarea}{A_{0}}
\newcommand{\Aareai}[1]{{\cal{A}}\lpar{#1}\rpar}
\newcommand{\sAarea}{{\cal{A}}}
\newcommand{\resolw}{\sigma_{\Energ}}
\newcommand{\chizer}{\chi_{_0}}
\newcommand{\ini}{\rm{in}}
\newcommand{\fin}{\rm{fin}}
\newcommand{\ifi}{\rm{if}}
\newcommand{\ipf}{\rm{i+f}}
\newcommand{\tot}{\rm{tot}}
\newcommand{\Bac}{Q}
\newcommand{\Res}{R}
\newcommand{\Int}{I}
\newcommand{\NRe}{NR}
\newcommand{\ratoe}{\delta}
\newcommand{\ratoes}{\delta^2}
%
%
\newcommand{\Fbox}[2]{f^{\rm{box}}_{#1}\lpar{#2}\rpar}
\newcommand{\Dbox}[2]{\delta^{\rm{box}}_{#1}\lpar{#2}\rpar}
\newcommand{\Bbox}[3]{{\cal{B}}_{#1}^{#2}\lpar{#3}\rpar}
%
%
\newcommand{\phm}{\lambda}
\newcommand{\phms}{\lambda^2}
\newcommand{\mV}{M_{_V}}
\newcommand{\mw}{M_{_W}}
\newcommand{\mX}{M_{_X}}
\newcommand{\mY}{M_{_Y}}
\newcommand{\LM}{M}
\newcommand{\mz}{M_{_Z}}
\newcommand{\bzm}{M_{_0}}
\newcommand{\mh}{M_{_H}}
\newcommand{\bhm}{M_{_{0H}}}
\newcommand{\mf}{m_f}
\newcommand{\mfp}{m_{f'}}
\newcommand{\mfh}{m_{h}}
\newcommand{\mt}{m_t}
\newcommand{\me}{m_e}
\newcommand{\mm}{m_{\mu}}
\newcommand{\mtau}{m_{\tau}}
\newcommand{\muq}{m_u}
\newcommand{\md}{m_d}
\newcommand{\muqp}{m'_u}
\newcommand{\mdqp}{m'_d}
\newcommand{\mc}{m_c}
\newcommand{\ms}{m_s}
\newcommand{\mb}{m_b}
\newcommand{\mup}{M_u}                              
\newcommand{\mdp}{M_d}
\newcommand{\mcp}{M_c}
\newcommand{\msp}{M_s}
\newcommand{\mbp}{M_b}
%
%
\newcommand{\mls}{m^2_l}
\newcommand{\mVs}{M^2_{_V}}
\newcommand{\mws}{M^2_{_W}}
\newcommand{\mwc}{M^3_{_W}}
\newcommand{\LMs}{M^2}
\newcommand{\LMc}{M^3}
\newcommand{\mzs}{M^2_{_Z}}
\newcommand{\mzc}{M^3_{_Z}}
\newcommand{\bzms}{M^2_{_0}}
\newcommand{\bzmc}{M^3_{_0}}
\newcommand{\bhms}{M^2_{_{0H}}}
\newcommand{\mhs}{M^2_{_H}}
\newcommand{\mfs}{m^2_f}
\newcommand{\mfc}{m^3_f}
\newcommand{\mfps}{m^2_{f'}}
\newcommand{\mfhs}{m^2_{h}}
\newcommand{\mfpc}{m^3_{f'}}
\newcommand{\mts}{m^2_t}
\newcommand{\mes}{m^2_e}
\newcommand{\mms}{m^2_{\mu}}
\newcommand{\mmc}{m^3_{\mu}}
\newcommand{\mmfour}{m^4_{\mu}}
\newcommand{\mmf}{m^5_{\mu}}
\newcommand{\mmfive}{m^5_{\mu}}
\newcommand{\mmsix}{m^6_{\mu}}
\newcommand{\mminv}{\frac{1}{m_{\mu}}}
\newcommand{\mtaus}{m^2_{\tau}}
\newcommand{\mus}{m^2_u}
\newcommand{\mds}{m^2_d}
\newcommand{\muqps}{m'^2_u}
\newcommand{\mdqps}{m'^2_d}
\newcommand{\mcs}{m^2_c}
\newcommand{\mss}{m^2_s}
\newcommand{\mbs}{m^2_b}
\newcommand{\mups}{M^2_u}
\newcommand{\mdps}{M^2_d}
\newcommand{\mcps}{M^2_c}
\newcommand{\msps}{M^2_s}
\newcommand{\mbps}{M^2_b}
%
%
\newcommand{\muf}{\mu_{_{\ff}}}
\newcommand{\mufs}{\mu^2_{_{\ff}}}
\newcommand{\mufq}{\mu^4_{_{\ff}}}
\newcommand{\muz}{\mu_{_{\zb}}}
\newcommand{\muzs}{\mu^2_{_{\zb}}}
\newcommand{\muw}{\mu_{_{\wb}}}
\newcommand{\muws}{\mu^2_{_{\wb}}}
\newcommand{\muSW}{\mu^2_{_{\wb}}}
\newcommand{\muwq}{\mu^4_{_{\wb}}}
\newcommand{\muwsx}{\mu^6_{_{\wb}}}
\newcommand{\muwms}{\mu^{-2}_{_{\wb}}}
\newcommand{\muhs}{\mu^2_{_{\hb}}}
\newcommand{\muhq}{\mu^4_{_{\hb}}}
\newcommand{\muhsx}{\mu^6_{_{\hb}}}
\newcommand{\muts}{\mu^2_{\ft}}
\newcommand{\mutq}{\mu^4_{_{\hb}}}   
\newcommand{\mutsx}{\mu^6_{_{\hb}}}  
\newcommand{\muL}{\mu}
\newcommand{\muS}{\mu^2}
\newcommand{\muQ}{\mu^4}
\newcommand{\muizs}{\mu^2_{0}}
\newcommand{\muizq}{\mu^4_{0}}
\newcommand{\muis}{\mu^2_{1}}
\newcommand{\muiis}{\mu^2_{2}}
\newcommand{\muiiis}{\mu^2_{3}}
\newcommand{\muii}[1]{\mu_{#1}}
\newcommand{\muisi}[1]{\mu^2_{#1}}
\newcommand{\muiqi}[1]{\mu^4_{#1}}
\newcommand{\muixi}[1]{\mu^6_{#1}}
\newcommand{\zm}{z_m}
\newcommand{\ri}[1]{r_{#1}}
\newcommand{\xw}{x_w}
\newcommand{\xws}{x^2_w}
\newcommand{\xwc}{x^3_w}
\newcommand{\xth}{x_t}
\newcommand{\xths}{x^2_t}
\newcommand{\xthc}{x^3_t}
\newcommand{\xthf}{x^4_t}
\newcommand{\xthv}{x^5_t}
\newcommand{\xthx}{x^6_t}
\newcommand{\xh}{x_h}
\newcommand{\xhs}{x^2_h}
\newcommand{\xhc}{x^3_h}
%
%
\newcommand{\mwq}{M^4_{_\wb}}
\newcommand{\mwf}{M^4_{_\wb}}
\newcommand{\LMq}{M^4}
\newcommand{\mzq}{M^4_{_Z}}
\newcommand{\bzmq}{M^4_{_0}}
\newcommand{\mhq}{M^4_{_H}}
\newcommand{\mfq}{m^4_f}
\newcommand{\mfpq}{m^4_{f'}}
\newcommand{\mtq}{m^4_t}
\newcommand{\meq}{m^4_e}
\newcommand{\mmq}{m^4_{\mu}}
\newcommand{\mtauq}{m^4_{\tau}}
\newcommand{\muqq}{m^4_u}
\newcommand{\mdq}{m^4_d}
\newcommand{\mcq}{m^4_c}
\newcommand{\msq}{m^4_s}
\newcommand{\mbq}{m^4_b}
\newcommand{\mupq}{M^4_u}
\newcommand{\mdpq}{M^4_d}
\newcommand{\mcpq}{M^4_c}
\newcommand{\mspq}{M^4_s}
\newcommand{\mbpq}{M^4_b}
%
%
\newcommand{\mwx}{M^6_{_W}}
\newcommand{\mzx}{M^6_{_Z}}
\newcommand{\mfx}{m^6_f}
\newcommand{\mfpx}{m^6_{f'}}
\newcommand{\LMx}{M^6}
%
%
\newcommand{\mer}{m_{er}}
\newcommand{\mlep}{m_l}
\newcommand{\mleps}{m^2_l}
\newcommand{\mone}{m_1}
\newcommand{\mtwo}{m_2}
\newcommand{\mtre}{m_3}
\newcommand{\mfor}{m_4}
\newcommand{\mlone}{m}
\newcommand{\mloneb}{\bar{m}}
\newcommand{\mind}[1]{m_{#1}}
\newcommand{\mones}{m^2_1}
\newcommand{\mtwos}{m^2_2}
\newcommand{\mtres}{m^2_3}
\newcommand{\mfors}{m^2_4}
\newcommand{\mlones}{m^2}
\newcommand{\minds}[1]{m^2_{#1}}
\newcommand{\moneq}{m^4_1}
\newcommand{\mtwoq}{m^4_2}
\newcommand{\mtreq}{m^4_3}
\newcommand{\mforq}{m^4_4}
\newcommand{\mloneq}{m^4}
\newcommand{\mindq}[1]{m^4_{#1}}
\newcommand{\mlonev}{m^5}
\newcommand{\mindv}[1]{m^5_{#1}}
\newcommand{\monex}{m^6_1}
\newcommand{\mtwox}{m^6_2}
\newcommand{\mtrex}{m^6_3}
\newcommand{\mforx}{m^6_4}
\newcommand{\mlonex}{m^6}
\newcommand{\mindx}[1]{m^6_{#1}}
\newcommand{\Mone}{M_1}
\newcommand{\Mtwo}{M_2}
\newcommand{\Mtre}{M_3}
\newcommand{\Mfor}{M_4}
\newcommand{\Mlone}{M}
\newcommand{\Mlonep}{M'}
\newcommand{\Miind}{M_i}
\newcommand{\Mind}[1]{M_{#1}}
\newcommand{\Minds}[1]{M^2_{#1}}
\newcommand{\Mindc}[1]{M^3_{#1}}
\newcommand{\Mindf}[1]{M^4_{#1}}
\newcommand{\Mones}{M^2_1}
\newcommand{\Mtwos}{M^2_2}
\newcommand{\Mtres}{M^2_3}
\newcommand{\Mfors}{M^2_4}
\newcommand{\Mlones}{M^2}
\newcommand{\Mloneps}{M'^2}
\newcommand{\Miinds}{M^2_i}
\newcommand{\Mlonec}{M^3}
\newcommand{\Monec}{M^3_1}
\newcommand{\Mtwoc}{M^3_2}
\newcommand{\Moneq}{M^4_1}
\newcommand{\Mtwoq}{M^4_2}
\newcommand{\Mtreq}{M^4_3}
\newcommand{\Mforq}{M^4_4}
\newcommand{\Mloneq}{M^4}
\newcommand{\Miindq}{M^4_i}
\newcommand{\Monex}{M^6_1}
\newcommand{\Mtwox}{M^6_2}
\newcommand{\Mtrex}{M^6_3}
\newcommand{\Mforx}{M^6_4}
\newcommand{\Mlonex}{M^6}
\newcommand{\Miindx}{M^6_i}
\newcommand{\meb}{m_0}
\newcommand{\mebs}{m^2_0}
%
%
\newcommand{\Mq }{M_q  }
\newcommand{\MqS}{M^2_q}
\newcommand{\Ms }{M_s  }
\newcommand{\MsS}{M^2_s}
\newcommand{\Mc }{M_c  }
\newcommand{\McS}{M^2_c}
\newcommand{\Mb }{M_b  }
\newcommand{\MbS}{M^2_b}
\newcommand{\Mt }{M_t  }
\newcommand{\MtS}{M^2_t}
%
%
\newcommand{\mq}{m_q}
\newcommand{\mqs}{m^2_q}
\newcommand{\mqS}{m^2_q}
\newcommand{\mqQ}{m^4_q}
\newcommand{\mqX}{m^6_q}
\newcommand{\mqp}{m'_q }
\newcommand{\mqpS}{m'^2_q}
\newcommand{\mqpQ}{m'^4_q}
%
%
\newcommand{\lL}{l}
\newcommand{\ls}{l^2}
\newcommand{\LL}{L}
\newcommand{\LS}{L^2}
\newcommand{\LC}{L^3}
\newcommand{\LQ}{L^4}
\newcommand{\lw}{l_w}
\newcommand{\Lw}{L_w}
\newcommand{\Lws}{L^2_w}
\newcommand{\Lz}{L_z}
\newcommand{\Lzs}{L^2_z}
\newcommand{\Li}[1]{L_{#1}}
\newcommand{\Lis}[1]{L^2_{#1}}
\newcommand{\Lic}[1]{L^3_{#1}}
%
%
\newcommand{\sman}{s}
\newcommand{\tman}{t}
\newcommand{\uman}{u}
\newcommand{\smani}[1]{s_{#1}}
\newcommand{\bsmani}[1]{{\bar{s}}_{#1}}
\newcommand{\smans}{s^2}
\newcommand{\tmans}{t^2}
\newcommand{\umans}{u^2}
\newcommand{\shat}{{\hat s}}
\newcommand{\that}{{\hat t}}
\newcommand{\uhat}{{\hat u}}
%
%
\newcommand{\smanp}{s'}
\newcommand{\smanpi}[1]{s'_{#1}}
\newcommand{\tmanp}{t'}
\newcommand{\umanp}{u'}
\newcommand{\kappi}[1]{\kappa_{#1}}
\newcommand{\zetai}[1]{\zeta_{#1}}
%
%
%
\newcommand{\Phaspi}[1]{\Gamma_{#1}}
\newcommand{\rbetai}[1]{\beta_{#1}}
\newcommand{\ralphai}[1]{\alpha_{#1}}
\newcommand{\rbetais}[1]{\beta^2_{#1}}
\newcommand{\Lambdi}[1]{\Lambda_{#1}}
\newcommand{\Nomini}[1]{N_{#1}}
\newcommand{\smlone}{\frac{-\sman-\ib\ep}{\mlones}}
%
%
\newcommand{\theti}[1]{\theta_{#1}}
\newcommand{\delti}[1]{\delta_{#1}}
\newcommand{\phigi}[1]{\phi_{#1}}
\newcommand{\acoli}[1]{\xi_{#1}}
\newcommand{\scatc}{c}
\newcommand{\scatcs}{c^2}
\newcommand{\scatci}[1]{c_{#1}}
\newcommand{\scatcis}[1]{c^2_{#1}}
\newcommand{\scatct}[2]{c_{#1}^{#2}}
\newcommand{\angamt}[2]{\gamma_{#1}^{#2}}
%
%
\newcommand{\Regia}{{\cal{R}}}
\newcommand{\Iconi}[2]{{\cal{I}}_{#1}\lpar{#2}\rpar}
\newcommand{\sIcon}[1]{{\cal{I}}_{#1}}
\newcommand{\betaf}{\beta_{\ff}}
\newcommand{\Kfact}[2]{{\cal{K}}_{#1}\lpar{#2}\rpar}
%
%
\newcommand{\Struf}[4]{{\cal D}^{#1}_{#2}\lpar{#3;#4}\rpar}
\newcommand{\sStruf}[2]{{\cal D}^{#1}_{#2}}
\newcommand{\Fluxf}[2]{H\lpar{#1;#2}\rpar}
\newcommand{\Fluxfi}[4]{H_{#1}^{#2}\lpar{#3;#4}\rpar}
\newcommand{\sFluxf}{H}
\newcommand{\Bflux}[2]{{\cal{B}}_{#1}\lpar{#2}\rpar}
\newcommand{\bflux}[2]{{\cal{B}}_{#1}\lpar{#2}\rpar}
\newcommand{\Fluxd}[2]{D_{#1}\lpar{#2}\rpar}
\newcommand{\fluxd}[2]{C_{#1}\lpar{#2}\rpar}
\newcommand{\Fluxh}[4]{{\cal{H}}_{#1}^{#2}\lpar{#3;#4}\rpar}
\newcommand{\Sluxh}[4]{{\cal{S}}_{#1}^{#2}\lpar{#3;#4}\rpar}
\newcommand{\Fluxhb}[4]{{\overline{{\cal{H}}}}_{#1}^{#2}\lpar{#3;#4}\rpar}
\newcommand{\sFluxhb}{{\overline{{\cal{H}}}}}
\newcommand{\Sluxhb}[4]{{\overline{{\cal{S}}}}_{#1}^{#2}\lpar{#3;#4}\rpar}
\newcommand{\sSluxhb}[2]{{\overline{{\cal{S}}}}_{#1}^{#2}}
\newcommand{\fluxh}[4]{h_{#1}^{#2}\lpar{#3;#4}\rpar}
\newcommand{\fluxhs}[3]{h_{#1}^{#2}\lpar{#3}\rpar}
\newcommand{\sfluxhs}[2]{h_{#1}^{#2}}
\newcommand{\fluxhb}[4]{{\overline{h}}_{#1}^{#2}\lpar{#3;#4}\rpar}
\newcommand{\Strufd}[2]{D\lpar{#1;#2}\rpar}
%
%
\newcommand{\rMQ}[1]{r^2_{#1}}
\newcommand{\rMQs}[1]{r^4_{#1}}
\newcommand{\rf}{w_{\ff}}
\newcommand{\zf}{z_{\ff}}
\newcommand{\rfs}{w^2_{\ff}}
\newcommand{\zfs}{z^2_{\ff}}
\newcommand{\rfc}{w^3_{\ff}}
\newcommand{\zfc}{z^3_{\ff}}
\newcommand{\df}{d_{\ff}}
\newcommand{\rfp}{w_{\ffp}}
\newcommand{\rfps}{w^2_{\ffp}}
\newcommand{\rfpc}{w^3_{\ffp}}
\newcommand{\rt}{w_{\ft}}
\newcommand{\rts}{w^2_{\ft}}
\newcommand{\dt}{d_{\ft}}
\newcommand{\dts}{d^2_{\ft}}
\newcommand{\rh}{r_{h}}
\newcommand{\Lnrt}{\ln{\rt}}
\newcommand{\Rw}{R_{_{\wb}}}
\newcommand{\Rws}{R^2_{_{\wb}}}
\newcommand{\Rz}{R_{_{\zb}}}
\newcommand{\Rzp}{R^{+}_{_{\zb}}}
\newcommand{\Rzm}{R^{-}_{_{\zb}}}
\newcommand{\Rzs}{R^2_{_{\zb}}}
\newcommand{\Rzc}{R^3_{_{\zb}}}
\newcommand{\Rv}{R_{_{\vb}}}
\newcommand{\rhw}{r_{_{\wb}}}
\newcommand{\rhz}{r_{_{\zb}}}
\newcommand{\rhws}{r^2_{_{\wb}}}
\newcommand{\rhzs}{r^2_{_{\zb}}}
%
%
\newcommand{\vqrato}{z}
\newcommand{\vqrats}{w}
\newcommand{\vqratq}{w^2}
\newcommand{\seyrat}{z}
\newcommand{\sexrat}{w}
\newcommand{\sehrat}{h}
\newcommand{\sewrat}{w}
\newcommand{\sezrat}{z}
\newcommand{\zetav}{\zeta}
\newcommand{\zetavi}[1]{\zeta_{#1}}
\newcommand{\bpo}{\beta^2}
\newcommand{\bpos}{\beta^4}
\newcommand{\bpt}{{\tilde\beta}^2}
\newcommand{\lap}{\kappa}
\newcommand{\hw}{h_{_{\wb}}}
\newcommand{\hz}{h_{_{\zb}}}
%
%
\newcommand{\ec}{e}
\newcommand{\ecs}{e^2}
\newcommand{\ect}{e^3}
\newcommand{\ecq}{e^4}
\newcommand{\ecb}{e_{_0}}
\newcommand{\ecbs}{e^2_{_0}}
\newcommand{\ecbq}{e^4_{_0}}
\newcommand{\eci}[1]{e_{#1}}
\newcommand{\ecis}[1]{e^2_{#1}}
\newcommand{\hate}{{\hat e}}
\newcommand{\gss}{g_{_S}}
\newcommand{\gsss}{g^2_{_S}}
\newcommand{\gssb}{g^2_{_{S_0}}}
\newcommand{\als}{\alpha_{_S}}
\newcommand{\as}{a_{_S}}
\newcommand{\ass}{a^2_{_S}}
\newcommand{\gf}{G_{\ssF}}
\newcommand{\gfs}{G^2_{\ssF}}
\newcommand{\gb}{g} 
\newcommand{\gbi}[1]{g_{#1}}
\newcommand{\gbb}{g_{0}}
\newcommand{\gbs}{g^2}
\newcommand{\gbc}{g^3}
\newcommand{\gbf}{g^4}
\newcommand{\gpb}{g'}
\newcommand{\gpbs}{g'^2}
\newcommand{\vc}[1]{v_{#1}}
\newcommand{\ac}[1]{a_{#1}}
\newcommand{\hatv}[1]{{\hat v}_{#1}}
\newcommand{\vcs}[1]{v^2_{#1}}
\newcommand{\acs}[1]{a^2_{#1}}
\newcommand{\gcp}[1]{g^{+}_{#1}}
\newcommand{\gcm}[1]{g^{-}_{#1}}
\newcommand{\vci}[2]{v^{#2}_{#1}}
\newcommand{\aci}[2]{a^{#2}_{#1}}
\newcommand{\vceff}[1]{v^{#1}_{\rm{eff}}}
\newcommand{\hvc}[1]{\hat{v}_{#1}}
\newcommand{\hvcs}[1]{\hat{v}^2_{#1}}
\newcommand{\Vc}[1]{V_{#1}}
\newcommand{\Ac}[1]{A_{#1}}
\newcommand{\Vcs}[1]{V^2_{#1}}
\newcommand{\Acs}[1]{A^2_{#1}}
\newcommand{\vpa}[2]{\sigma_{#1}^{#2}}
\newcommand{\vma}[2]{\delta_{#1}^{#2}}
\newcommand{\vfw}{\sigma^{a}_{\ff}}
\newcommand{\vfpw}{\sigma^{a}_{\ffp}}
\newcommand{\vfwi}[1]{\sigma^{a}_{#1}}
\newcommand{\vfwsi}[1]{\lpar\sigma^{a}_{#1}\rpar^2}
\newcommand{\vvfw}{\sigma^{a}_{\ff}}
\newcommand{\vvew}{\sigma^{a}_{\fe}}
\newcommand{\gv}{g_{_V}}
\newcommand{\ga}{g_{_A}}
\newcommand{\gve}{g^{\fe}_{_{V}}}
\newcommand{\gae}{g^{\fe}_{_{A}}}
\newcommand{\gvf}{g^{\ff}_{_{V}}}
\newcommand{\gaf}{g^{\ff}_{_{A}}}
\newcommand{\gvaf}{g^{\ff}_{_{V,A}}}
\newcommand{\phenGV}[1]{{\cal{G}}^{#1}_{_V}}
\newcommand{\phenGA}[1]{{\cal{G}}^{#1}_{_A}}
\newcommand{\sGv}{{\cal{G}}_{_V}}
\newcommand{\cGa}{{\cal{G}}^{*}_{_A}}
\newcommand{\cGv}{{\cal{G}}^{*}_{_V}}
\newcommand{\sGa}{{\cal{G}}_{_A}}
\newcommand{\Gvf}{{\cal{G}}^{\ff}_{_{V}}}
\newcommand{\Gaf}{{\cal{G}}^{\ff}_{_{A}}}
\newcommand{\Gvaf}{{\cal{G}}^{\ff}_{_{V,A}}}
\newcommand{\Gve}{{\cal{G}}^{\fe}_{_{V}}}
\newcommand{\Gae}{{\cal{G}}^{\fe}_{_{A}}}
\newcommand{\Gvae}{{\cal{G}}^{\fe}_{_{V,A}}}
\newcommand{\gvl}{g^{\fl}_{_{V}}}
\newcommand{\gal}{g^{\fl}_{_{A}}}
\newcommand{\gval}{g^{\fl}_{_{V,A}}}
\newcommand{\gvb}{g^{\ffb}_{_{V}}}
\newcommand{\gab}{g^{\ffb}_{_{A}}}
\newcommand{\fvf}{F_{_V}^{\ff}}
\newcommand{\faf}{F_{_A}^{\ff}}
\newcommand{\fvl}{F_{_V}^{\fl}}
\newcommand{\fal}{F_{_A}^{\fl}}
\newcommand{\corat}{\kappa}
\newcommand{\corats}{\kappa^2}
%
%
\newcommand{\dr}{\Delta r}
\newcommand{\drl}{\Delta r_{_L}}
\newcommand{\drh}{\Delta{\hat r}}
\newcommand{\drhw}{\Delta{\hat r}_{_W}}
\newcommand{\rhou}{\rho_{_U}}
\newcommand{\rhoz}{\rho_{_\zb}}
\newcommand{\rZ}{\rho_{_\zb}}
\newcommand{\rhob}{\rho_{_0}}
\newcommand{\rZf}{\rho^{\ff}_{_\zb}}
\newcommand{\kZf}{\kappa^{\ff}_{_\zb}}
\newcommand{\rWf}{\rho^{\ff}_{_\wb}}
\newcommand{\brWf}{{\bar{\rho}}^{\ff}_{_\wb}}
\newcommand{\rHf}{\rho^{\ff}_{_\hb}}
\newcommand{\brHf}{{\bar{\rho}}^{\ff}_{_\hb}}
\newcommand{\rhoR}{\rho^R_{_{\zb}}}
\newcommand{\hatrh}{{\hat\rho}}
\newcommand{\ku}{\kappa_{_U}}
\newcommand{\rZdf}[1]{\rho^{#1}_{_\zb}}
\newcommand{\kZdf}[1]{\kappa^{#1}_{_\zb}}
\newcommand{\rdfL}[1]{\rho^{#1}_{_L}}
\newcommand{\kdfL}[1]{\kappa^{#1}_{_L}}
\newcommand{\rdfR}[1]{\rho^{#1}_{\rm{rem}}}
\newcommand{\kdfR}[1]{\kappa^{#1}_{\rm{rem}}}
\newcommand{\bark}{\overline\kappa}
%
%
\newcommand{\stw}{s_{\theta}}             
\newcommand{\ctw}{c_{\theta}}
\newcommand{\stws}{s_{\theta}^2}
\newcommand{\stwc}{s_{\theta}^3}
\newcommand{\stwf}{s_{\theta}^4}
\newcommand{\stwx}{s_{\theta}^6}
\newcommand{\ctws}{c_{\theta}^2}
\newcommand{\ctwc}{c_{\theta}^3}
\newcommand{\ctwf}{c_{\theta}^4}
\newcommand{\ctwx}{c_{\theta}^6}
\newcommand{\stwfiv}{s_{\theta}^5}
\newcommand{\ctwfiv}{c_{\theta}^5}
\newcommand{\stwsix}{s_{\theta}^6}
\newcommand{\ctwsix}{c_{\theta}^6}
%
%
\newcommand{\siw}{s_{_W}}           
\newcommand{\cow}{c_{_W}}
\newcommand{\siws}{s^2_{_W}}
\newcommand{\cows}{c^2_{_W}}
\newcommand{\siwc}{s^3_{_W}}
\newcommand{\cowc}{c^3_{_W}}
\newcommand{\siwf}{s^4_{_W}}
\newcommand{\cowf}{c^4_{_W}}
\newcommand{\siwx}{s^6_{_W}}
\newcommand{\cowx}{c^6_{_W}}
\newcommand{\sons}{s_{_W}}
\newcommand{\sonss}{s^2_{_W}}
\newcommand{\cons}{c_{_W}}
\newcommand{\cooss}{c^2_{_W}}
%
%
\newcommand{\szs}{{\overline s}^2}
\newcommand{\szq}{{\overline s}^4}
\newcommand{\czs}{{\overline c}^2}
\newcommand{\sbs}{s_{_0}^2}
\newcommand{\cbs}{c_{_0}^2}
\newcommand{\dss}{\Delta s^2}
\newcommand{\snes}{s_{\nu e}^2}
\newcommand{\cnes}{c_{\nu e}^2}
\newcommand{\shs}{{\hat s}^2}
\newcommand{\chs}{{\hat c}^2}
\newcommand{\chl}{{\hat c}}
\newcommand{\seffs}{s^2_{\rm{eff}}}
\newcommand{\seffsf}[1]{\sin^2\theta^{#1}_{\rm{eff}}}
\newcommand{\sress}{s^2_{\rm res}}                
\newcommand{\sR}{s_{_R}}
\newcommand{\sRs}{s^2_{_R}}
\newcommand{\ctwe}{c_{\theta}^6}
\newcommand{\sany}{s}
\newcommand{\cany}{c}
\newcommand{\sanys}{s^2}
\newcommand{\canys}{c^2}
%
%
\newcommand{\sip}{u}                             
\newcommand{\siap}{{\bar{v}}}                    
\newcommand{\sop}{{\bar{u}}}                     
\newcommand{\soap}{v}                            
\newcommand{\ip}[1]{u\lpar{#1}\rpar}             
\newcommand{\iap}[1]{{\bar{v}}\lpar{#1}\rpar}    
\newcommand{\op}[1]{{\bar{u}}\lpar{#1}\rpar}     
\newcommand{\oap}[1]{v\lpar{#1}\rpar}            
%
%
\newcommand{\ipp}[2]{u\lpar{#1,#2}\rpar}         
\newcommand{\ipap}[2]{{\bar v}\lpar{#1,#2}\rpar} 
\newcommand{\opp}[2]{{\bar u}\lpar{#1,#2}\rpar}  
\newcommand{\opap}[2]{v\lpar{#1,#2}\rpar}        
\newcommand{\upspi}[1]{u\lpar{#1}\rpar}
\newcommand{\vpspi}[1]{v\lpar{#1}\rpar}
\newcommand{\wpspi}[1]{w\lpar{#1}\rpar}
\newcommand{\ubpspi}[1]{{\bar{u}}\lpar{#1}\rpar}
\newcommand{\vbpspi}[1]{{\bar{v}}\lpar{#1}\rpar}
\newcommand{\wbpspi}[1]{{\bar{w}}\lpar{#1}\rpar}
\newcommand{\udpspi}[1]{u^{\dagger}\lpar{#1}\rpar}
\newcommand{\vdpspi}[1]{v^{\dagger}\lpar{#1}\rpar}
\newcommand{\wdpspi}[1]{w^{\dagger}\lpar{#1}\rpar}
\newcommand{\Ubilin}[1]{U\lpar{#1}\rpar}
\newcommand{\Vbilin}[1]{V\lpar{#1}\rpar}
\newcommand{\Xbilin}[1]{X\lpar{#1}\rpar}
\newcommand{\Ybilin}[1]{Y\lpar{#1}\rpar}
\newcommand{\up}[2]{u_{#1}\lpar #2\rpar}
\newcommand{\vp}[2]{v_{#1}\lpar #2\rpar}
\newcommand{\ubp}[2]{{\overline u}_{#1}\lpar #2\rpar}
\newcommand{\vbp}[2]{{\overline v}_{#1}\lpar #2\rpar}
\newcommand{\Pje}[1]{\frac{1}{2}\lpar 1 + #1\,\gfd\rpar}
\newcommand{\Pj}[1]{\Pi_{#1}}
\newcommand{\trace}{\mbox{Tr}}
%
%
\newcommand{\Poper}[2]{P_{#1}\lpar{#2}\rpar}
\newcommand{\Loper}[2]{\Lambda_{#1}\lpar{#2}\rpar}
\newcommand{\proj}[3]{P_{#1}\lpar{#2,#3}\rpar}
\newcommand{\sproj}[1]{P_{#1}}
\newcommand{\Nden}[3]{N_{#1}^{#2}\lpar{#3}\rpar}
\newcommand{\sNden}[1]{N_{#1}}
\newcommand{\nden}[2]{n_{#1}^{#2}}
%
%
\newcommand{\vwf}[2]{e_{#1}\lpar#2\rpar}             
\newcommand{\vwfb}[2]{{\overline e}_{#1}\lpar#2\rpar}
\newcommand{\pwf}[2]{\epsilon_{#1}\lpar#2\rpar}      
\newcommand{\sla}[1]{/\!\!\!#1}
\newcommand{\slac}[1]{/\!\!\!\!#1}
%
%
\newcommand{\iemom}{p_{_-}}                    
\newcommand{\ipmom}{p_{_+}}
\newcommand{\oemom}{q_{_-}}                    
\newcommand{\opmom}{q_{_+}}
%
%
\newcommand{\spro}[2]{{#1}\cdot{#2}}
%
%
\newcommand{\gfour}{\gamma_4}                    
\newcommand{\gfd}{\gamma_5}                    
\newcommand{\gap}{\lpar 1+\gamma_5\rpar}
\newcommand{\gam}{\lpar 1-\gamma_5\rpar}
\newcommand{\gdp}{\gamma_+}
\newcommand{\gdm}{\gamma_-}
\newcommand{\gdpm}{\gamma_{\pm}}
\newcommand{\gad}{\gamma}
\newcommand{\gapm}{\lpar 1\pm\gamma_5\rpar}
\newcommand{\gadi}[1]{\gamma_{#1}}
\newcommand{\gadu}[1]{\gamma_{#1}}
\newcommand{\gapu}[1]{\gamma^{#1}}
\newcommand{\sigd}[2]{\sigma_{#1#2}}
\newcommand{\sumsp}{\sum_{\mbox{spins}}}
%
%
\newcommand{\li}[2]{\mathrm{Li}_{#1}\lpar\displaystyle{#2}\rpar} 
\newcommand{\etaf}[2]{\eta\lpar#1,#2\rpar}
\newcommand{\lkall}[3]{\lambda\lpar#1,#2,#3\rpar}       
\newcommand{\slkall}[3]{\lambda^{1/2}\lpar#1,#2,#3\rpar}
\newcommand{\egam}[1]{\Gamma\lpar#1\rpar}               
\newcommand{\ebe}[2]{B\lpar#1,#2\rpar}                  
\newcommand{\ddel}[1]{\delta\lpar#1\rpar}               
\newcommand{\drii}[2]{\delta_{#1#2}}                    
\newcommand{\driv}[4]{\delta_{#1#2#3#4}}                
\newcommand{\intmomi}[2]{\int\,d^{#1}#2}
\newcommand{\intmomii}[3]{\int\,d^{#1}#2\,\int\,d^{#1}#3}
\newcommand{\intfx}[1]{\int_{\scriptstyle 0}^{\scriptstyle 1}\,d#1}
\newcommand{\intfxy}[2]{\int_{\scriptstyle 0}^{\scriptstyle 1}\,d#1\,
                        \int_{\scriptstyle 0}^{\scriptstyle #1}\,d#2}
\newcommand{\intfxyz}[3]{\int_{\scriptstyle 0}^{\scriptstyle 1}\,d#1\,
                         \int_{\scriptstyle 0}^{\scriptstyle #1}\,d#2\,
                         \int_{\scriptstyle 0}^{\scriptstyle #2}\,d#3}
\newcommand{\Beta}[2]{{\rm{B}}\lpar #1,#2\rpar}
\newcommand{\sBeta}{\rm{B}}
\newcommand{\sign}[1]{{\rm{sign}}\lpar{#1}\rpar}
%
%
\newcommand{\gn}{\Gamma_{\nu}}
\newcommand{\gel}{\Gamma_{\fe}}
\newcommand{\gmu}{\Gamma_{\mu}}
\newcommand{\gff}{\Gamma_{\ff}}
\newcommand{\gt}{\Gamma_{\tau}}
\newcommand{\gl}{\Gamma_{\fl}}
\newcommand{\gq}{\Gamma_{q}}
\newcommand{\gu}{\Gamma_{u}}
\newcommand{\gd}{\Gamma_{d}}
\newcommand{\gc}{\Gamma_{c}}
\newcommand{\gs}{\Gamma_{s}}
\newcommand{\gbq}{\Gamma_{b}}
\newcommand{\gz}{\Gamma_{_{\zb}}}
\newcommand{\gw}{\Gamma_{_{\wb}}}
\newcommand{\gh}{\Gamma_{_{\hb}}}
\newcommand{\gi}{\Gamma_{\rm{inv}}}
\newcommand{\gzs}{\Gamma^2_{_{\zb}}}
%
%
\newcommand{\tcie}{I^{(3)}_{\fe}}
\newcommand{\tcim}{I^{(3)}_{\flm}}
\newcommand{\tcif}{I^{(3)}_{\ff}}
\newcommand{\tciq}{I^{(3)}_{\fq}}
\newcommand{\tcib}{I^{(3)}_{\ffb}}
\newcommand{\tcih}{I^{(3)}_h}
\newcommand{\tcii}{I^{(3)}_i}
\newcommand{\tcift}{I^{(3)}_{\tilde f}}
\newcommand{\tcifp}{I^{(3)}_{f'}}
\newcommand{\wispt}[1]{I^{(3)}_{#1}}
\newcommand{\ql}{Q_l}
\newcommand{\qe}{Q_e}
\newcommand{\qu}{Q_u}
\newcommand{\qd}{Q_d}
\newcommand{\qb}{Q_b}
\newcommand{\qt}{Q_t}
\newcommand{\qup}{Q'_u}
\newcommand{\qdp}{Q'_d}
\newcommand{\qmu}{Q_{\mu}}
\newcommand{\qes}{Q^2_e}
\newcommand{\qec}{Q^3_e}
\newcommand{\qus}{Q^2_u}
\newcommand{\qds}{Q^2_d}
\newcommand{\qbs}{Q^2_b}
\newcommand{\qts}{Q^2_t}
\newcommand{\qbc}{Q^3_b}
\newcommand{\qf}{Q_f}
\newcommand{\qfs}{Q^2_f}
\newcommand{\qfc}{Q^3_f}
\newcommand{\qff}{Q^4_f}
\newcommand{\qep}{Q_{e'}}
\newcommand{\qfp}{Q_{f'}}
\newcommand{\qfps}{Q^2_{f'}}
\newcommand{\qfpc}{Q^3_{f'}}
\newcommand{\qq}{Q_q}
\newcommand{\qqs}{Q^2_q}
\newcommand{\qi}{Q_i}
\newcommand{\qis}{Q^2_i}
\newcommand{\qj}{Q_j}
\newcommand{\qjs}{Q^2_j}
\newcommand{\QW}{Q_{_\wb}}
\newcommand{\QWs}{Q^2_{_\wb}}
\newcommand{\Qd}{Q_d}
\newcommand{\Qds}{Q^2_d}
\newcommand{\Qu}{Q_u}
\newcommand{\Qus}{Q^2_u}
\newcommand{\vi}{v_i}
\newcommand{\vis}{v^2_i}
\newcommand{\ai}{a_i}
\newcommand{\ais}{a^2_i}
%
%
\newcommand{\piv}{\Pi_{_V}}
\newcommand{\pia}{\Pi_{_A}}
\newcommand{\piva}{\Pi_{_{V,A}}}
\newcommand{\pivi}[1]{\Pi^{({#1})}_{_V}}
\newcommand{\piai}[1]{\Pi^{({#1})}_{_A}}
\newcommand{\pivai}[1]{\Pi^{({#1})}_{_{V,A}}}
\newcommand{\pih}{{\hat\Pi}}
\newcommand{\sgh}{{\hat\Sigma}}
\newcommand{\Pgg}{\Pi_{\ph\ph}}
\newcommand{\Ptg}{\Pi_{_{3Q}}}
\newcommand{\Ptt}{\Pi_{_{33}}}
\newcommand{\Pzg}{\Pi_{_{\zb\ab}}}
\newcommand{\Pzga}[2]{\Pi^{#1}_{_{\zb\ab}}\lpar#2\rpar}
\newcommand{\Pf}{\Pi_{_F}}
\newcommand{\Sgg}{\Sigma_{_{\ab\ab}}}
\newcommand{\Szg}{\Sigma_{_{\zb\ab}}}
\newcommand{\SVV}{\Sigma_{_{\vb\vb}}}
\newcommand{\USvv}{{\hat\Sigma}_{_{\vb\vb}}}
\newcommand{\Sww}{\Sigma_{_{\wb\wb}}}
\newcommand{\Swwg}{\Sigma^{_G}_{_{\wb\wb}}}
\newcommand{\Szz}{\Sigma_{_{\zb\zb}}}
\newcommand{\Shh}{\Sigma_{_{\hb\hb}}}
\newcommand{\Spzz}{\Sigma'_{_{\zb\zb}}}
\newcommand{\Stg}{\Sigma_{_{3Q}}}
\newcommand{\Stt}{\Sigma_{_{33}}}
\newcommand{\bSww}{{\overline\Sigma}_{_{WW}}}
\newcommand{\bStg}{{\overline\Sigma}_{_{3Q}}}
\newcommand{\bStt}{{\overline\Sigma}_{_{33}}}
\newcommand{\Sssn}{\Sigma_{_{\hkn\hkn}}}
\newcommand{\Sssc}{\Sigma_{_{\phi\phi}}}
\newcommand{\Szn}{\Sigma_{_{\zb\hkn}}}
\newcommand{\Swc}{\Sigma_{_{\wb\hkg}}}
\newcommand{\mix}[2]{{\cal{M}}^{#1}\lpar{#2}\rpar}
\newcommand{\bmix}[2]{\Pi^{{#1},F}_{_{\zb\ab}}\lpar{#2}\rpar}
\newcommand{\hPgg}[2]{{\hat{\Pi}^{{#1},F}}_{_{\ph\ph}}\lpar{#2}\rpar}
\newcommand{\hmix}[2]{{\hat{\Pi}^{{#1},F}}_{_{\zb\ab}}\lpar{#2}\rpar}
\newcommand{\Dz}[2]{{\cal{D}}_{_{\zb}}^{#1}\lpar{#2}\rpar}
\newcommand{\bDz}[2]{{\cal{D}}^{{#1},F}_{_{\zb}}\lpar{#2}\rpar}
\newcommand{\hDz}[2]{{\hat{\cal{D}}}^{{#1},F}_{_{\zb}}\lpar{#2}\rpar}
\newcommand{\Szzd}[2]{\Sigma'^{#1}_{_{\zb\zb}}\lpar{#2}\rpar}
\newcommand{\Swwd}[2]{\Sigma'^{#1}_{_{\wb\wb}}\lpar{#2}\rpar}
\newcommand{\Shhd}[2]{\Sigma'^{#1}_{_{\hb\hb}}\lpar{#2}\rpar}
\newcommand{\ZFren}[2]{{\cal{Z}}^{#1}\lpar{#2}\rpar}
\newcommand{\WFren}[2]{{\cal{W}}^{#1}\lpar{#2}\rpar}
\newcommand{\HFren}[2]{{\cal{H}}^{#1}\lpar{#2}\rpar}
%
%
\newcommand{\cf}{c_f}
\newcommand{\Cf}{C_{_F}}
\newcommand{\Nf}{N_f}
\newcommand{\Nc}{N_c}
\newcommand{\Ncs}{N^2_c}
\newcommand{\nf }{n_f}
\newcommand{\nfs}{n^2_f}
\newcommand{\nfc}{n^3_f}
\newcommand{\MSB}{\overline{MS}}
\newcommand{\LMSB}{\Lambda_{\overline{\mathrm{MS}}}}
\newcommand{\LMSBp}{\Lambda'_{\overline{\mathrm{MS}}}}
\newcommand{\LMSBS}{\Lambda^2_{\overline{\mathrm{MS}}}}
\newcommand{\LMSBv }{\mbox{$\Lambda^{(5)}_{\overline{\mathrm{MS}}}$}}
\newcommand{\LMSBvS}{\mbox{$\left(\Lambda^{(5)}_{\overline{\mathrm{MS}}}\right)^2$}}
\newcommand{\LMSBt }{\mbox{$\Lambda^{(3)}_{\overline{\mathrm{MS}}}$}}
\newcommand{\LMSBtS}{\mbox{$\left(\Lambda^{(3)}_{\overline{\mathrm{MS}}}\right)^2$}}
\newcommand{\LMSBf }{\mbox{$\Lambda^{(4)}_{\overline{\mathrm{MS}}}$}}
\newcommand{\LMSBfS}{\mbox{$\left(\Lambda^{(4)}_{\overline{\mathrm{MS}}}\right)^2$}}
\newcommand{\LMSBn }{\mbox{$\Lambda^{(\nf)}_{\overline{\mathrm{MS}}}$}}
\newcommand{\LMSBnS}{\mbox{$\left(\Lambda^{(\nf)}_{\overline{\mathrm{MS}}}\right)^2$}}
\newcommand{\LMSBnml }{\mbox{$\Lambda^{(\nf-1)}_{\overline{\mathrm{MS}}}$}}
\newcommand{\LMSBnmlS}{\mbox{$\left(\Lambda^{(\nf-1)}_{\overline{\mathrm{MS}}}\right)^2$}}
\newcommand{\Bnf}{\lpar\nf \rpar}
\newcommand{\Bnfm}{\lpar\nf-1 \rpar}
\newcommand{\LuM}{L_{_M}}
\newcommand{\bef}{\beta_{\ff}}
\newcommand{\befs}{\beta^2_{\ff}}
\newcommand{\befc}{\beta^3_{f}}
\newcommand{\alsp}{\alpha'_{_S}}
\newcommand{\api}{\displaystyle \frac{\als(s)}{\pi}}
\newcommand{\alss}{\alpha^2_{_S}}
\newcommand{\ztwo}{\zeta(2)}
\newcommand{\ztri}{\zeta(3)}
\newcommand{\zfor}{\zeta(4)}
\newcommand{\zfiv}{\zeta(5)}
\newcommand{\bi}[1]{b_{#1}}
\newcommand{\ci}[1]{c_{#1}}
\newcommand{\Ci}[1]{C_{#1}}
\newcommand{\bip}[1]{b'_{#1}}
\newcommand{\cip}[1]{c'_{#1}}
%
%
\newcommand{\osps}{16\,\pi^2}
\newcommand{\srt}{\sqrt{2}}
\newcommand{\ospsi}{\displaystyle{\frac{i}{16\,\pi^2}}}
%
%
\newcommand{\tfpromu}{\mbox{$e^+e^-\to \mu^+\mu^-$}}
\newcommand{\tfprotau}{\mbox{$e^+e^-\to \tau^+\tau^-$}}
\newcommand{\tfproe}{\mbox{$e^+e^-\to e^+e^-$}}
\newcommand{\tfpronu}{\mbox{$e^+e^-\to \barnu\nu$}}
\newcommand{\tfproqq}{\mbox{$e^+e^-\to \barq q$}}
\newcommand{\tfprohad}{\mbox{$e^+e^-\to\,$} hadrons}
%
%
\newcommand{\bpromu}{\mbox{$e^+e^-\to \mu^+\mu^-\ph$}}
\newcommand{\bprotau}{\mbox{$e^+e^-\to \tau^+\tau^-\ph$}}
\newcommand{\bproe}{\mbox{$e^+e^-\to e^+e^-\ph$}}
\newcommand{\bpronu}{\mbox{$e^+e^-\to \barnu\nu\ph$}}
\newcommand{\bproqq}{\mbox{$e^+e^-\to \barq q \ph$}}
%
%
\newcommand{\tbprow} {\mbox{$e^+e^-\to \wbp \wbm $}}
\newcommand{\tbproz} {\mbox{$e^+e^-\to \zb  \zb  $}}
\newcommand{\tbproh} {\mbox{$e^+e^-\to \zb  \hb  $}}
\newcommand{\tbprozg}{\mbox{$e^+e^-\to \zb  \ph  $}}
\newcommand{\tbprog} {\mbox{$e^+e^-\to \ph  \ph  $}}
%
%
\newcommand{\Fermionline}[1]{
\vcenter{\hbox{
  \begin{picture}(60,20)(0,{#1})
  \SetScale{2.}
    \ArrowLine(0,5)(30,5)
  \end{picture}}}
}
\newcommand{\AntiFermionline}[1]{
\vcenter{\hbox{
  \begin{picture}(60,20)(0,{#1})
  \SetScale{2.}
    \ArrowLine(30,5)(0,5)
  \end{picture}}}
}
\newcommand{\Photonline}[1]{
\vcenter{\hbox{
  \begin{picture}(60,20)(0,{#1})
  \SetScale{2.}
    \Photon(0,5)(30,5){2}{6.5}
  \end{picture}}}
}
\newcommand{\Gluonline}[1]{
\vcenter{\hbox{
  \begin{picture}(60,20)(0,{#1})
  \SetScale{2.}
    \Gluon(0,5)(30,5){2}{6.5}
  \end{picture}}}
}
\newcommand{\Wbosline}[1]{
\vcenter{\hbox{
  \begin{picture}(60,20)(0,{#1})
  \SetScale{2.}
    \Photon(0,5)(30,5){2}{4}
    \ArrowLine(13.3,3.1)(16.9,7.2)
  \end{picture}}}
}
\newcommand{\Zbosline}[1]{
\vcenter{\hbox{
  \begin{picture}(60,20)(0,{#1})
  \SetScale{2.}
    \Photon(0,5)(30,5){2}{4}
  \end{picture}}}
}
\newcommand{\Philine}[1]{
\vcenter{\hbox{
  \begin{picture}(60,20)(0,{#1})
  \SetScale{2.}
    \DashLine(0,5)(30,5){2}
  \end{picture}}}
}
\newcommand{\Phicline}[1]{
\vcenter{\hbox{
  \begin{picture}(60,20)(0,{#1})
  \SetScale{2.}
    \DashLine(0,5)(30,5){2}
    \ArrowLine(14,5)(16,5)
  \end{picture}}}
}
\newcommand{\Ghostline}[1]{
\vcenter{\hbox{
  \begin{picture}(60,20)(0,{#1})
  \SetScale{2.}
    \DashLine(0,5)(30,5){.5}
    \ArrowLine(14,5)(16,5)
  \end{picture}}}
}
%
%
\newcommand{\gauge}{g}
\newcommand{\gpar}{\xi}
\newcommand{\gparA}{\xi_{_A}}
\newcommand{\gparZ}{\xi_{_Z}}
\newcommand{\gpari}[1]{\gpar_{#1}}
\newcommand{\gparis}[1]{\gpar^2_{#1}}
\newcommand{\gpariq}[1]{\gpar^4_{#1}}
\newcommand{\gpars}{\xi^2}
\newcommand{\dgpar}{\Delta\gpar}
\newcommand{\dgparA}{\Delta\gparA}
\newcommand{\dgparZ}{\Delta\gparZ}
\newcommand{\gparq}{\xi^4}
\newcommand{\gparAs}{\xi^2_{_A}}
\newcommand{\gparAq}{\xi^4_{_A}}
\newcommand{\gparZs}{\xi^2_{_Z}}
\newcommand{\gparZq}{\xi^4_{_Z}}
\newcommand{\Rxi}{R_{\gpar}}
\newcommand{\hxi}{\chi}
%
%
\newcommand{\LSM}{{\cal{L}}_{_{\rm{SM}}}}
\newcommand{\LSMr}{{\cal{L}}^{\rm{ren}}_{_{\rm{SM}}}}
\newcommand{\LYM}{{\cal{L}}_{_{YM}}}
\newcommand{\Lzer}{{\cal{L}}_{_{0}}}
\newcommand{\Lone}{{\cal{L}}^{{\bos},I}}
\newcommand{\Lpro}{{\cal{L}}_{\rm{prop}}}
\newcommand{\Ls  }{{\cal{L}}_{_{S}}}
\newcommand{\Lsi }{{\cal{L}}^{I}_{_{S}}}
\newcommand{\Lgf }{{\cal{L}}_{gf  }}
\newcommand{\Lgfi}{{\cal{L}}^{I}_{gf}}
\newcommand{\Lf  }{{\cal{L}}^{{\fer},I}_{\ssV}}
\newcommand{\LHf }{{\cal{L}}^{\fer}_{\ssS}}
\newcommand{\LHfi}{{\cal{L}}^{{\fer},I}_{\ssS}}
\newcommand{\Lren}{{\cal{L}}_{\rm{ren}}}
\newcommand{\Lct}{{\cal{L}}_{\rm{ct}}}
\newcommand{\Lcti}[1]{{\cal{L}}^{#1}_{\rm{ct}}}
\newcommand{\LctI}{{\cal{L}}^{(2)}_{\rm{ct}}}
\newcommand{\Llone}{{\cal{L}}}
\newcommand{\LQED}{{\cal{L}}_{_{\rm{QED}}}}
\newcommand{\LQEDr}{{\cal{L}}^{\rm{ren}}_{_{\rm{QED}}}}
\newcommand{\FST}[3]{F_{#1#2}^{#3}}
\newcommand{\cD}[1]{D_{#1}}
\newcommand{\pd}[1]{\partial_{#1}}
\newcommand{\tgen}[1]{\tau^{#1}}
\newcommand{\gbl}{g_1}
\newcommand{\lctt}[3]{\varepsilon_{#1#2#3}}
\newcommand{\lctf}[4]{\varepsilon_{#1#2#3#4}}
\newcommand{\lctfb}[4]{\varepsilon\lpar{#1#2#3#4}\rpar}
\newcommand{\slct}{\varepsilon}
\newcommand{\cgfi}[1]{{\cal{C}}^{#1}}
\newcommand{\cgfZ}{{\cal{C}}_{_Z}}
\newcommand{\cgfA}{{\cal{C}}_{_A}}
\newcommand{\cgfZs}{{\cal{C}}^2_{_Z}}
\newcommand{\cgfAs}{{\cal{C}}^2_{_A}}
\newcommand{\hpms}{\mu^2}
\newcommand{\hpal}{\alpha_{_H}}
\newcommand{\hpals}{\alpha^2_{_H}}
\newcommand{\hpbe}{\beta_{_H}}
\newcommand{\hpbep}{\beta^{'}_{_H}}
\newcommand{\hpla}{\lambda}
\newcommand{\hpalf}{\alpha_{f}}
\newcommand{\hpbef}{\beta_{f}}
\newcommand{\tpar}[1]{\Lambda^{#1}}
\newcommand{\Mop}[2]{{\rm{M}}^{#1#2}}
\newcommand{\Lop}[2]{{\rm{L}}^{#1#2}}
\newcommand{\Lgen}[1]{T^{#1}}
\newcommand{\Rgen}[1]{t^{#1}}
\newcommand{\fpari}[1]{\lambda_{#1}}
\newcommand{\fQ}[1]{Q_{#1}}
\newcommand{\unm}{I}
\newcommand{\cDsla}{/\!\!\!\!D}
%
%
\newcommand{\saff}[1]{A_{#1}}                    
\newcommand{\aff}[2]{A_{#1}\lpar #2\rpar}                   
\newcommand{\sbff}[1]{B_{#1}}                    
\newcommand{\sfbff}[1]{B^{F}_{#1}}
\newcommand{\bff}[4]{B_{#1}\lpar #2;#3,#4\rpar}             
\newcommand{\bfft}[3]{B_{#1}\lpar #2,#3\rpar}             
\newcommand{\fbff}[4]{B^{F}_{#1}\lpar #2;#3,#4\rpar}        
\newcommand{\cdbff}[4]{\Delta B_{#1}\lpar #2;#3,#4\rpar}             
\newcommand{\sdbff}[4]{\delta B_{#1}\lpar #2;#3,#4\rpar}             
\newcommand{\cdbfft}[3]{\Delta B_{#1}\lpar #2,#3\rpar}             
\newcommand{\sdbfft}[3]{\delta B_{#1}\lpar #2,#3\rpar}             
\newcommand{\scff}[1]{C_{#1}}                    
\newcommand{\scffo}[2]{C_{#1}\lpar{#2}\rpar}                
\newcommand{\cff}[7]{C_{#1}\lpar #2,#3,#4;#5,#6,#7\rpar}    
\newcommand{\sdff}[1]{D_{#1}}                    
\newcommand{\dffp}[7]{D_{#1}\lpar #2,#3,#4,#5,#6,#7;}       
\newcommand{\dffm}[4]{#1,#2,#3,#4\rpar}                     
\newcommand{\bzfa}[2]{B^{F}_{_{#2}}\lpar{#1}\rpar}
\newcommand{\bzfaa}[3]{B^{F}_{_{#2#3}}\lpar{#1}\rpar}
\newcommand{\shcff}[4]{C_{_{#2#3#4}}\lpar{#1}\rpar}
\newcommand{\shdff}[6]{D_{_{#3#4#5#6}}\lpar{#1,#2}\rpar}
\newcommand{\caldff}[2]{{\cal{D}}_{#1}\lpar{#2}\rpar}
%
%
\newcommand{\slaff}[1]{a_{#1}}                        
\newcommand{\slbff}[1]{b_{#1}}                        
\newcommand{\slbffh}[1]{{\hat{b}}_{#1}}    
\newcommand{\ssldff}[1]{d_{#1}}                        
\newcommand{\sslcff}[1]{c_{#1}}                        
\newcommand{\slcff}[2]{c_{#1}^{(#2)}}                        
\newcommand{\sldff}[2]{d_{#1}^{(#2)}}                        
\newcommand{\lbff}[3]{b_{#1}\lpar #2;#3\rpar}         
\newcommand{\lbffh}[2]{{\hat{b}}_{#1}\lpar #2\rpar}   
\newcommand{\lcff}[8]{c_{#1}^{(#2)}\lpar  #3,#4,#5;#6,#7,#8\rpar}         
\newcommand{\ldffp}[8]{d_{#1}^{(#2)}\lpar #3,#4,#5,#6,#7,#8;}
\newcommand{\ldffm}[4]{#1,#2,#3,#4\rpar}                   
%
%
\newcommand{\Iff}[4]{I_{#1}\lpar #2;#3,#4 \rpar}
\newcommand{\Jff}[4]{J_{#1}\lpar #2;#3,#4 \rpar}
\newcommand{\Jds}[5]{{\bar{J}}_{#1}\lpar #2,#3;#4,#5 \rpar}
%
\newcommand{\nhmt}{\frac{n}{2}-2}
\newcommand{\nhmts}{{n}/{2}-2}
\newcommand{\omnh}{1-\frac{n}{2}}
\newcommand{\nhmo}{\frac{n}{2}-1}
\newcommand{\fmon}{4-n}
\newcommand{\lpi}{\ln\pi}
\newcommand{\lmass}[1]{\ln #1}
\newcommand{\egnh}{\egam{\frac{n}{2}}}
\newcommand{\egomnh}{\egam{1-\frac{n}{2}}}
\newcommand{\egtmnh}{\egam{2-\frac{n}{2}}}
\newcommand{\Ddr}{{\ds\frac{1}{{\bar{\varepsilon}}}}}
\newcommand{\Ddrs}{{\ds\frac{1}{{\bar{\varepsilon}^2}}}}
\newcommand{\Ddrd}{{\bar{\varepsilon}}}
\newcommand{\ept}{\hat\varepsilon}
\newcommand{\Ddrh}{{\ds\frac{1}{\hat{\varepsilon}}}}
\newcommand{\Ddrp}{{\ds\frac{1}{\varepsilon'}}}
\newcommand{\Ddrps}{\lpar{\ds{\frac{1}{\varepsilon'}}}\rpar^2}
\newcommand{\dre}{\varepsilon}
\newcommand{\drei}[1]{\varepsilon_{#1}}
\newcommand{\epp}{\varepsilon'}
\newcommand{\eps}{\varepsilon^*}
\newcommand{\ep}{\epsilon}
\newcommand{\propbt}[6]{{{#1_{#2}#1_{#3}}\over{\lpar #1^2 + #4 
-\ib\ep\rpar\lpar\lpar #5\rpar^2 + #6 -\ib\ep\rpar}}}
\newcommand{\propbo}[5]{{{#1_{#2}}\over{\lpar #1^2 + #3 - \ib\ep\rpar
\lpar\lpar #4\rpar^2 + #5 -\ib\ep\rpar}}}
\newcommand{\propc}[6]{{1\over{\lpar #1^2 + #2 - \ib\ep\rpar
\lpar\lpar #3\rpar^2 + #4 -\ib\ep\rpar
\lpar\lpar #5\rpar^2 + #6 -\ib\ep\rpar}}}
\newcommand{\propa}[2]{{1\over {#1^2 + #2^2 - \ib\ep}}}
\newcommand{\propb}[4]{{1\over {\lpar #1^2 + #2 - \ib\ep\rpar
\lpar\lpar #3\rpar^2 + #4 -\ib\ep\rpar}}}
\newcommand{\propbs}[4]{{1\over {\lpar\lpar #1\rpar^2 + #2 - \ib\ep\rpar
\lpar\lpar #3\rpar^2 + #4 -\ib\ep\rpar}}}
\newcommand{\propat}[4]{{#3_{#1}#3_{#2}\over {#3^2 + #4^2 - \ib\ep}}}
\newcommand{\propaf}[6]{{#5_{#1}#5_{#2}#5_{#3}#5_{#4}\over 
{#5^2 + #6^2 -\ib\ep}}}
\newcommand{\momeps}[1]{#1^2 - \ib\ep}
\newcommand{\mopeps}[1]{#1^2 + \ib\ep}
\newcommand{\propz}[1]{{1\over{#1^2 + \mzs - \ib\ep}}}
\newcommand{\propw}[1]{{1\over{#1^2 + \mws - \ib\ep}}}
\newcommand{\proph}[1]{{1\over{#1^2 + \mhs - \ib\ep}}}
\newcommand{\propf}[2]{{1\over{#1^2 + #2}}}
\newcommand{\propzrg}[3]{{{\delta_{#1#2}}\over{{#3}^2 + \mzs - \ib\ep}}}
\newcommand{\propwrg}[3]{{{\delta_{#1#2}}\over{{#3}^2 + \mws - \ib\ep}}}
\newcommand{\propzug}[3]{{
      {\delta_{#1#2} + \displaystyle{{{#3}^{#1}{#3}^{#2}}\over{\mzs}}}
                         \over{{#3}^2 + \mzs - \ib\ep}}}
\newcommand{\propwug}[3]{{
      {\delta_{#1#2} + \displaystyle{{{#3}^{#1}{#3}^{#2}}\over{\mws}}}
                        \over{{#3}^2 + \mws - \ib\ep}}}
\newcommand{\thf}[1]{\theta\lpar #1\rpar}
\newcommand{\epf}[1]{\varepsilon\lpar #1\rpar}
\newcommand{\singp}{\stackrel{sing}{\rightarrow}}
\newcommand{\aint}[3]{\int_{#1}^{#2}\,d #3}
\newcommand{\aroot}[1]{\sqrt{#1}}
\newcommand{\gramc}{\Delta_3}
\newcommand{\gramd}{\Delta_4}
\newcommand{\pinch}[2]{P^{(#1)}\lpar #2\rpar}
\newcommand{\pinchc}[2]{C^{(#1)}_{#2}}
\newcommand{\pinchd}[2]{D^{(#1)}_{#2}}
\newcommand{\loarg}[1]{\ln\lpar #1\rpar}
\newcommand{\loargr}[1]{\ln\lrbr #1\rrbr}
\newcommand{\lsoarg}[1]{\ln^2\lpar #1\rpar}
\newcommand{\ltarg}[2]{\ln\lpar #1\rpar\lpar #2\rpar}
\newcommand{\rfun}[2]{R\lpar #1,#2\rpar}
\newcommand{\pinchb}[3]{B_{#1}\lpar #2,#3\rpar}
\newcommand{\lga}{\ph}
\newcommand{\lzga}{\ssZ\ph}
%
%
\newcommand{\afa}[5]{A_{#1}^{#2}\lpar #3;#4,#5\rpar}
\newcommand{\bfa}[5]{B_{#1}^{#2}\lpar #3;#4,#5\rpar} 
\newcommand{\hfa}[5]{H_{#1}^{#2}\lpar #3;#4,#5\rpar}
\newcommand{\rfa}[5]{R_{#1}^{#2}\lpar #3;#4,#5\rpar}
\newcommand{\afao}[3]{A_{#1}^{#2}\lpar #3\rpar}
\newcommand{\bfao}[3]{B_{#1}^{#2}\lpar #3\rpar}
\newcommand{\hfao}[3]{H_{#1}^{#2}\lpar #3\rpar}
\newcommand{\rfao}[3]{R_{#1}^{#2}\lpar #3\rpar}
\newcommand{\afas}[2]{A_{#1}^{#2}}
\newcommand{\bfas}[2]{B_{#1}^{#2}}
\newcommand{\hfas}[2]{H_{#1}^{#2}}
\newcommand{\rfas}[2]{R_{#1}^{#2}}
\newcommand{\tfas}[2]{T_{#1}^{#2}}
\newcommand{\afaR}[6]{A_{#1}^{\gpar}\lpar #2;#3,#4,#5,#6 \rpar}
\newcommand{\bfaR}[6]{B_{#1}^{\gpar}\lpar #2;#3,#4,#5,#6 \rpar}
\newcommand{\hfaR}[6]{H_{#1}^{\gpar}\lpar #2;#3,#4,#5,#6 \rpar}
\newcommand{\shfaR}[1]{H_{#1}^{\gpar}}
\newcommand{\rfaR}[6]{R_{#1}^{\gpar}\lpar #2;#3,#4,#5,#6 \rpar}
\newcommand{\srfaR}[1]{R_{#1}^{\gpar}}
\newcommand{\afaRg}[5]{A_{#1 \lga}^{\gpar}\lpar #2;#3,#4,#5 \rpar}
\newcommand{\bfaRg}[5]{B_{#1 \lga}^{\gpar}\lpar #2;#3,#4,#5 \rpar}
\newcommand{\hfaRg}[5]{H_{#1 \lga}^{\gpar}\lpar #2;#3,#4,#5 \rpar}
\newcommand{\shfaRg}[1]{H_{#1\lga}^{\gpar}}
\newcommand{\rfaRg}[5]{R_{#1 \lga}^{\gpar}\lpar #2;#3,#4,#5 \rpar}
\newcommand{\srfaRg}[1]{R_{#1\lga}^{\gpar}}
\newcommand{\afaRt}[3]{A_{#1}^{\gpar}\lpar #2,#3 \rpar}
\newcommand{\hfaRt}[3]{H_{#1}^{\gpar}\lpar #2,#3 \rpar}
\newcommand{\hfaRf}[4]{H_{#1}^{\gpar}\lpar #2,#3,#4 \rpar}
\newcommand{\afasm}[4]{A_{#1}^{\lpar #2,#3,#4 \rpar}}
\newcommand{\bfasm}[4]{B_{#1}^{\lpar #2,#3,#4 \rpar}}
\newcommand{\color}[1]{c_{#1}}
\newcommand{\htf}[2]{H_2\lpar #1,#2\rpar}
\newcommand{\rof}[2]{R_1\lpar #1,#2\rpar}
\newcommand{\rtf}[2]{R_3\lpar #1,#2\rpar}
\newcommand{\rtrans}[2]{R_{#1}^{#2}}
\newcommand{\momf}[2]{#1^2_{#2}}
\newcommand{\Scalvert}[8][70]{
  \vcenter{\hbox{
  \SetScale{0.8}
  \begin{picture}(#1,50)(15,15)
    \Line(25,25)(50,50)      \Text(34,20)[lc]{#6} \Text(11,20)[lc]{#3}
    \Line(50,50)(25,75)      \Text(34,60)[lc]{#7} \Text(11,60)[lc]{#4}
    \Line(50,50)(90,50)      \Text(11,40)[lc]{#2} \Text(55,33)[lc]{#8}
    \GCirc(50,50){10}{1}          \Text(60,48)[lc]{#5} 
  \end{picture}}}
  }
%
%
\newcommand{\tHs}{\mu}
\newcommand{\tHsz}{\mu_{_0}}
\newcommand{\tHss}{\mu^2}
\newcommand{\Reb}{{\rm{Re}}}
\newcommand{\Imb}{{\rm{Im}}}
%
%
\newcommand{\spd}{\partial}
\newcommand{\fun}[1]{f\lpar{#1}\rpar}
\newcommand{\ffun}[2]{F_{#1}\lpar #2\rpar}
\newcommand{\gfun}[2]{G_{#1}\lpar #2\rpar}
\newcommand{\sffun}[1]{F_{#1}}
\newcommand{\csffun}[1]{{\cal{F}}_{#1}}
\newcommand{\sgfun}[1]{G_{#1}}
\newcommand{\tpfi}{\lpar 2\pi\rpar^4\ib}
\newcommand{\ffv}{F_{_V}}
\newcommand{\fga}{G_{_A}}
\newcommand{\ffm}{F_{_M}}
\newcommand{\ffs}{F_{_S}}
\newcommand{\fgp}{G_{_P}}
\newcommand{\fge}{G_{_E}}
\newcommand{\ffa}{F_{_A}}
\newcommand{\ffps}{F_{_P}}
\newcommand{\ffe}{F_{_E}}
\newcommand{\gacom}[2]{\lpar #1 + #2\gfd\rpar}
\newcommand{\mft}{m_{\tilde f}}
\newcommand{\qft}{Q_{f'}}
\newcommand{\vft}{v_{\tilde f}}
\newcommand{\subb}[2]{b_{#1}\lpar #2 \rpar}
\newcommand{\fwfr}[5]{\Sigma\lpar #1,#2,#3;#4,#5 \rpar}
\newcommand{\slim}[2]{\lim_{#1 \to #2}}
\newcommand{\sprop}[3]{
{#1\over {\lpar q^2\rpar^2\lpar \lpar q+ #2\rpar^2+#3^2\rpar }}}
%
%
\newcommand{\xroot}[1]{x_{#1}}
\newcommand{\yroot}[1]{y_{#1}}
\newcommand{\zroot}[1]{z_{#1}}
\newcommand{\rvar}{r}
\newcommand{\xvar}{x}
\newcommand{\yvar}{y}
\newcommand{\zvar}{z}
\newcommand{\uvar}{u}
\newcommand{\yvarp}{y'}
\newcommand{\rvars}{r^2}
\newcommand{\xvars}{x^2}
\newcommand{\yvars}{y^2}
\newcommand{\zvars}{z^2}
\newcommand{\rvarc}{r^3}
\newcommand{\xvarc}{x^3}
\newcommand{\yvarc}{y^3}
\newcommand{\zvarc}{z^3}
\newcommand{\rvarq}{r^4}
\newcommand{\xvarq}{x^4}
\newcommand{\yvarq}{y^4}
\newcommand{\zvarq}{z^4}
\newcommand{\avar}{a}
\newcommand{\avars}{a^2}
\newcommand{\avarc}{a^3}
\newcommand{\avari}[1]{a_{#1}}
\newcommand{\avart}[2]{a_{#1}^{#2}}
\newcommand{\delvari}[1]{\delta_{#1}}
\newcommand{\rvari}[1]{r_{#1}}
\newcommand{\xvari}[1]{x_{#1}}
\newcommand{\yvari}[1]{y_{#1}}
\newcommand{\zvari}[1]{z_{#1}}
\newcommand{\rvart}[2]{r_{#1}^{#2}}
\newcommand{\xvart}[2]{x_{#1}^{#2}}
\newcommand{\yvart}[2]{y_{#1}^{#2}}
\newcommand{\zvart}[2]{z_{#1}^{#2}}
\newcommand{\rvaris}[1]{r^2_{#1}}
\newcommand{\xvaris}[1]{x^2_{#1}}
\newcommand{\yvaris}[1]{y^2_{#1}}
\newcommand{\zvaris}[1]{z^2_{#1}}
\newcommand{\Xvar}{X}
\newcommand{\Xvars}{X^2}
\newcommand{\Xvari}[1]{X_{#1}}
\newcommand{\Xvaris}[1]{X^2_{#1}}
\newcommand{\Yvar}{Y}
\newcommand{\Yvars}{Y^2}
\newcommand{\Yvari}[1]{Y_{#1}}
\newcommand{\Yvaris}[1]{Y^2_{#1}}
\newcommand{\lnx}{\ln\xvar}
\newcommand{\lnz}{\ln\zvar}
\newcommand{\lnsx}{\ln^2\xvar}
\newcommand{\lnsz}{\ln^2\zvar}
\newcommand{\lncz}{\ln^3\zvar}
\newcommand{\lnomz}{\ln\lpar 1-\zvar\rpar}
\newcommand{\lnsomz}{\ln^2\lpar 1-\zvar\rpar}
\newcommand{\Xmat}[1]{X_{#1}}
\newcommand{\XmatI}[1]{X^{-1}_{#1}}
\newcommand{\unitmat}{I}
\newcommand{\Kmat}{{C}}
\newcommand{\Kmatc}{{C}^{\dagger}}
\newcommand{\Kmati}[1]{{C}_{#1}}
\newcommand{\Kmatci}[1]{{C}^{\dagger}_{#1}}
\newcommand{\ffac}[2]{f_{#1}^{#2}}
\newcommand{\Ffac}[1]{F_{#1}}
\newcommand{\Rvec}[2]{R^{(#1)}_{#2}}
\newcommand{\momfl}[2]{#1_{#2}}
\newcommand{\momfs}[2]{#1^2_{#2}}
\newcommand{\fpseZ}{A^{^{FP,Z}}}
\newcommand{\fpseA}{A^{^{FP,A}}}
\newcommand{\fptZ}{T^{^{FP,Z}}}
\newcommand{\fptA}{T^{^{FP,A}}}
\newcommand{\dprop}{\overline\Delta}
\newcommand{\dpropi}[1]{d_{#1}}
\newcommand{\dpropic}[1]{d^{c}_{#1}}
\newcommand{\dpropii}[2]{d_{#1}\lpar #2\rpar}
\newcommand{\dpropis}[1]{d^2_{#1}}
\newcommand{\dproppi}[1]{d'_{#1}}
\newcommand{\psf}[4]{P\lpar #1;#2,#3,#4\rpar}
\newcommand{\ssf}[5]{S^{(#1)}\lpar #2;#3,#4,#5\rpar}
\newcommand{\csf}[5]{C_{_S}^{(#1)}\lpar #2;#3,#4,#5\rpar}
%
%
\newcommand{\lvec}{l}
\newcommand{\lvecs}{l^2}
\newcommand{\lveci}[1]{l_{#1}}
\newcommand{\mvec}{m}
\newcommand{\mvecs}{m^2}
\newcommand{\mveci}[1]{m_{#1}}
\newcommand{\nvec}{n}
\newcommand{\nvecs}{n^2}
\newcommand{\nveci}[1]{n_{#1}}
\newcommand{\epi}[1]{\epsilon_{#1}}
\newcommand{\phep}[1]{\ep_{#1}}
\newcommand{\sphep}{\ep}
\newcommand{\vbep}[1]{e_{#1}}
\newcommand{\svbep}{e}
%
%
\newcommand{\lpol}{\lambda}
\newcommand{\spol}{\sigma}
\newcommand{\rpol}{\rho  }
\newcommand{\kpol}{\kappa}
\newcommand{\lpols}{\lambda^2}
\newcommand{\spols}{\sigma^2}
\newcommand{\rpols}{\rho^2}
\newcommand{\kpols}{\kappa^2}
\newcommand{\lpoli}[1]{\lambda_{#1}}
\newcommand{\spoli}[1]{\sigma_{#1}}
\newcommand{\rpoli}[1]{\rho_{#1}}
\newcommand{\kpoli}[1]{\kappa_{#1}}
%
%
\newcommand{\uvec}{u}
\newcommand{\uveci}[1]{u_{#1}}
%
%
\newcommand{\imom}{q}
\newcommand{\imomi}[1]{q_{#1}}
\newcommand{\imoms}{q^2}
\newcommand{\pmom}{p}
\newcommand{\pmomp}{p'}
\newcommand{\pmoms}{p^2}
\newcommand{\pmomq}{p^4}
\newcommand{\pmomx}{p^6}
\newcommand{\pmomi}[1]{p_{#1}}
\newcommand{\pmomis}[1]{p^2_{#1}}
\newcommand{\Pmom}{P}
\newcommand{\Pmoms}{P^2}
\newcommand{\Pmomi}[1]{P_{#1}}
\newcommand{\Pmomis}[1]{P^2_{#1}}
\newcommand{\Kmom}{K}
\newcommand{\Kmoms}{K^2}
\newcommand{\Kmomi}[1]{K_{#1}}
\newcommand{\Kmomis}[1]{K^2_{#1}}
\newcommand{\kmom}{k}
\newcommand{\kmoms}{k^2}
\newcommand{\kmomi}[1]{k_{#1}}
\newcommand{\lmom}{l}
\newcommand{\lmoms}{l^2}
\newcommand{\lmomi}[1]{l_{#1}}
\newcommand{\qmom}{q}
\newcommand{\qmoms}{q^2}
\newcommand{\qmomi}[1]{q_{#1}}
\newcommand{\qmomis}[1]{q^2_{#1}}
\newcommand{\smom}{s}
\newcommand{\smoms}{s^2}
\newcommand{\smomi}[1]{s_{#1}}
\newcommand{\tmom}{t}
\newcommand{\tmoms}{t^2}
\newcommand{\tmomi}[1]{t_{#1}}
\newcommand{\Trmom}{Q}
\newcommand{\Prmom}{P}
\newcommand{\gmv}{Q^2}
\newcommand{\Trmoms}{Q^2}
\newcommand{\Prmoms}{P^2}
\newcommand{\Ptmoms}{T^2}
\newcommand{\Pumoms}{U^2}
\newcommand{\Trmomq}{Q^4}
\newcommand{\Prmomq}{P^4}
\newcommand{\Ptmomq}{T^4}
\newcommand{\Pumomq}{U^4}
\newcommand{\Trmomx}{Q^6}
\newcommand{\Trmomi}[1]{Q_{#1}}
\newcommand{\Trmomis}[1]{Q^2_{#1}}
\newcommand{\Prmomi}[1]{P_{#1}}
\newcommand{\pone}{p_1}
\newcommand{\ptwo}{p_2}
\newcommand{\ptre}{p_3}
\newcommand{\pfor}{p_4}
\newcommand{\pones}{p_1^2}
\newcommand{\ptwos}{p_2^2}
\newcommand{\ptres}{p_3^2}
\newcommand{\pfors}{p_4^2}
\newcommand{\poneq}{p_1^4}
\newcommand{\ptwoq}{p_2^4}
\newcommand{\ptreq}{p_3^4}
\newcommand{\pforq}{p_4^4}
\newcommand{\modmom}[1]{\mid{\vec{#1}}\mid}
\newcommand{\modmomi}[2]{\mid{\vec{#1}}_{#2}\mid}
\newcommand{\vect}[1]{{\vec{#1}}}
\newcommand{\Energ}{E}
\newcommand{\Energp}{E'}
\newcommand{\Energpp}{E''}
\newcommand{\Energs}{E^2}
\newcommand{\Energc}{E^3}
\newcommand{\Energf}{E^4}
\newcommand{\Energv}{E^5}
\newcommand{\Energx}{E^6}
\newcommand{\Energi}[1]{E_{#1}}
\newcommand{\Energt}[2]{E_{#1}^{#2}}
\newcommand{\Energis}[1]{E^2_{#1}}
\newcommand{\energ}{e}
\newcommand{\energp}{e'}
\newcommand{\energpp}{e''}
\newcommand{\energs}{e^2}
\newcommand{\energi}[1]{e_{#1}}
\newcommand{\energt}[2]{e_{#1}^{#2}}
\newcommand{\energis}[1]{e^2_{#1}}
\newcommand{\wenerg}{w}
\newcommand{\wenergs}{w^2}
\newcommand{\wenergi}[1]{w_{#1}}
\newcommand{\wenergp}{w'}
\newcommand{\wenergpp}{w''}
%
%
\newcommand{\ecut}{e}
\newcommand{\ecuts}{e^2}
\newcommand{\ecuti}[1]{e^{#1}}
\newcommand{\ccut}{c_m}
\newcommand{\ccuti}[1]{c_{#1}}
\newcommand{\ccuts}{c^2_m}
\newcommand{\scuts}{s^2_m}
\newcommand{\ccutis}[1]{c^2_{#1}}
\newcommand{\ccutic}[1]{c^3_{#1}}
\newcommand{\ccutc}{c^3_m}
\newcommand{\rcut}{\varrho}
\newcommand{\rcuts}{\varrho^2}
\newcommand{\rcuti}[1]{\varrho_{#1}}
\newcommand{\rcutu}[1]{\varrho^{#1}}
\newcommand{\Dcut}{\Delta}
%
\newcommand{\dwf}{\delta_{_{WF}}}
\newcommand{\gbar}{\overline g}
\newcommand{\PP}{\mbox{PP}}
\newcommand{\mv}{m_{_V}}
\newcommand{\bGv}{{\overline\Gamma}_{_V}}
\newcommand{\Umuv}{\hat{\mu}_\ssV}
\newcommand{\Svv}{{\Sigma}_\ssV}
\newcommand{\muv}{p_\ssV}
\newcommand{\muvb}{\mu_{\ssV_{0}}}
\newcommand{\URPvv}{{P}_\ssV}
\newcommand{\RPvv}{{P}_\ssV}
\newcommand{\Svvrem}{{\Sigma}_\ssV^{\mathrm{rem}}}
\newcommand{\USvvrem}{\hat{\Sigma}_\ssV^{\mathrm{rem}}}
\newcommand{\Gv}{\Gamma_{_V}}
%
%
\newcommand{\param}{p}
\newcommand{\parami}[1]{p^{#1}}
\newcommand{\paramb}{p_{0}}
\newcommand{\Zcon}{Z}
\newcommand{\Zconi}[1]{Z_{#1}}
\newcommand{\zconi}[1]{z_{#1}}
\newcommand{\Zconim}[1]{{Z^-_{#1}}}
\newcommand{\zconim}[1]{{z^-_{#1}}}
\newcommand{\Zcont}[2]{Z_{#1}^{#2}}
\newcommand{\zcont}[2]{z_{#1}^{#2}}
\newcommand{\zcontm}[2]{z_{#1}^{{#2}-}}
\newcommand{\sZconi}[2]{\sqrt{Z_{#1}}^{\;#2}}
\newcommand{\Smat}{{\cal{S}}}
\newcommand{\Mmat}{{\cal{M}}}
\newcommand{\php}[3]{e^{#1}_{#2}\lpar #3 \rpar}
\newcommand{\gacome}[1]{\lpar #1 - \gfd\rpar}
\newcommand{\sPj}[2]{\Lambda^{#1}_{#2}}
\newcommand{\sPjs}[2]{\Lambda_{#1,#2}}
\newcommand{\amos}{\mbox{$M^2_{_1}$}}
\newcommand{\amts}{\mbox{$M^2_{_2}$}}
\newcommand{\er}{e_{_{R}}}
\newcommand{\epr}{e'_{_{R}}}
\newcommand{\ers}{e^2_{_{R}}}
\newcommand{\erc}{e^3_{_{R}}}
\newcommand{\erq}{e^4_{_{R}}}
\newcommand{\erf}{e^5_{_{R}}}
\newcommand{\sour}{J}
\newcommand{\sourb}{\overline J}
\newcommand{\lrm}{M_{_R}}
%
%
\newcommand{\vlami}[1]{\lambda_{#1}}
\newcommand{\vlamis}[1]{\lambda^2_{#1}}
\newcommand{\Vvert}{V}
\newcommand{\Avert}{A}
\newcommand{\Svert}{S}
\newcommand{\Pvert}{P}
\newcommand{\vvert}{F}
\newcommand{\Cvert}{\cal{V}}
\newcommand{\Bvert}{\cal{B}}
\newcommand{\Vveri}[2]{V_{#1}^{#2}}
\newcommand{\Fveri}[1]{{\cal{F}}^{#1}}
\newcommand{\Cveri}[1]{{\cal{V}}\lpar{#1}\rpar}
\newcommand{\Bveri}[1]{{\cal{B}}\lpar{#1}\rpar}
\newcommand{\Vverti}[3]{V_{#1}^{#2}\lpar{#3}\rpar}
\newcommand{\Averti}[3]{A_{#1}^{#2}\lpar{#3}\rpar}
\newcommand{\Gverti}[3]{G_{#1}^{#2}\lpar{#3}\rpar}
\newcommand{\Zverti}[3]{Z_{#1}^{#2}\lpar{#3}\rpar}
\newcommand{\Hverti}[2]{H^{#1}\lpar{#2}\rpar}
\newcommand{\Wverti}[3]{W_{#1}^{#2}\lpar{#3}\rpar}
\newcommand{\Cverti}[2]{{\cal{V}}_{#1}^{#2}}
\newcommand{\vverti}[3]{F^{#1}_{#2}\lpar{#3}\rpar}
\newcommand{\averti}[3]{{\overline{F}}^{#1}_{#2}\lpar{#3}\rpar}
\newcommand{\fveone}[1]{f_{#1}}
\newcommand{\fvetri}[3]{f^{#1}_{#2}\lpar{#3}\rpar}
\newcommand{\gvetri}[3]{g^{#1}_{#2}\lpar{#3}\rpar}
\newcommand{\cvetri}[3]{{\cal{F}}^{#1}_{#2}\lpar{#3}\rpar}
\newcommand{\hvetri}[3]{{\hat{\cal{F}}}^{#1}_{#2}\lpar{#3}\rpar}
\newcommand{\avetri}[3]{{\overline{\cal{F}}}^{#1}_{#2}\lpar{#3}\rpar}
\newcommand{\fverti}[2]{F^{#1}_{#2}}
\newcommand{\cverti}[2]{{\cal{F}}_{#1}^{#2}}
\newcommand{\fV}{f_{_{\Vvert}}}
\newcommand{\gA}{g_{_{\Avert}}}
\newcommand{\fVi}[1]{f^{#1}_{_{\Vvert}}}
\newcommand{\seai}[1]{a_{#1}}
\newcommand{\seapi}[1]{a'_{#1}}
\newcommand{\seAi}[2]{A_{#1}^{#2}}
\newcommand{\sewi}[1]{w_{#1}}
\newcommand{\seWi}[1]{W_{#1}}
\newcommand{\seWsi}[1]{W^{*}_{#1}}
\newcommand{\seWti}[2]{W_{#1}^{#2}}
\newcommand{\sewti}[2]{w_{#1}^{#2}}
\newcommand{\seSig}[1]{\Sigma_{#1}\lpar\sla{\pmom}\rpar}
\newcommand{\ww}{w}
%
%
\newcommand{\bbff}[1]{{\overline B}_{#1}}
\newcommand{\sW}{p_{_W}}
\newcommand{\sZ}{p_{_Z}}
\newcommand{\ssp}{s_p}
\newcommand{\fW}{f_{_W}}
\newcommand{\fZ}{f_{_Z}}
\newcommand{\tabn}[1]{Tab.(\ref{#1})}
\newcommand{\subMSB}[1]{{#1}_{\mbox{$\overline{\scriptscriptstyle MS}$}}}
\newcommand{\supMSB}[1]{{#1}^{\mbox{$\overline{\scriptscriptstyle MS}$}}}
\newcommand{\redMSB}{{\mbox{$\overline{\scriptscriptstyle MS}$}}}
\newcommand{\gpbb}{g'_{0}}
\newcommand{\Zconip}[1]{Z'_{#1}}
\newcommand{\bpff}[4]{B'_{#1}\lpar #2;#3,#4\rpar}             
\newcommand{\xidf}{\xi^2-1}
\newcommand{\tDdr}{1/{\bar{\varepsilon}}}
\newcommand{\cRz}{{\cal R}_{_Z}}
\newcommand{\cRg}{{\cal R}_{\gamma}}
\newcommand{\Sz}{\Sigma_{_Z}}
\newcommand{\alh}{{\hat\alpha}}
\newcommand{\alhz}{\alpha_{_Z}}
\newcommand{\Phzg}{{\hat\Pi}_{_{\zb\ab}}}
\newcommand{\fvvert}{F^{\rm vert}_{_V}}
\newcommand{\gavert}{G^{\rm vert}_{_A}}
\newcommand{\bmv}{{\overline m}_{_V}}
\newcommand{\Sgn}{\Sigma_{\gamma\hkn}}
\newcommand{\tabns}[2]{Tabs.(\ref{#1}--\ref{#2})}
\newcommand{\rmboxd}{{\rm Box}_d\lpar s,t,u;M_1,M_2,M_3,M_4\rpar}
\newcommand{\rmboxc}{{\rm Box}_c\lpar s,t,u;M_1,M_2,M_3,M_4\rpar}
%
%
\newcommand{\Afaci}[1]{A_{#1}}
\newcommand{\Afacis}[1]{A^2_{#1}}
\newcommand{\upar}[1]{u}
\newcommand{\upari}[1]{u_{#1}}
\newcommand{\vpari}[1]{v_{#1}}
\newcommand{\lpari}[1]{l_{#1}}
\newcommand{\Lpari}[1]{l_{#1}}
\newcommand{\Nff}[2]{N^{(#1)}_{#2}}
\newcommand{\Sff}[2]{S^{(#1)}_{#2}}
\newcommand{\sSff}{S}
\newcommand{\FQED}[2]{F_{#1#2}}
\newcommand{\fbpsif}{{\overline{\psi}_f}}
\newcommand{\fpsif}{\psi_f}
\newcommand{\etafd}[2]{\eta_d\lpar#1,#2\rpar}
\newcommand{\sigdu}[2]{\sigma_{#1#2}}
\newcommand{\scalc}[4]{c_{_0}\lpar #1;#2,#3,#4\rpar}
\newcommand{\scald}[2]{d_{_0}\lpar #1,#2\rpar}
\newcommand{\pir}[1]{\Pi^{\rm ren}\lpar #1\rpar}
\newcommand{\sigh}{\sigma_{\rm had}}
\newcommand{\dah}{\Delta\alpha^{(5)}_{\rm had}}
\newcommand{\dat}{\Delta\alpha_{\rm top}}
\newcommand{\Vqed}[3]{V_1^{\rm sub}\lpar#1;#2,#3\rpar}
\newcommand{\thetah}{{\hat\theta}}
\newcommand{\mtsix}{m^6_t}
\newcommand{\smlon}{\frac{\mlones}{s}}
\newcommand{\lntwo}{\ln 2}
\newcommand{\wmin}{w_{\rm min}}
\newcommand{\kmin}{k_{\rm min}}
\newcommand{\scaldi}[3]{d_{_0}^{#1}\lpar #2,#3\rpar}
\newcommand{\mdls}{\Big|}
\newcommand{\smf}{\frac{\mfs}{s}}
\newcommand{\bint}{\beta_{\rm int}}
\newcommand{\IRv}{V_{_{\rm IR}}}
\newcommand{\IRr}{R_{_{\rm IR}}}
\newcommand{\fssts}{\frac{s^2}{t^2}}
\newcommand{\fssus}{\frac{s^2}{u^2}}
\newcommand{\optM}{1+\frac{t}{M^2}}
\newcommand{\opuM}{1+\frac{u}{M^2}}
\newcommand{\ftM}{\lpar -\frac{t}{M^2}\rpar}
\newcommand{\fuM}{\lpar -\frac{u}{M^2}\rpar}
\newcommand{\omsM}{1-\frac{s}{M^2}}
\newcommand{\xsf}{\sigma_{_{\rm F}}}
\newcommand{\xsb}{\sigma_{_{\rm B}}}
\newcommand{\afb}{A_{_{\rm FB}}}
\newcommand{\rsoft}{\rm soft}
\newcommand{\rms}{\rm s}
\newcommand{\rsmx}{\sqrt{s_{\rm max}}}
\newcommand{\rspm}{\sqrt{s_{\pm}}}
\newcommand{\rsp}{\sqrt{s_{+}}}
\newcommand{\rsm}{\sqrt{s_{-}}}
\newcommand{\sigmx}{\sigma_{\rm max}}
\newcommand{\gG}[2]{G_{#1}^{#2}}
\newcommand{\gacomm}[2]{\lpar #1 - #2\gfd\rpar}
\newcommand{\fcsx}{\frac{1}{\ctwsix}}
\newcommand{\fcq}{\frac{1}{\ctwf}}
\newcommand{\fcs}{\frac{1}{\ctws}}
\newcommand{\affs}[2]{{\cal A}_{#1}\lpar #2\rpar}                   
\newcommand{\stwei}{s_{\theta}^8}
\def\mdan{\vspace{1mm}\mpar{\hfil$\downarrow$new\hfil}\vspace{-1mm}
          \ignorespaces}
\def\muan{\vspace{-1mm}\mpar{\hfil$\uparrow$new\hfil}\vspace{1mm}\ignorespaces}
\def\mlan{\vspace{-1mm}\mpar{\hfil$\rightarrow$new\hfil}\vspace{1mm}\ignorespaces}
\def\mnnew{\mpar{\hfil NEWNEW \hfil}\ignorespaces}
%
%
\newcommand{\boxc}[2]{{\cal{B}}_{#1}^{#2}}
\newcommand{\boxct}[3]{{\cal{B}}_{#1}^{#2}\lpar{#3}\rpar}
\newcommand{\hboxc}[3]{\hat{{\cal{B}}}_{#1}^{#2}\lpar{#3}\rpar}
\newcommand{\vev}{\langle v \rangle}
\newcommand{\vevs}{\langle v^2 \rangle}
\newcommand{\fwfrV}[5]{\Sigma_{_V}\lpar #1,#2,#3;#4,#5 \rpar}
\newcommand{\fwfrS}[7]{\Sigma_{_S}\lpar #1,#2,#3;#4,#5;#6,#7 \rpar}
\newcommand{\fSi}[1]{f^{#1}_{_{\Svert}}}
\newcommand{\fPi}[1]{f^{#1}_{_{\Pvert}}}
\newcommand{\mXs}{m_{_X}}
\newcommand{\mXss}{m^2_{_X}}
\newcommand{\mYs}{M^2_{_Y}}
\newcommand{\xik}{\xi_k}
\newcommand{\xiks}{\xi^2_k}
\newcommand{\mpls}{m^2_+}
\newcommand{\mmis}{m^2_-}
%
\newcommand{\SN}{\Sigma_{_N}}
\newcommand{\SC}{\Sigma_{_C}}
\newcommand{\SPN}{\Sigma'_{_N}}
\newcommand{\PFf}{\Pi^{\fer}_{_F}}
\newcommand{\PFb}{\Pi^{\bos}_{_F}}
\newcommand{\dPZ}{\Delta{\hat\Pi}_{_Z}}
\newcommand{\Sfin}{\Sigma_{_F}}
\newcommand{\Sfir}{\Sigma_{_R}}
\newcommand{\Sfinh}{{\hat\Sigma}_{_F}}
\newcommand{\Sfinf}{\Sigma^{\fer}_{_F}}
\newcommand{\Sfinbh}{\Sigma^{\bos}_{_F}}
\newcommand{\alf}{\alpha^{\fer}}
\newcommand{\alhfz}{\alpha^{\fer}\lpar{\ssZ}\rpar}
\newcommand{\alhfs}{\alpha^{\fer}\lpar{\sman}\rpar}
\newcommand{\gfQ}{g^f_{_{Q}}}
\newcommand{\gfL}{g^f_{_{L}}}
\newcommand{\ccf}{\frac{\gbs}{16\,\pi^2}}
\newcommand{\chq}{{\hat c}^4}
\newcommand{\muuq}{m_{u'}}
\newcommand{\muus}{m^2_{u'}}
\newcommand{\mdd}{m_{d'}}
\newcommand{\clf}[2]{\mathrm{Cli}_{_#1}\lpar\displaystyle{#2}\rpar}
\def\stes{\sin^2\theta}
\def\acal{\cal A}
\def\alr{A_{_{\rm{LR}}}}
\newcommand{\barQ}{\overline Q}
\newcommand{\Sptg}{\Sigma'_{_{3Q}}}
\newcommand{\Sptt}{\Sigma'_{_{33}}}
\newcommand{\Ppgg}{\Pi'_{\ph\ph}}
\newcommand{\Pww}{\Pi_{_{\wb\wb}}}
\newcommand{\capV}[2]{{\cal F}^{#2}_{_{#1}}}
\newcommand{\bt}{\beta_t}
\newcommand{\mhsix}{M^6_{_H}}
\newcommand{\topt}{{\cal T}_{33}}
\newcommand{\topq}{{\cal T}_4}
\newcommand{\Phzgf}{{\hat\Pi}^{\fer}_{_{\zb\ab}}}
\newcommand{\Phzgb}{{\hat\Pi}^{\bos}_{_{\zb\ab}}}
\newcommand{\Sfirh}{{\hat\Sigma}_{_R}}
\newcommand{\Szgh}{{\hat\Sigma}_{_{\zb\ab}}}
\newcommand{\Szghb}{{\hat\Sigma}^{\bos}_{_{\zb\ab}}}
\newcommand{\Szghf}{{\hat\Sigma}^{\fer}_{_{\zb\ab}}}
\newcommand{\Szgb}{\Sigma^{\bos}_{_{\zb\ab}}}
\newcommand{\Szgf}{\Sigma^{\fer}_{_{\zb\ab}}}
\newcommand{\chig}{\chi_{_{\ph}}}
\newcommand{\chiz}{\chi_{_{\zb}}}
\newcommand{\Sfih}{{\hat\Sigma}}
\newcommand{\Szzh}{\hat{\Sigma}_{_{\zb\zb}}}
\newcommand{\dPZf}{\Delta{\hat\Pi}^f_{_{\zb}}}
\newcommand{\khZdf}[1]{{\hat\kappa}^{#1}_{_{\zb}}}
\newcommand{\chf}{{\hat c}^4}
\newcommand{\amp}[2]{{\cal{A}}_{_{#1}}^{\rm{#2}}}
\newcommand{\hatvm}[1]{{\hat v}^-_{#1}}
\newcommand{\hatvp}[1]{{\hat v}^+_{#1}}
\newcommand{\hatvpm}[1]{{\hat v}^{\pm}_{#1}}
\newcommand{\kvz}[1]{\kappa^{\zb #1}_{_V}}
\newcommand{\barp}{\overline p}                
\newcommand{\delw}{\Delta_{_{\wb}}}
\newcommand{\bdelw}{{\bar{\Delta}}_{_{\wb}}}
\newcommand{\bdelf}{{\bar{\Delta}}_{\ff}}
\newcommand{\delz}{\Delta_{_\zb}}
\newcommand{\deli}[1]{\Delta\lpar{#1}\rpar}
\newcommand{\chizb}{\chi_{_\zb}}
\newcommand{\Swwp}{\Sigma'_{_{\wb\wb}}}
\newcommand{\epph}{\varepsilon'/2}
\newcommand{\sbffp}[1]{B'_{#1}}                    
\newcommand{\epss}{\varepsilon^*}
\newcommand{\Ddrhs}{{\ds\frac{1}{\hat{\varepsilon}^2}}}
\newcommand{\lnmsb}{L_{_\wb}}
\newcommand{\lnsmsb}{L^2_{_\wb}}
\newcommand{\tpni}{\lpar 2\pi\rpar^n\ib}
\newcommand{\tpn}{2^n\,\pi^{n-2}}
\newcommand{\cmf}{M_f}
\newcommand{\cmfs}{M^2_f}
\newcommand{\toDdr}{{\ds\frac{2}{{\bar{\varepsilon}}}}}
\newcommand{\troDdr}{{\ds\frac{3}{{\bar{\varepsilon}}}}}
\newcommand{\totDdr}{{\ds\frac{3}{{2\,\bar{\varepsilon}}}}}
\newcommand{\foDdr}{{\ds\frac{4}{{\bar{\varepsilon}}}}}
\newcommand{\smh}{m_{_H}}
\newcommand{\smhs}{m^2_{_H}}
\newcommand{\Ph}{\Pi_{_\hb}}
\newcommand{\ghb}{\Gamma_{_\hb}}
\newcommand{\Sphh}{\Sigma'_{_{\hb\hb}}}
\newcommand{\bh}{\beta}
\newcommand{\alsn}{\alpha^{(n_f)}_{_S}}
\newcommand{\smq}{m_q}
\newcommand{\smqp}{m_{q'}}
\newcommand{\shb}{h}
\newcommand{\hab}{A}
\newcommand{\hbpm}{H^{\pm}}
\newcommand{\hbp}{H^{+}}
\newcommand{\hbm}{H^{-}}
\newcommand{\msh}{M_h}
\newcommand{\mha}{M_{_A}}
\newcommand{\mhc}{M_{_{H^{\pm}}}}
\newcommand{\mshs}{M^2_h}
\newcommand{\mhas}{M^2_{_A}}
\newcommand{\barfp}{\overline{f'}}                
\newcommand{\chiii}{{\hat c}^3}
\newcommand{\chiv}{{\hat c}^4}
\newcommand{\chv}{{\hat c}^5}
\newcommand{\chvi}{{\hat c}^6}
\newcommand{\alsvi}{\alpha^{6}_{_S}}
\newcommand{\tww}{t_{_W}}
\newcommand{\ti}{t_{_1}}
\newcommand{\tii}{t_{_2}}
\newcommand{\tiii}{t_{_3}}
\newcommand{\tiv}{t_{_4}}
\newcommand{\psla}{\hbox{\rlap/p}}
\newcommand{\qsla}{\hbox{\rlap/q}}
\newcommand{\nsla}{\hbox{\rlap/n}}
\newcommand{\lsla}{\hbox{\rlap/l}}
\newcommand{\msla}{\hbox{\rlap/m}}
\newcommand{\cnsla}{\hbox{\rlap/N}}
\newcommand{\clsla}{\hbox{\rlap/L}}
\newcommand{\cmsla}{\hbox{\rlap/M}}
\newcommand{\blmt}{\lrbr - 3\rrbr}
\newcommand{\blfo}{\lrbr 4 1\rrbr}
\newcommand{\bltp}{\lrbr 2 +\rrbr}
\newcommand{\clitwo}[1]{{\rm{Li}}_{2}\lpar{#1}\rpar}
\newcommand{\clitri}[1]{{\rm{Li}}_{3}\lpar{#1}\rpar}
\newcommand{\xt}{x_{\ft}}
\newcommand{\zt}{z_{\ft}}
\newcommand{\Ht}{h_{\ft}}
\newcommand{\xts}{x^2_{\ft}}
\newcommand{\zts}{z^2_{\ft}}
\newcommand{\Hts}{h^2_{\ft}}
\newcommand{\ztc}{z^3_{\ft}}
\newcommand{\Htc}{h^3_{\ft}}
\newcommand{\ztq}{z^4_{\ft}}
\newcommand{\Htq}{h^4_{\ft}}
\newcommand{\ztv}{z^5_{\ft}}
\newcommand{\Htv}{h^5_{\ft}}
\newcommand{\ztx}{z^6_{\ft}}
\newcommand{\Htx}{h^6_{\ft}}
\newcommand{\ztz}{z^7_{\ft}}
\newcommand{\Htz}{h^7_{\ft}}
\newcommand{\sht}{\sqrt{\Ht}}
\newcommand{\atan}[1]{{\rm{arctan}}\lpar{#1}\rpar}
\newcommand{\dbff}[3]{{\hat{B}}_{_{{#2}{#3}}}\lpar{#1}\rpar}
\newcommand{\ztbs}{{\bar{z}}^{2}_{\ft}}
\newcommand{\ztb}{{\bar{z}}_{\ft}}
\newcommand{\Htbs}{{\bar{h}}^{2}_{\ft}}
\newcommand{\Htb}{{\bar{h}}_{\ft}}
\newcommand{\Hztb}{{\bar{hz}}_{\ft}}
\newcommand{\Ln}[1]{{\rm{Ln}}\lpar{#1}\rpar}
\newcommand{\Lns}[1]{{\rm{Ln}}^2\lpar{#1}\rpar}
\newcommand{\wt}{w_{\ft}}
\newcommand{\wts}{w^2_{\ft}}
\newcommand{\wtb}{\overline{w}}
\newcommand{\fra}{\frac{1}{2}}
\newcommand{\frb}{\frac{1}{4}}
\newcommand{\frc}{\frac{3}{2}}
\newcommand{\frd}{\frac{3}{4}}
\newcommand{\fre}{\frac{9}{2}}
\newcommand{\frf}{\frac{9}{4}}
\newcommand{\frg}{\frac{5}{4}}
\newcommand{\frh}{\frac{5}{2}}
\newcommand{\fri}{\frac{1}{8}}
\newcommand{\frj}{\frac{7}{4}}
\newcommand{\frl}{\frac{7}{8}}
\newcommand{\Spzzh}{\hat{\Sigma}'_{_{\zb\zb}}}
\newcommand{\sss}{s\sqrt{s}}
\newcommand{\sqs}{\sqrt{s}}
\newcommand{\Rtg}{R_{_{3Q}}}
\newcommand{\Rtt}{R_{_{33}}}
\newcommand{\Rww}{R_{_{\wb\wb}}}
\newcommand{\ssZ}{{\scriptscriptstyle{\zb}}}
\newcommand{\ssW}{{\scriptscriptstyle{\wb}}}
\newcommand{\ssH}{{\scriptscriptstyle{\hb}}}
\newcommand{\ssV}{{\scriptscriptstyle{\vb}}}
\newcommand{\ssA}{{\scriptscriptstyle{A}}}
\newcommand{\ssB}{{\scriptscriptstyle{B}}}
\newcommand{\ssF}{{\scriptscriptstyle{F}}}
\newcommand{\ssG}{{\scriptscriptstyle{G}}}
\newcommand{\ssL}{{\scriptscriptstyle{L}}}
\newcommand{\ssM}{{\scriptscriptstyle{M}}}
\newcommand{\ssN}{{\scriptscriptstyle{N}}}
\newcommand{\ssQ}{{\scriptscriptstyle{Q}}}
\newcommand{\ssR}{{\scriptscriptstyle{R}}}
\newcommand{\ssS}{{\scriptscriptstyle{S}}}
\newcommand{\ssU}{{\scriptscriptstyle{U}}}
\newcommand{\ssX}{{\scriptscriptstyle{X}}}
\newcommand{\ssY}{{\scriptscriptstyle{Y}}}
\newcommand{\ssWF}{{\scriptscriptstyle{WF}}}
\newcommand{\DiagramFermionToBosonFullWithMomenta}[8][70]{
  \vcenter{\hbox{
  \SetScale{0.8}
  \begin{picture}(#1,50)(15,15)
    \put(27,22){$\nearrow$}      
    \put(27,54){$\searrow$}    
    \put(59,29){$\to$}    
    \ArrowLine(25,25)(50,50)      \Text(34,20)[lc]{#6} \Text(11,20)[lc]{#3}
    \ArrowLine(50,50)(25,75)      \Text(34,60)[lc]{#7} \Text(11,60)[lc]{#4}
    \Photon(50,50)(90,50){2}{8}   \Text(80,40)[lc]{#2} \Text(55,33)[ct]{#8}
    \Vertex(50,50){2,5}          \Text(60,48)[cb]{#5} 
    \Vertex(90,50){2}
  \end{picture}}}
  }
\newcommand{\DiagramFermionToBosonPropagator}[4][85]{
  \vcenter{\hbox{
  \SetScale{0.8}
  \begin{picture}(#1,50)(15,15)
    \ArrowLine(25,25)(50,50)
    \ArrowLine(50,50)(25,75)
    \Photon(50,50)(105,50){2}{8}   \Text(90,40)[lc]{#2}
    \Vertex(50,50){0.5}         \Text(80,48)[cb]{#3}
    \GCirc(82,50){8}{1}            \Text(55,48)[cb]{#4}
    \Vertex(105,50){2}
  \end{picture}}}
  }
\newcommand{\DiagramFermionToBosonEffective}[3][70]{
  \vcenter{\hbox{
  \SetScale{0.8}
  \begin{picture}(#1,50)(15,15)
    \ArrowLine(25,25)(50,50)
    \ArrowLine(50,50)(25,75)
    \Photon(50,50)(90,50){2}{8}   \Text(80,40)[lc]{#2}
    \BBoxc(50,50)(5,5)            \Text(55,48)[cb]{#3}
    \Vertex(90,50){2}
  \end{picture}}}
  }
\newcommand{\DiagramFermionToBosonFull}[3][70]{
  \vcenter{\hbox{
  \SetScale{0.8}
  \begin{picture}(#1,50)(15,15)
    \ArrowLine(25,25)(50,50)
    \ArrowLine(50,50)(25,75)
    \Photon(50,50)(90,50){2}{8}   \Text(80,40)[lc]{#2}
    \Vertex(50,50){2.5}          \Text(60,48)[cb]{#3}
    \Vertex(90,50){2}
  \end{picture}}}
  }
\newcommand{\expgw}{\frac{\gf\mws}{2\srt\,\pi^2}}
\newcommand{\expgz}{\frac{\gf\mzs}{2\srt\,\pi^2}}
\newcommand{\Spww}{\Sigma'_{_{\wb\wb}}}
\newcommand{\shf}{{\hat s}^4}
\newcommand{\acz}{\scff{0}}
\newcommand{\acoo}{\scff{11}}
\newcommand{\acod}{\scff{12}}
\newcommand{\acdo}{\scff{21}}
\newcommand{\acdd}{\scff{22}}
\newcommand{\acdt}{\scff{23}}
\newcommand{\acdq}{\scff{24}}
\newcommand{\acto}{\scff{31}}
\newcommand{\actd}{\scff{32}}
\newcommand{\actt}{\scff{33}}
\newcommand{\actq}{\scff{34}}
\newcommand{\actc}{\scff{35}}
\newcommand{\acts}{\scff{36}}
\newcommand{\acoA}{\scff{1A}}
\newcommand{\acdA}{\scff{2A}}
\newcommand{\acdB}{\scff{2B}}
\newcommand{\acdC}{\scff{2C}}
\newcommand{\acdD}{\scff{2D}}
\newcommand{\actA}{\scff{3A}}
\newcommand{\actB}{\scff{3B}}
\newcommand{\actC}{\scff{3C}}
\newcommand{\ada}{\sdff{0}}
\newcommand{\adb}{\sdff{11}}
\newcommand{\adc}{\sdff{12}}
\newcommand{\add}{\sdff{13}}
\newcommand{\ade}{\sdff{21}}
\newcommand{\adf}{\sdff{22}}
\newcommand{\adg}{\sdff{23}}
\newcommand{\adh}{\sdff{24}}
\newcommand{\adi}{\sdff{25}}
\newcommand{\adj}{\sdff{26}}
\newcommand{\adl}{\sdff{27}}
\newcommand{\adm}{\sdff{31}}
\newcommand{\adn}{\sdff{32}}
\newcommand{\ado}{\sdff{33}}
\newcommand{\adp}{\sdff{34}}
\newcommand{\adq}{\sdff{35}}
\newcommand{\adr}{\sdff{36}}
\newcommand{\ads}{\sdff{37}}
\newcommand{\adt}{\sdff{38}}
\newcommand{\adu}{\sdff{39}}
\newcommand{\adw}{\sdff{310}}
\newcommand{\adv}{\sdff{311}}
\newcommand{\ady}{\sdff{312}}
\newcommand{\adz}{\sdff{313}}
\newcommand{\admt}{\frac{\tman}{\sman}}
\newcommand{\admu}{\frac{\uman}{\sman}}
\newcommand{\frm}{\frac{3}{8}}
\newcommand{\frn}{\frac{5}{8}}
\newcommand{\fro}{\frac{15}{8}}
\newcommand{\frp}{\frac{3}{16}}
\newcommand{\frq}{\frac{5}{16}}
\newcommand{\frr}{\frac{1}{16}}
\newcommand{\frs}{\frac{7}{2}}
\newcommand{\frt}{\frac{7}{16}}
\newcommand{\fru}{\frac{1}{3}}
\newcommand{\frw}{\frac{2}{3}}
\newcommand{\frz}{\frac{4}{3}}
\newcommand{\fry}{\frac{13}{3}}
\newcommand{\fraa}{\frac{11}{4}}
\newcommand{\bee}{\beta_{e}}
\newcommand{\beW}{\beta_{_\wb}}
\newcommand{\toDdrh}{{\ds\frac{2}{{\hat{\varepsilon}}}}}
\newcommand{\bqas}{\begin{eqnarray*}}
\newcommand{\eqas}{\end{eqnarray*}}
\newcommand{\mhcub}{M^3_{_H}}
\newcommand{\adComA}{\sdff{A}}
\newcommand{\adComB}{\sdff{B}}
\newcommand{\adComC}{\sdff{C}}
\newcommand{\adComD}{\sdff{D}}
\newcommand{\adComE}{\sdff{E}}
\newcommand{\adComF}{\sdff{F}}
\newcommand{\adComG}{\sdff{G}}
\newcommand{\adComH}{\sdff{H}}
\newcommand{\adComI}{\sdff{I}}
\newcommand{\adComJ}{\sdff{J}}
\newcommand{\adComL}{\sdff{L}}
\newcommand{\adComM}{\sdff{M}}
\newcommand{\adComN}{\sdff{N}}
\newcommand{\adComO}{\sdff{O}}
\newcommand{\adComP}{\sdff{P}}
\newcommand{\adComQ}{\sdff{Q}}
\newcommand{\adComR}{\sdff{R}}
\newcommand{\adComS}{\sdff{S}}
\newcommand{\adComT}{\sdff{T}}
\newcommand{\adComU}{\sdff{U}}
\newcommand{\adComAc}{\sdff{A}^c}
\newcommand{\adComBc}{\sdff{B}^c}
\newcommand{\adComCc}{\sdff{C}^c}
\newcommand{\adComDc}{\sdff{D}^c}
\newcommand{\adComEc}{\sdff{E}^c}
\newcommand{\adComFc}{\sdff{F}^c}
\newcommand{\adComGc}{\sdff{G}^c}
\newcommand{\adComHc}{\sdff{H}^c}
\newcommand{\adComIc}{\sdff{I}^c}
\newcommand{\adComJc}{\sdff{J}^c}
\newcommand{\adComLc}{\sdff{L}^c}
\newcommand{\adComMc}{\sdff{M}^c}
\newcommand{\adComNc}{\sdff{N}^c}
\newcommand{\adComOc}{\sdff{O}^c}
\newcommand{\adComPc}{\sdff{P}^c}
\newcommand{\adComQc}{\sdff{Q}^c}
\newcommand{\adComRc}{\sdff{R}^c}
\newcommand{\adComSc}{\sdff{S}^c}
\newcommand{\adComTc}{\sdff{T}^c}
\newcommand{\adComUc}{\sdff{U}^c}
\newcommand{\adComAf}{\sdff{A}^f}
\newcommand{\adComBf}{\sdff{B}^f}
\newcommand{\adComCf}{\sdff{F}^f}
\newcommand{\adComDf}{\sdff{D}^f}
\newcommand{\adComEf}{\sdff{E}^f}
\newcommand{\adComFf}{\sdff{F}^f}
\newcommand{\adComGf}{\sdff{G}^f}
\newcommand{\adComHf}{\sdff{H}^f}
\newcommand{\adComIf}{\sdff{I}^f}
\newcommand{\adComJf}{\sdff{J}^f}
\newcommand{\adComLf}{\sdff{L}^f}
\newcommand{\adComMf}{\sdff{M}^f}
\newcommand{\adComNf}{\sdff{N}^f}
\newcommand{\adComOf}{\sdff{O}^f}
\newcommand{\adComPf}{\sdff{P}^f}
\newcommand{\adComQf}{\sdff{Q}^f}
\newcommand{\adComRf}{\sdff{R}^f}
\newcommand{\adComSf}{\sdff{S}^f}
\newcommand{\adComTf}{\sdff{T}^f}
\newcommand{\adComUf}{\sdff{U}^f}
\newcommand{\adComBfc}{\sdff{B}^{fc}} 
\newcommand{\adComCfco}{\sdff{C}^{fc1}}
\newcommand{\adComCfcd}{\sdff{C}^{fc2}} 
\newcommand{\adComCfct}{\sdff{C}^{fc3}} 
\newcommand{\adComDfc}{\sdff{D}^{fc}}
\newcommand{\adComEfc}{\sdff{E}^{fc}}
\newcommand{\adComFfc}{\sdff{F}^{fc}}
\newcommand{\adComGfc}{\sdff{G}^{fc}}
\newcommand{\adComHfc}{\sdff{H}^{fc}}
\newcommand{\adComLfc}{\sdff{L}^{fc}}
\newcommand{\afba}[1]{A^{#1}_{_{\rm FB}}}
\newcommand{\alra}[1]{A^{#1}_{_{\rm LR}}}
\newcommand{\adComAt}{\sdff{A}^t}
\newcommand{\adComBt}{\sdff{B}^t}
\newcommand{\adComCt}{\sdff{T}^t}
\newcommand{\adComDt}{\sdff{D}^t}
\newcommand{\adComEt}{\sdff{E}^t}
\newcommand{\adComFt}{\sdff{T}^t}
\newcommand{\adComGt}{\sdff{G}^t}
\newcommand{\adComHt}{\sdff{H}^t}
\newcommand{\adComIt}{\sdff{I}^t}
\newcommand{\adComJt}{\sdff{J}^t}
\newcommand{\adComLt}{\sdff{L}^t}
\newcommand{\adComMt}{\sdff{M}^t}
\newcommand{\adComNt}{\sdff{N}^t}
\newcommand{\adComOt}{\sdff{O}^t}
\newcommand{\adComPt}{\sdff{P}^t}
\newcommand{\adComQt}{\sdff{Q}^t}
\newcommand{\adComRt}{\sdff{R}^t}
\newcommand{\adComSt}{\sdff{S}^t}
\newcommand{\adComTt}{\sdff{T}^t}
\newcommand{\adComUt}{\sdff{U}^t}
\newcommand{\adComAtt}{\sdff{A}^{\tau}}
\newcommand{\adComBtt}{\sdff{B}^{\tau}}
\newcommand{\adComCtt}{\sdff{T}^{\tau}}
\newcommand{\adComDtt}{\sdff{D}^{\tau}}
\newcommand{\adComEtt}{\sdff{E}^{\tau}}
\newcommand{\adComFtt}{\sdff{T}^{\tau}}
\newcommand{\adComGtt}{\sdff{G}^{\tau}}
\newcommand{\adComHtt}{\sdff{H}^{\tau}}
\newcommand{\adComItt}{\sdff{I}^{\tau}}
\newcommand{\adComJtt}{\sdff{J}^{\tau}}
\newcommand{\adComLtt}{\sdff{L}^{\tau}}
\newcommand{\adComMtt}{\sdff{M}^{\tau}}
\newcommand{\adComNtt}{\sdff{N}^{\tau}}
\newcommand{\adComOtt}{\sdff{O}^{\tau}}
\newcommand{\adComPtt}{\sdff{P}^{\tau}}
\newcommand{\adComQtt}{\sdff{Q}^{\tau}}
\newcommand{\adComRtt}{\sdff{R}^{\tau}}
\newcommand{\adComStt}{\sdff{S}^{\tau}}
\newcommand{\adComTtt}{\sdff{T}^{\tau}}
\newcommand{\adComUtt}{\sdff{U}^{\tau}}
\newcommand{\etavz}[1]{\eta^{\zb #1}_{_V}}
\newcommand{\phanst}{$\hphantom{\sigma^{s+t}\ }$}
\newcommand{\phanat}{$\hphantom{A_{FB}^{s+t}\ }$}
\newcommand{\phanss}{$\hphantom{\sigma^{s}\ }$}
\newcommand{\phanas}{$\hphantom{A_{FB}^{s}\ }$} 
\newcommand{\pbb}{\,\mbox{\bf pb}}
\newcommand{\pe}{\,\%\:}
\newcommand{\pc}{\,\%}
\newcommand{\temiv}{10^{-4}}
\newcommand{\temv}{10^{-5}}
\newcommand{\temvi}{10^{-6}}
\newcommand{\di}[1]{d_{#1}}
\newcommand{\delip}[1]{\Delta_+\lpar{#1}\rpar}
\newcommand{\propbb}[5]{{{#1}\over {\lpar #2^2 + #3 - \ib\varepsilon\rpar
\lpar\lpar #4\rpar^2 + #5 -\ib\varepsilon\rpar}}}
\newcommand{\cfft}[5]{C_{#1}\lpar #2;#3,#4,#5\rpar}    
\newcommand{\ppl}[1]{p_{+{#1}}}
\newcommand{\pmi}[1]{p_{-{#1}}}
\newcommand{\bpox}{\beta^2_{\xi}}
\newcommand{\bffdiff}[5]{B_{\rm d}\lpar #1;#2,#3;#4,#5\rpar}             
\newcommand{\cffdiff}[7]{C_{\rm d}\lpar #1;#2,#3,#4;#5,#6,#7\rpar}    
\newcommand{\affdiff}[2]{A_{\rm d}\lpar #1;#2\rpar}             
\newcommand{\Dqf}{\Delta\qf}
\newcommand{\bposx}{\beta^4_{\xi}}
\newcommand{\svverti}[3]{f^{#1}_{#2}\lpar{#3}\rpar}
\newcommand{\Mods}{\mbox{$M^2_{12}$}}
\newcommand{\Mots}{\mbox{$M^2_{13}$}}
\newcommand{\Motq}{\mbox{$M^4_{13}$}}
\newcommand{\Mdts}{\mbox{$M^2_{23}$}}
\newcommand{\Mdos}{\mbox{$M^2_{21}$}}
\newcommand{\Mtds}{\mbox{$M^2_{32}$}}
\newcommand{\dffpt}[3]{D_{#1}\lpar #2,#3;}           
\newcommand{\quu}{Q_{uu}}
\newcommand{\qdd}{Q_{dd}}
\newcommand{\qud}{Q_{ud}}
\newcommand{\qdu}{Q_{du}}
\newcommand{\msPj}[6]{\Lambda^{#1#2#3}_{#4#5#6}}
\newcommand{\bdiff}[4]{B_{\rm d}\lpar #1,#2;#3,#4\rpar}             
\newcommand{\bdifff}[7]{B_{\rm d}\lpar #1;#2;#3;#4,#5;#6,#7\rpar}             
\newcommand{\adiff}[3]{A_{\rm d}\lpar #1;#2;#3\rpar}  
\newcommand{\aw}{a_{_\wb}}
\newcommand{\az}{a_{_\zb}}
\newcommand{\sct}[1]{sect.~\ref{#1}}
\newcommand{\gcpm}[1]{g^{\pm}_{#1}}
\newcommand{\dreim}[1]{\varepsilon^{\rm M}_{#1}}
\newcommand{\drem}{\varepsilon^{\rm M}}
\newcommand{\hcapV}[2]{{\hat{\cal F}}^{#2}_{_{#1}}}
\newcommand{\swww}{{\scriptscriptstyle \wb\wb\wb}}
\newcommand{\szhz}{{\scriptscriptstyle \zb\hb\zb}}
\newcommand{\shzh}{{\scriptscriptstyle \hb\zb\hb}}
\newcommand{\bwith}[3]{\beta^{#3}_{#1}\lpar #2\rpar}
\newcommand{\Shhh}{{\hat\Sigma}_{_{\hb\hb}}}
\newcommand{\Sphhh}{{\hat\Sigma}'_{_{\hb\hb}}}
\newcommand{\seWilc}[1]{w_{#1}}
\newcommand{\seWtilc}[2]{w_{#1}^{#2}}
\newcommand{\eilc}{\gamma}
\newcommand{\eilcs}{\gamma^2}
\newcommand{\eilcb}{{\overline{\gamma}}}
\newcommand{\eilcbs}{{\overline{\gamma}^2}}
\newcommand{\Sttww}{\Sigma_{_{33;\wb\wb}}}
\newcommand{\bSttww}{{\overline\Sigma}_{_{33;\wb\wb}}}
\newcommand{\Pggtg}{\Pi_{\ph\ph;3Q}}
\newcommand{\hpmomi}[1]{{\hat p}_{#1}}
\newcommand{\hQ}{{\hat Q}}
\newcommand{\hQs}{{\hat Q}^2}
\newcommand{\hQq}{{\hat Q}^4}
\newcommand{\hy}{{\hat y}}
\newcommand{\bk}{{\bar k}}
%
%
\begin{titlepage}

\begin{center}
\baselineskip25pt

{\Large\sc The Single $\wb$ Production Case}

\end{center}

\setcounter{footnote}{3}

\vspace{1cm}

\begin{center}
\baselineskip12pt

{\large\sc 
Giampiero Passarino} 

\vspace{1cm}

Dipartimento di Fisica Teorica, Universit\`a di Torino, Italy \\
INFN, Sezione di Torino, Italy$^{\dagger}$

\vspace{0.3cm}

\end{center}

\vspace{2cm}

\begin{abstract}
  \normalsize \noindent 
\end{abstract}
The process $e^+e^- \to e^- \barnu_e f_1 \barf_2$, belonging to the so-called
CC20 family, has been extensively analyzed in the literature.
It is a sensitive probe of anomalous electromagnetic couplings of the $\wb$ 
boson and represents a background to searches for new physics beyond the 
standard model.
Moreover, it represents a contribution to the $e^+e^- \to \wbp\wbm$ total 
cross section, used to derive a value for $\mw$, the $\wb$ boson mass.
The issue of gauge invariance in the CC20 family has been solved by the
introduction of the Fermion-Loop scheme but several subtleties remain,
connected with the region of vanishing scattering angle of the electron
and with the limit of massless final state fermions in a fully extrapolated
setup. A satisfactory solution for computing the total cross section is given 
in the context of the equivalent photon or Weizs\"acker-Williams approximation 
which factorizes the flux of quasi-real photons emitted by the electron from 
the interaction rate between the positron and the photon assumed to be real.
The correct kinematics for the inclusion of initial state QED radiation
is established.
QCD corrections to the process are discussed and numerical results are shown 
and commented.

\vskip 60pt
\noindent 
\rule[.1in]{12cm}{.002in}

\noindent
$^{\dagger)}$ e-mail address: giampiero@to.infn.it.
\vspace*{1.0cm}

\end{titlepage}

\def\thefootnote{\arabic{footnote}} \setcounter{footnote}{0}

\setcounter{page}{1}
\section{Introduction\label{sect_1}}

The so-called single $\wb$ production process is $e^+e^- \to e^-\barnu_e \wbp$
for small scattering angles of the outgoing electron
and it has been measured at LEP 2 at centre-of-mass energies $130\,\GeV \le 
\sqrt{s} \le 183\,\GeV$ using both leptonic and hadronic decays of $\wb$ bosons
\cite{kn:l3}.
The signal is, therefore, defined as
\bqa
e^+e^- &\to& e^-\, \barnu_e\, l^-\, \barnu_l ,  \nl
e^+e^- &\to& e^-\, \barnu_e\, u\, \bard,  
\eqa
where $u(d)$ stands for a generic up-(down-)quark. In the terminology of
\cite{kn:lep2ww} it is a four-fermion process belonging to the 
CC20-family. The charge conjugate reactions are always understood to
be included.
The cross section for single $\wb$ production is expected to be small at 
LEP 2 energies, of the order of $0.5\,\pb$. However, this process constitutes
a very interesting case both theoretically and experimentally.
It is a sensitive probe of anomalous electromagnetic couplings of the $\wb$ 
boson and represents a background to searches for new physics beyond the 
standard model.
\vspace{0.2cm}
\bqas
\ba{ccc}
\vcenter{\hbox{
  \begin{picture}(110,100)(0,0)
  \ArrowLine(50,50)(0,100)
  \ArrowLine(100,100)(50,50)
  \ArrowLine(0,0)(50,50)
  \ArrowLine(50,50)(100,0)
  \ArrowLine(100,70)(50,50)
  \ArrowLine(50,50)(100,30)
  \GCirc(50,50){15}{0.5}
  \Text(10,100)[lc]{$e^+$}
  \Text(10,0)[lc]{$e^-$}
  \Text(110,100)[lc]{$\barnu_e$}
  \Text(110,70)[lc]{$\barf_2$}
  \Text(110,30)[lc]{$f_1$}
  \Text(110,0)[lc]{$e^-$}
  \end{picture}}}
&\quad+&
\vcenter{\hbox{
  \begin{picture}(110,100)(0,0)
  \ArrowLine(50,50)(0,100)
  \ArrowLine(100,100)(50,50)
  \ArrowLine(100,70)(50,50)
  \ArrowLine(50,50)(100,30)
  \Photon(50,50)(50,10){2}{7}
  \ArrowLine(0,0)(50,10)
  \ArrowLine(50,10)(100,0)
  \GCirc(50,50){15}{0.5}
  \Text(10,100)[lc]{$e^+$}
  \Text(10,0)[lc]{$e^-$}
  \Text(110,100)[lc]{$\barnu_e$}
  \Text(110,70)[lc]{$\barf_2$}
  \Text(110,30)[lc]{$f_1$}
  \Text(110,0)[lc]{$e^-$}
  \Text(60,20)[lc]{$Q^2$}
  \end{picture}}}
\ea
\eqas
\vspace{-2mm}
\begin{figure}[h]
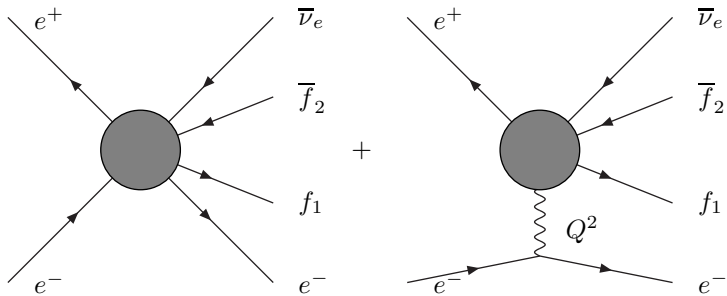

\caption[]{The CC20 family of diagrams with the explicit component containing
a $t$-channel photon.}
\label{basic}
\end{figure}
\vskip 5 pt
The  CC20 process of \fig{basic} is sensitive to the breaking of U(1)
gauge invariance in the collinear limit.
For $e^+e^- \to e^-\fbnue f_1\barf_2$, the U(1) gauge 
invariance becomes essential in the region of phase space where the angle 
between the incoming and outgoing electrons is small, see the work of
\cite{kn:fl} and also alternative formulations in \cite{kn:kuri,kn:zepp}.
In this limit the superficial $1/Q^4$ divergence of the  
propagator structure is reduced to $1/Q^2$ by U(1) gauge  
invariance. In the presence of light fermion masses this gives raise  
to the familiar $\ln(\mes/s)$ large logarithms.
The correct way of handling CC20 is represented by the so-called 
Fermion-Loop (FL)scheme \cite{kn:fl}, the gauge-invariant treatment of the 
finite-width effects of $\wb$ and $\zb$ bosons in LEP2 processes.
Briefly, this scheme consists in including all fermionic one-loop 
corrections in tree-level amplitudes and re-summing the self-energies. 
However, for practical applications, it has been shown that the 
fixed-width scheme is also satisfactory. 
Here the cross-section is computed using the tree-level amplitude. The massive 
gauge-boson propagators are given by  $1/(\pmoms+\mlones-\ib\Glone\mlone)$.
This gives an unphysical width for $\pmoms>0$, but retains U(1) gauge 
invariance in the  CC20 process~\cite{kn:ofl}.

The CC20 process is usually considered in two regimes, $|\cos\theta(e^-)|
\ge c$ or LACC20 and $|\cos\theta(e^-)| \le c$ or SACC20.
Strictly speaking the single $\wb$ production is defined by those events that
satisfy $|\cos\theta(e^-)| \ge 0.997$ and, therefore is a SACC20.

The LACC20 cross section has been computed by many authors and references can 
be found in \cite{kn:lep2ww}. It
represents a contribution to the $e^+e^- \to \wbp\wbm$ total cross section, in 
turn used to derive a value for $\mw$, the $\wb$ boson mass.
This point deserves a comment: by $e^+e^- \to \wbp\wbm$ it is meant the
ideal cross section obtained with the three double-resonant CC03
diagrams and therefore the background, FULL -  CC03, is
evaluated with the help of some MonteCarlo, estimating the error on the 
subtraction by comparing with some other MonteCarlo.
Then $\mw$ is derived from a fit to $\sigma$(CC03) with the help of a third 
calculation. From a theoretical point of view the evaluation of LACC20
is free of ambiguity, even in the approximation of massless fermions, as long
as a gauge-preserving scheme is applied and $\theta(e^-)$ is not too small.

For SACC20 instead, one cannot employ the massless approximation anymore
and this fact makes the calculation unaccessible to most of the 
MonteCarlos used by the experimental collaborations, with the noticeable 
exception of GRC4F \cite{kn:grace}. 
EXCALIBUR \cite{kn:exca} is often used in this context with a version where
a fudge is put so that for one-electron final states on can go down to 
zero scattering angle.\footnote{R.~Pittau private communication.}

Actually, constructing a CC20 calculation with
unconstrained electron scattering angle is not a problem from the point of
view of writing the fully massive amplitudes but it is, instead, a question
of stability in the numerical integration. Moreover the goal of this paper
is to show that several subtleties arise in CC20 for a fully extrapolated
setup.

Single $\wb$ production and $\mw$ measurement are, therefore, complementary.
Indeed, the phase space requirement $|\cos\theta(e^-)| > c$ eliminates events
predominantly consisting of $\wb$ pair production since 
single $\wb$ production peaks strongly at zero scattering angle.

There is another place where the electron angle cannot be constrained.
Experimentally  events of the type $e^+e^- \to \fu\bard$
plus a neutrino and an electron, possibly in the beam pipe, are not
excluded from the hadronic $\zb$ lineshape.
Hadronic events are selected based on final state particle
multiplicity in the detector, so both genuine high-energy $q\barq$
events, radiative return events and hadronic four-fermion events are
selected for the $e^+e^- \to \barq q$ lineshape.
This gives the total sample. The background is subtracted on MonteCarlo basis,
using $\wb$ double-resonant  CC03 diagrams, i.e. $\wb\wb \to\,$ all, 
$\zb$ double-resonant NC02 diagrams, i.e. $\zb\zb \to\,$ all
and hadronic two-photon collisions (e.g. using PHOJET \cite{kn:twop}).

Therefore double-resonant $\wb$'s, which are dominant, are treated correctly
in the experimental procedure and, moreover, single-resonant $\wb$'s only 
represent a small contribution. The latter 
could, however, be treated correctly by using a CC20 MonteCarlo rather 
than a CC03 one. In this case the electron is again unconstrained and
one needs a full angle, massive CC20: the so-called full-CC20, or FCC20, regime.

However, keeping a finite electron mass through the calculation is not enough.
One of the main results of this paper is to show that
there are subtleties in CC20 also associated with the zero mass limit
for the remaining fermions.

The outline of the paper will be as follows. In Sect. \ref{sect_2} we 
introduce the general problem of defining the total hadronic cross section at
LEP2 energies and beyond, and describe how a calculation of FCC20 is to
be seen in this context. In Sect. \ref{sect_3} we give a description
of FCC20 in the presence of initial state QED radiation and show how to 
implement the correct kinematics for the process.
In Sect. \ref{sect_4} we introduce and discuss the Weizs\"acker-Williams  
approximation for a small scattering angle of the outgoing electron.
The sub-process $e^+ \gamma \to \barnu_e u \bard$, arising in the discussion
of the WW-approximation is analyzed in Sects. \ref{sect_5}-\ref{sect_6}.
The fully extrapolated setup with massless quarks and QCD corrections
are presented in Sect. \ref{sect_7}. Finally, numerical results and 
conclusions are shown in Sect. \ref{sect_8}.

\section{The region of vanishing $\theta(e^-)$\label{sect_2}}

There are at least three applications of CC20 which require an analysis at
vanishing scattering angle of the outgoing electron, $\theta(e^-)$. They are:

\begin{enumerate}

\item the true single $\wb$ production, i.e. CC20 with $|\cos\theta(e^-)|
\ge c$ where, usually, $c = 0.997$,

\item the evaluation of background for the total $e^+e^- \to \wbp\wbm$
cross section,

\item the evaluation of the inclusive hadronic cross section at LEP2 energies.

\end{enumerate}

Let us consider in more detail the last application.
The FCC20 process is not the only background for the total
hadronic cross section $\sigma(\barq q X)$ defined as the cross section 
for $\barq q$ plus anything. Here, we would like to illustrate
the general problem, to return in the next section to the study of FCC20.

The total hadronic cross section that we have defined is an inclusive 
measurement of 
hadron production in $e^+e^-$-annihilation in which production thresholds can 
be seen, e.g. $\wb$-pair production with at least one of the $\wb$ bosons 
decaying hadronically, or $\zb\zb$ or other background \cite{kn:opal}.

Let us repeat what has been done so far in the experimental Collaborations.
Hadronic events are selected based on final state particle multiplicity
in the detector, so both genuine high energy $\barq q$ events,
radiative returns and four-fermion hadronic events 
are selected for the hadronic lineshape.
This gives the total sample: the background is subtracted on
a MonteCarlo, using CC03, NC02 and hadronic two-photon collisions.
Clearly the above strategy is good enough for the present precision, but wrong
in principle.
Let us consider the relation between (radiatively corrected) two-fermion (2F)
and four-fermion (4F) final states in $e^+e^-$ annihilation at LEP~2.
There are several components in the radiative corrections to
fermion pair production: among them there is initial-state (or final-state) 
fermion-pair production.
For definiteness consider $e^+e^- \to \barb b$ with radiation of an $e^+e^-$ 
pair \cite{kn:epi}.
The background is represented by the full four-fermion process,
the so-called NC48 process, which is built out of 48 Feynman diagrams. 
For studies around the $\zb$ resonance the default \cite{kn:ocyr95} was to 
included pairs from 
initial state and a cut was selected so that $M(\barb b) > 0.25\,s$.
At LEP2 energies or higher one needs a more precise separation 
between radiative corrections to 2F production and real 4F events 
\cite{kn:narr}.

We will denote the evaluation of any one-loop corrected
cross section, e.g. $e^+e^-\to b\barb$ as a 2F-calculation.
By 4F-calculation we mean a tree level evaluation, e.g. $e^+e^-\to b\barb 
e^+e^-$.
Note that the soft pairs, $\gamma^*\to e^+e^-$ are divergent in the limit of 
zero $e^+e^-$ invariant mass and therefore any simulation of very soft pairs 
with massless 4F-calculations is bound to produce wrong results.

But also a massive 4F-calculation is not enough,
because if pairs are soft enough we must include virtual pairs as well,
and all $e^+e^-$ pairs are allowed down to $M(e^+e^-) = 2\,m_e$.

Also soft+virtual initial/final pairs in a 2F-calculation are not enough
because no upper cut is imposed on $M(e^+e^-)$, 
so that all pairs compatible with the request $M(\barb b) >$ (some thresholds) 
are accepted. 
Thus there is more than S+V pairs, there are 
many topologies for hard pairs and some of them require a finite $\me$ also 
for hard pairs. Indeed in NC48 there are multi-peripheral diagrams which 
diverge for $\me \to 0$. 

The evaluation of $\sigma(\barq q X)$ requires \cite{kn:epi}

\begin{itemize}

\item[A] Include virtual+soft (up to some invariant mass $\Delta$) 
I/F state pairs with a complete 2F-calculation.

\item[B] The contributions to $e^+e^- \to \barb b e^+e^-$ not in [A]
are then included with a constraint on $M(\barb b)$, 
but with no further restriction on $M(e^+e^-)$, with a complete 
(i.e. fully massive) 4F-calculation.

\item[C] The contributions to $e^+e^- \to \barb b e^+e^-$that are 
already contained in [A], are included with $M(e^+e^-) > \Delta$.

\end{itemize}

Step B requires evaluation of the following 4F-processes:
Fully hadronic, with at least one invariant mass passing the cut,
semi-leptonic, with $ M(q_j\barq_i)$ passing the cut.

To summarize we may say that around the $\zb$ resonance
the rate for real and virtual radiation is known \cite{kn:pair} and included
in the existing calculations.
Both are enhanced by large logarithms but they cancel
to a large extent, leading to a small contribution to the inclusive
decay rates.

The complete evaluation of $\sigma(\barq q X)$ would be
relatively easy if we could separate sub-classes of diagrams, e.g.
primary from secondary production.
For that it is necessary that the interference between them be 
zero, or very small or, at least non-singular.
In the limit of massless fermions, singularities will arise from:

\begin{enumerate}
\item $s$-channel $ \gamma(g)$
\item $t$-channel $\gamma$ with outgoing $ e^{\pm}$ lost in the beam pipe.
\end{enumerate}

This consideration suggests the appropriate strategy:
classes of diagrams showing a mass singularity 
must be included through some analytical calculation which also accounts for 
$\ord{\alpha^2}$ virtual radiation,
all interferences exhibiting mass singularities belong to this
category while the rest, including most of the interferences, are accounted for
by some (numerical) massless 4F-calculation.

The goal of this paper is to investigate in more detail the class 2) introduced
above, of which CC20 is a prototype.

\section{Kinematics and structure functions\label{sect_3}}

The inclusion of QED initial state radiation in $e^+e^-$-annihilation is
based on renormalization group ideas and on factorization of mass
singularities. The corresponding cross section may be cast into the 
following form:
\bq
\sigma(s) = \intfx{x_1}\intfx{x_2} \,\Theta_{\rm cut} \asums{f=e^+,e^-}
D^f_{e^-}(x_1,s)D^{\barf}_{e^+}(x_2,s)\,{\hat\sigma}_{f\barf}(x_1x_2s),
\eq
where the structure function $D^f_e(x,s)$ is the probability density to find a
{\em parton} $f$ with energy fraction $x$. The restriction on the region of
integration, given by $\Theta_{\rm cut}$, reflects the presence of kinematical 
cuts.

Let $p$ be the four-momentum of the incoming electron in the laboratory
system,
\bq
p = \frac{1}{2}\sqrt{s}\,\lpar 0,0,\beta,1\rpar, \qquad \beta^2= 1 - 4\,
\frac{\mes}{s}.
\label{defbeta}
\eq
The electron, before interacting, emits soft and collinear photons. Let $k=
k_1+k_2+\dots$ be the total four-momentum of the radiated photons. Thus
\bq
k= \frac{1}{2}\sqrt{s}\,(1-x)\,\lpar 0,0,1,1\rpar,
\eq
so that $k^2= 0$, as requested by collinear, massless, photons.
Usually one can work with the massless approximation for the electron taking
part in the hard scattering, thus an on-shell (massless) electron can
emit a bunch of massless, collinear, photons and remain on its (massless)
mass shell. But the electron mass cannot be neglected in the hard CC20
scattering and, after radiation, the electron finds itself in a
virtual state having four-momentum
\bq
{\hat p} = p - k = \frac{1}{2}\sqrt{s}\,\lpar 0,0,\beta-1+x,x\rpar,
\eq
with $x$ being the fraction of energy remaining after radiation. As a 
consequence, the electron is put off its mass shell,
\bq
{\hat p}^2 = - \mes + \frac{1}{2}(1-\beta)(1-x)\,s \sim -x\,\mes \quad
\mbox{for} \quad \me \to 0.
\eq
When considering the whole process we introduce $p_{\pm}$ for the incoming 
$e^{\pm}$ in the laboratory system. Once radiation has been emitted
the momenta will be denoted by ${\hat p}_{\pm}$ with
\bq
{\hat p}_{\pm} = \frac{1}{2}\sqrt{s}\,\lpar 0,0,\mp(\beta-1+x_{\pm}),
x_{\pm}\rpar.
\eq
The total four-momentum becomes
\bq
{\hat P} = \hpmomi{+} + \hpmomi{-} = \frac{1}{2}\sqrt{s}\,\lpar 0,0,
x_- - x_+, x_- + x_+\rpar,
\eq
with a corresponding invariant mass
\bq
{\hat P}^2 = - x_+x_-\,s = \shat
\eq
In the following we will be able to discuss the effects of a correct treatment
of QED initial-state radiation (ISR) on the processes under consideration.
\section{Weizs\"acker-Williams approximation for CC20\label{sect_4}}

\begin{figure}[h]
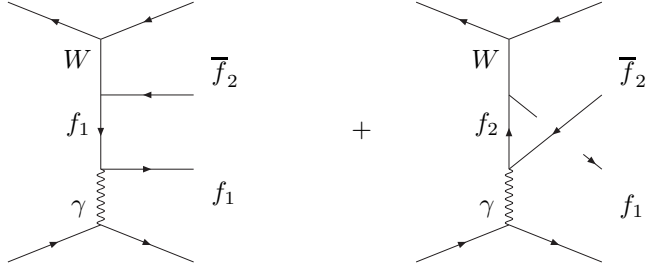

\caption[]{The CC20$_{\gamma}$ family of diagrams.}
\label{cc20g}
\end{figure}

The strategy for the calculation of the CC20 process will be as follows. First,
we split the 20 Feynman diagrams of the CC20 family into the four diagrams 
of \fig{cc20g}, characterized by the presence of a $t$-channel photon, and 
the rest
\bq
\mbox{CC20} = \mbox{CC20}_{\gamma} + \mbox{CC20}_{\rm R}.
\label{splitCC20}
\eq
Then we introduce $\theta_c$, the angle separating the SACC20 from the LACC20 
regions. The total cross section will be computed as
\bq
|\mbox{CC20}^{<}_{\gamma}(\me)|^2 + |\mbox{CC20}^{>}(0)|^2 +
2\,\Bigl[ \mbox{CC20}^{<}_{\gamma}(0)\Bigr]^{\dagger}\,
\mbox{CC20}^{<}_{\rm R}(0) + |\mbox{CC20}^{<}_{\rm R}(0)|^2 
\label{split}
\eq
where CC20$^{>}$(CC20$^{<}$) implies $\theta>\theta_c(\theta<\theta_c)$ and
the argument $\me(0)$ implies a finite(zero) electron mass. For the first term
in \eqn{split} we need an analytical calculation which keeps $\me\ne 0$ while 
the remaining terms can be treated numerically with the approximation of 
$\me = 0$.
The square of the four diagrams of \fig{cc20g} will be computed within the
improved Weizs\"acker-Williams - approximation (WW), provided that $\theta_c$ 
is not too large. This application of the WW-approximation is very similar
to the one applied in \cite{kn:wwa}.
\vspace{0.2cm}
\bqas
\ba{ccc}
\vcenter{\hbox{
  \SetScale{0.7}
  \begin{picture}(110,100)(0,0)
  \ArrowLine(50,120)(0,140)
  \ArrowLine(100,140)(50,120)
  \ArrowLine(0,0)(50,20)
  \ArrowLine(50,20)(100,0)
  \ArrowLine(100,110)(80,70)
  \ArrowLine(80,70)(100,30)
  \Photon(50,20)(50,70){2}{7}
  \Line(50,70)(50,120)
  \Line(50,70)(80,70)
  \Text(-14,98)[lc]{$e^+$}
  \Text(77,98)[lc]{$\barnu_e$}
  \Text(-14,0)[lc]{$e^-$}
  \Text(77,0)[lc]{$e^-$}
  \Text(77,77)[lc]{$\barf_2$}
  \Text(77,21)[lc]{$f_1$}
  \Text(22,60)[lc]{$\wb$}
  \Text(25,26)[lc]{$\gamma$}
  \end{picture}}}
&\quad+&
\vcenter{\hbox{
  \SetScale{0.7}
  \begin{picture}(110,100)(0,0)
  \ArrowLine(50,120)(0,140)
  \ArrowLine(65,126)(50,120)
  \ArrowLine(100,140)(65,126)
  \ArrowLine(0,0)(50,20)
  \ArrowLine(50,20)(100,0)
  \ArrowLine(100,110)(80,70)
  \ArrowLine(80,70)(100,30)
  \Photon(50,20)(50,120){2}{7}
  \Line(65,126)(80,70)
  \Text(77,77)[lc]{$\barf_2$}
  \Text(77,21)[lc]{$f_1$}
  \Text(54,76)[lc]{$\wb$}
  \Text(25,66)[lc]{$\gamma$}
  \end{picture}}}
\ea
\eqas
\bqas
\ba{ccc}
\vcenter{\hbox{
  \SetScale{0.7}
  \begin{picture}(110,100)(0,0)
  \ArrowLine(50,120)(0,140)
  \ArrowLine(100,140)(50,120)
  \ArrowLine(0,0)(50,20)
  \ArrowLine(50,20)(100,0)
  \ArrowLine(100,90)(50,90)
  \ArrowLine(50,90)(50,50)
  \ArrowLine(50,50)(100,50)
  \Photon(50,20)(50,50){2}{7}
  \Line(50,90)(50,120)
  \Text(77,72)[lc]{$\barf_2$}
  \Text(77,26)[lc]{$f_1$}
  \Text(22,52)[lc]{$f_1$}
  \Text(24,20)[lc]{$\gamma$}
  \Text(21,77)[lc]{$\wb$}
  \end{picture}}}
&\quad+&
\vcenter{\hbox{
  \SetScale{0.7}
  \begin{picture}(110,100)(0,0)
  \ArrowLine(50,120)(0,140)
  \ArrowLine(100,140)(50,120)
  \ArrowLine(0,0)(50,20)
  \ArrowLine(50,20)(100,0)
  \ArrowLine(100,90)(50,50)
  \Line(50,90)(65,78)
  \ArrowLine(90,58)(100,50)
  \ArrowLine(50,50)(50,90)
  \Photon(50,20)(50,50){2}{7}
  \Line(50,90)(50,120)
  \Text(77,72)[lc]{$\barf_2$}
  \Text(77,21)[lc]{$f_1$}
  \Text(22,52)[lc]{$f_2$}
  \Text(24,20)[lc]{$\gamma$}
  \Text(21,77)[lc]{$\wb$}
  \end{picture}}}
\ea
\eqas
\vspace{-2mm}
\vskip 5 pt
The advantage of using the improved WW-approximation is in the possibility
of performing an analytical integration over the momentum transferred to the
photon, allowing to obtain the exact logarithmic enhancement as well as the 
first, constant, correction to it.

The kernel cross section for the process CC20$_{\gamma}(\me)$, 
\bq
e^+(\hpmomi{+}) e^-(\hpmomi{-}) \to e^-(q_-) \barnu_e(q_+) u(k) 
\bard(\bk), 
\eq
can be written as
\bq
{\hat\sigma} = \frac{\gbs\stws}{(2\,\pi)^8}\,\frac{N_c}{2\,\shat}\,
\int d^4q_-\,\delta^+\lpar q^2_-+\mes\rpar\,\int d\Phi_3\,
\frac{1}{\hQq}\,{\hat L}_{\mu\nu}{\hat W}_{\mu\nu},
\label{sigmag}
\eq
where $\hQ = \hpmomi{-} - q_-$ and $N_c = 1$ for a fully leptonic final state 
and $3$ otherwise. Furthermore $\stw$ is the sine of the weak mixing angle.
In this equation, $d\Phi_3$ is the phase space integral for the
$\barnu_e u\bard$ system, $\delta^+(p^2+m^2) = \theta(E)\delta(p^2+m^2)$
and ${\hat L},{\hat W}$ are the leptonic tensor and the SACC20 tensor.
A straightforward calculation gives
\bq
{\hat L}_{\mu\nu} = \frac{1}{2}\,\Bigl[ \hQs - \lpar 1-x_-\rpar\,\mes
\Bigr]\,\drii{\mu}{\nu} + {\hat p}_{-\mu}q_{-\nu} + {\hat p}_{-\nu}q_{-\mu}.
\eq
The four diagrams of \fig{cc20g} form a $U(1)$ gauge-invariant set and therefore
$\hQ_{\mu}{\hat W}_{\mu\nu} = \hQ_{\nu}{\hat W}_{\mu\nu} = 0$.
The ${\hat W}$-tensor admits a decomposition into three form factors
\bqa
\int\,d\Phi_3\,{\hat W}_{\mu\nu} &=& {\hat W}_1\,\lpar -\drii{\mu}{\nu} + 
\frac{\hQ_{\mu}\hQ_{\nu}}{\hQs}\rpar -
{\hat W}_2\,\frac{\hQs}{\lpar \spro{\hpmomi{+}}{\hQ}\rpar^2}\,
{\cal P}_{\mu}{\cal P}_{\nu} + {\hat W}_3\,\varepsilon_{\mu\nu\alpha\beta}\,
\frac{\hQ^{\alpha}\hpmomi{+}^{\beta}}{\spro{\hpmomi{+}}{\hQ}},  \nl
{\cal P}_{\mu} &=& {\hat p}_{+\mu} - \frac{\spro{\hpmomi{+}}{\hQ}}
{\hQs}\,\hQ_{\mu}.
\eqa
Note that the ${\hat W}_3$ form factor gives zero contribution in this case.
Actually the gauge invariance of CC20$_{\gamma}$ poses a problem with a 
well-known solution: a complete treatment would require the application of 
the Fermion-Loop scheme, but for our purposes it is enough to introduce a 
fixed width for the $\wb$, both for the $s$-channel and the $t$-channel, i.e. 
the fixed-width scheme. For a complete discussion we refer to the work in
\cite{kn:fl}.

In the limit $\hQs \to 0, {\hat W}^{\mu\nu}$ must be an
analytical function of $\hQs$.
By requiring that $\hQs\int d\Phi_3 {\hat W}^{\mu\nu} = 0$
for $\hQs = 0$ one obtains
\bq
{\hat W}_2 \lpar\hQs,\hy\rpar = {\hat W}_1(0,\hy) + 
\ord{\hQs},  
\eq
where we have introduced the variable
\bq
\hy = {{\spro{\hpmomi{+}}{\hQ}}\over  {\spro{\hpmomi{+}}{\hpmomi{-}}}}.
\label{defhy}
\eq
The WW-approximation is defined by the following equation:
\bqa
{}&{}& \int d\Phi_3\,{\hat L}_{\mu\nu}\,{\hat W}_{\mu\nu} \Longrightarrow
- \frac{1}{4}\,{\hat W}_1 \lpar 0,\hy\rpar\,
f_{\gamma}\lpar \hQs,\hy,x_-\rpar  \nl
{}&{}& {\hat W}_1 \lpar 0,\hy\rpar = - \frac{1}{2}\,\int d\Phi_3\,
{\hat W}_{\mu\mu}\mdls_{\hQs = 0}.
\label{defww}
\eqa
${\hat W}_1$ is therefore proportional to the cross section
for $e^+\gamma \to \barnu_e u \bard$ (with real $\gamma$) and
$f_{\gamma}$ is the photon density, which in the presence of QED
initial state radiation reads as follows:
\bq
f_{\gamma} = - \mes\,\lpar 1 + x_- + 2\,\frac{1-x_-}{\hy}\rpar +
\hQs\,\lpar 1 - \frac{2}{\hy} + \frac{2}{\hy^2}\rpar.
\eq
Thanks to \eqn{defww} the $\hQs$ integration can be performed analytically. 
Note that the integrand is the sum of two terms, proportional to
\bq
\frac{1}{\hQs}, \qquad \frac{\mes}{\hQq}.
\eq
To integrate over $\hQs$ we need the kinematics of the process which is 
specified, in the laboratory system, by
\bq
p_- = \frac{1}{2}\sqrt{s}\,\lpar 0,0,\beta,1\rpar, \quad
q_- = E_f\,\lpar \beta_f\sin\theta,0,\beta_f\cos\theta,1\rpar,
\eq
with $\beta^2_f = 1-\mes/E^2_f$ and $\beta$ defined in \eqn{defbeta}. Let 
$\hQ(Q)$ be the momentum transfer with (without) inclusion of ISR, then
\bqa
\hQs &=& x_-Q^2 -(1-x_-)\mes +(1-\beta)(1-x_-)\,E_f\beta_f\sqrt{s}\cos\theta \nl
{}&=& x_-Q^2 - (1-x_-)\,\lpar 1 - 2\,\frac{E_f}{\sqrt{s}}\cos\theta\rpar
\mes + \ord{\meq/s}.
\label{hqs}
\eqa
If we introduce the variable $y$, equivalent to the fraction of the electron
energy carried by the photon in absence of ISR and defined by
\bq
y = {{\spro{p_+}{Q}}\over {\spro{p_+}{p_-}}},
\eq
then the following relations hold for $\me = 0$,
\bq
Q^2 = (1-\cos\theta)\,E_f\sqrt{s}, \qquad
y = 1 - (1+\cos\theta)\,\frac{E_f}{\sqrt{s}}.
\eq
Using this result in \eqn{hqs} one derives
\bq
\hQs = x_-Q^2 - (1-x_-)\lpar y + \frac{Q^2}{s}\rpar\,\mes + 
\ord{\frac{\meq}{s}}.
\eq
With ISR one uses $\hy$ defined in \eqn{defhy} and the relation between 
$\hy$ and $y$ is obtained from
\bqa
2\spro{\hpmomi{+}}{\hQ} &=& - \shat -2\,x_+\spro{p_+}{q_-} + (x_++x_-)\mes +
(1-x_+)\,\lpar y - 1 + \frac{Q^2}{s}\rpar,  \nl
2\,\spro{p_+}{q_-} &=& (1-y)\,(2\,\mes-s).
\eqa
For finite electron mass the relations linking $\hQs,\hy$ to $Q^2,y$ read as
follows:
\bq
\hQs = a\,Q^2 + b\,\mes\,y, \qquad \hy = c\,\frac{Q^2}{s} + d\,y + e,
\eq
where the $a,\,\dots,\,e$ coefficients are
\bqa
a &=& x_- - (1-x_-)\,\frac{\mes}{s}, \qquad b = - (1-x_-),  \nl
c &=& \frac{1-x_+}{D}\,\frac{\mes}{s}, \qquad d = \frac{-x_+}{D} +
\frac{1+x_+}{D}\,\frac{\mes}{s},  \nl
e &=& \frac{1-x_-}{D}\,\lpar x_+ - \frac{\mes}{s}\rpar, \qquad
D = -x_+x_- + (X_+ + x_-)\,\frac{\mes}{s}.
\eqa
We have introduced $Q^2$ and $y$ because they are natural variables for 
describing the outgoing electron in absence of ISR. Indeed, one can show that
\bq\frac{d^3q_-}{E_f} = 2\,\pi\beta_f E_fdE_fd\cos\theta = \pi\,
\frac{1+\beta^2}{2\,\beta}\,dQ^2dy.
\eq
The transition to hatted variables, to be used with ISR, is completed by 
deriving the jacobian of the transformation,
\bq
d\hQs d\hy = \Bigl[ 1 + \ord{\frac{\mes}{s}}\Bigr]\,dQ^2dy.
\eq
Having specified the relevant variables we now proceed to deriving the 
boundaries of the phase space.
First we derive the boundaries for $Q^2$. We start from the relations
\bqa
Q^2 &=& -2\,\mes + \lpar 1 - \beta\beta_f\cos\theta\rpar\,E_f\sqrt{s},  \nl
y &=& 1 - 2\,\frac{1 + \beta\beta_f\cos\theta}{1+\beta^2}\,\frac{E_f}{\sqrt{s}},
\label{defQy}
\eqa
and introduce a new variable $\chi$ defined by
\bq
E_f = \frac{\chi^2+\mes}{2\,\chi}, \qquad \beta_f = \frac{\chi^2-\mes}
{\chi^2+\mes}.
\label{defef}
\eq
From \eqn{defQy} a solution for $\chi$ is
\bq
\chi = \frac{1}{2}\,\frac{1+\beta^2}{1+\beta c}\,\Biggl[
1 - y + \sqrt{
(1-y)^2 - 4\,\frac{1-\beta^2 c^2}{(1+\beta^2)^2}\,\frac{\mes}{s}}\;\Biggr]\,
\sqrt{s},
\eq
where $c = \cos\theta$. At zero scattering angle for the outgoing electron,
$c = 1$, one obtains
\bqa
\chi(c=1) &=& \frac{1}{2}\,\frac{1+\beta^2}{1+\beta}\,\Biggl[
1 - y + \sqrt{
(1-y)^2 - 4\,\frac{1-\beta^2}{(1+\beta^2)^2}\,\frac{\mes}{s}\,}\;\Biggr]
\sqrt{s}  \nl
{}&=& 1 - y + \ord{\frac{\mes}{s}}.
\eqa
Inserting $E_f$ from \eqn{defef} into \eqn{defQy} one derives
\bq
Q^2 = - 2\,\mes + \frac{1}{2}\,\frac{\sqrt{s}}{\chi}\,\Bigl[\lpar
1 - \beta c\rpar\,\chi^2 + \lpar 1 + \beta c\rpar\,\mes\Bigr],
\eq
and, therefore the lower limit for the square of the momentum transfer is set 
by
\bqa
Q^2(c=1) &=& -2\,\mes + \frac{1}{2}\,\frac{\sqrt{s}}{\chi}\,\Bigl[
(1-\beta)\,\chi^2 + (1+\beta)\,\mes\Bigr]  \nl
{}&=& \mes\,\frac{y^2}{1-y} + \ord{\frac{\meq}{s}}.
\eqa
If we now require that $\theta \le \theta_c$, with $\theta_c \ll 1$, the limits 
for $Q^2$ are as follows:
\bqa
Q^2_0 &\le& Q^2 \le Q^2_c,  \nl
Q^2_0 &=& \mes\,\frac{y^2}{1-y}, \quad Q^2_c = Q^2_0 + \frac{1}{4}\,
\frac{\chi^2_c - \mes}{\chi_c}\,\theta^2_c\,\sqrt{s} +
\ord{\frac{\meq}{s},\mes\theta^2_c},
\label{defbound}
\eqa
where $\chi_c$ is 
\bq
\frac{\chi_c}{\sqrt{s}} = \lpar 1 - \frac{\mes}{s} + \frac{\theta^2_c}{4}\rpar\,
(1-y) + \ord{\frac{\meq}{s^2},\frac{\mes}{s}\,\theta^2_c},
\eq
giving an upper limit of integration for $Q^2$
\bq
Q^2_c = Q^2_0 + \frac{1}{4}\,(1-y)\,\theta^2_c s + \ord{\frac{\meq}{s},\mes
\theta^2_c}.
\eq
The limits for $Q^2$ can be immediately translated into limits for $\hQs$,
and one finds
\bqa
\hQs_0 &\le& \hQs \le \hQs_c,  \nl
\hQs_0 &=& \mes\,\lpar x_-\hy + 1 - x_-\rpar\,\frac{\hy}{1-\hy} + 
\ord{\frac{\meq}{s}},  \nl
\hQs_c &=& \hQs_0 + \frac{1}{4}\,x^2_-(1-\hy)\,\theta^2_c s + \ord{\frac{\meq}
{s},\mes\theta^2_c}.
\eqa
The photon flux-function, an essential ingredient of the WW-approximation, is 
now defined by
\bq
{\cal F}_{\gamma} = \int_{\hQs_0}^{\hQs_c}\,d\hQs\,\frac{f_{\gamma}}{\hQq}.
\label{defff}
\eq
Next we discuss the limits of integration for $y$ and assume that the fermions 
in the final state are massless, apart from the electron. As we will see this 
can be the origin of new mass singularities. From this point of view ISR is
inessential. Let us introduce variables $\rho$ and $\kappa$ by
\bq
\rho = \frac{1+\beta^2}{2}\,(1-y), \qquad \kappa = \frac{Q^2+\mes}{s},
\eq
such that the electron energy and scattering angle become
\bq
E_f = \frac{1}{2}\,(\rho+\kappa)\,\sqrt{s}, \qquad \beta\beta_f\cos\theta = 
\frac{\rho-\kappa}{\rho+\kappa}.
\label{rk}
\eq
For $\theta = 0$, after squaring the second relation in \eqn{rk}
and substituting $E_f$ from the first one, one obtains
\bq
\kappa^2 - 2\,\frac{1+\beta^2}{1-\beta^2}\,\rho\kappa - \beta^2 + \rho^2 = 0.
\eq
In this way the allowed region of the phase space for the outgoing electron is 
completely specified. It is seen that for $\rho > \beta$
one has $\kappa < 0$ or $Q^2 < - \mes$, i.e. the square of the momentum
transfer is not positive definite and crosses the zero independently of the 
finite electron mass. The variable $\kappa$ becomes negative for $\rho > 
\beta$ or
\bq
y \le \frac{(1-\beta)^2}{1+\beta^2} \sim 2\,\frac{\meq}{s^2}.
\eq
This simple fact is better illustrated by considering the process
$e^+e^- \to e^- X$ with $X = \{\barnu_e u \bard\}$. Let the cluster $X$ be 
characterized by having four-momentum $q_X$ and mass $M$, i.e. $q^2_X = - M^2$.
The $2 \to 2$ process $p_+ + p_- \to q_- + q_X$ is described in terms of
Mandelstam invariants
\bq
s = - \lpar p_+ + p_-\rpar^2, \qquad t = - \lpar p_- - q_-\rpar^2,
\eq
so that $Q^2 = -t$ and
\bq
y = {{Q^2-\mes+M^2}\over {s - 2\,\mes}}.
\eq
The physical portion of the phase space must satisfy the condition $X \ge 0$
with
\bq
X = \frac{1}{4}\,\lkall{s}{\mes}{\mes}\,\lkall{s}{\mes}{M^2} -
s^2\,\Bigl[  t-\mes-M^2 + \frac{1}{2}\,\lpar s+M^2-\mes\rpar\Bigr]^2,
\eq
where $\lambda$ is the usual K\"allen-function.
When $M = 0$ the condition $X \ge 0$ is equivalent to
\bq
t_- \le t \le t_+,
\eq
where one easily finds that
\bq
t_- \sim \frac{27}{4}\,\frac{m^6_e}{s^2}, \qquad t_+ \sim -s, \quad
\mbox{for} \quad \me \to 0.
\eq
Therefore, for $M=0$, $t$ is not negative definite. The amplitude squared is 
proportional to $1/Q^2$ or to $\mes/Q^4$ and massless quarks induce a
singularity, even for finite $\me$, if a cut is not imposed on the invariant 
mass $M(u\bard)$\footnote{This fact was firstly pointed out in a private 
communication of A.~Ballestrero.}.
The singularity is, in any case, avoided by requiring a cut such that
\bq
y \ge \frac{(1-\beta)^2}{1+\beta^2}.
\eq
An upper limit on $y$ is derived by considering again
\bq
p_+ + p_- \to q_- + q_X, \qquad q_x = q_+ + k + {\bar k} = p_+ + Q.
\eq
Next we introduce the invariant mass of the quark-antiquark system,
\bq
M^2 = - (p_+ + Q)^2 = \mes - Q^2 +(s-2\,\mes)\,y,
\eq
and require the constraint
\bq
\sqrt{s} \ge \me + M,
\eq
equivalent to
\bq
(s-2\,\mes)\,y \le s - 2\,\me\sqrt{s} + Q^2.
\eq
This inequality is satisfied for
\bq
y \le 1 - \frac{\me}{\sqrt{s}},
\eq
following from the relation giving $y_{\rm max}$ in terms of $Q^2_{\rm min}$,
\bq
(s-2\,\mes)\,y_{\rm max} = s - 2\,\me\sqrt{s} + Q^2_{\rm min}, \qquad 
Q^2_{\rm min} = Q^2_0.
\eq
Here $Q^2_0$ is taken from \eqn{defbound}. The equivalent bound for $\hy$ 
follows as
\bq
\hy \le 1 - \frac{\me}{\sqrt{\shat}},
\eq
With a cut on $M(\bard u)$ the singularity at $Q^2 = 0$ (or $\hQs = 0$ 
with ISR) is avoided but we 
still have additional singularities.
There are two multi-peripheral diagrams contributing to the CC20 process
$e^+e^- \to e^- \barnu_e f_1 \barf_2$, the last two in \fig{cc20g}.
When $Q^2 = 0$, i.e. the electron is lost in the beam pipe, and the
(massless) $f_1(f_2)$-fermion is emitted parallel to the (quasi-real) photon
then the internal fermion propagator will produce an enhancement in the cross 
section. Taking into account a $\ln \mes$ from the photon flux-function, $3$
options follow:

\begin{enumerate}
\item to consider massive fermions, giving a result proportional to 
$\ln\mes\ln\mfs$,
\item to use massless fermions, giving instead $\ln^2\mes$,
\item to introduce an angular cut on the outgoing $f_1$ and $\barf_2$
fermions with respect to the beam axis, $\theta(f_1,{\bar f}_2) \ge 
\theta_{\rm cut}$, giving $\ln\mes\ln\theta_{\rm cut}$.
\end{enumerate}

The first option is clean but ambiguous when the final state fermions are 
light quarks, what to use for $m_u, m_d$?
The second one presents no problems for a fully leptonic CC20 final state
but completely fails to describe quarks, as we will show in discussing QCD 
corrections.
The last option is also theoretically clean and can be used to give 
differential distributions for the final state jets. It is, however,
disliked by the experimentalists when computing the total sample of events: 
hadronized jets are seen and not isolated quarks. Even if the quark is 
parallel to the beam axis the jet could be broad enough and the event selected.

These events are also interesting since they correspond to a situation where
the electron and one of the quarks are lost in the beam pipe, while
the other quark is recoiling against the neutrino, i.e. one has a totally 
imbalanced mono-jet structure, background to new particle searches.

The singularity induced by massless quarks in $e^+e^- \to e^- \barnu_e u \bard$
can only be treated within the context of QCD final state corrections and of 
the photon hadronic structure function (PHSF) scenario. We will come back to the
problem later in the paper. In the next section we discuss, instead the cross
section for the sub-process $e^+ \gamma \to \barnu_e u \bard$ in the two
regimes, massless and massive quarks.
It will be seen that one can use different parametrizations for the
corresponding phase space, depending on the presence of kinematical cuts.
\section{The sub-process $e^+ \gamma \to \barnu_e u \bard$ with cuts
\label{sect_5}}

As a consequence of \eqn{defww}, the WW-approximation, we write the result
for a CC20 cross section as the convolution of the photon flux-function of
\eqn{defff} with the cross section for $e^+ \gamma \to \barnu_e u \bard$.
The process $e^+(\hpmomi{+}) \gamma(\hQ) \to \barnu_e(q_+) u(k) \bard(\bk)$ is 
illustrated in \fig{subp} and is described by the following invariants:
\bqa
\spro{\hpmomi{+}}{\hQ} &=& - \frac{1}{2}\,\hy\,\shat, \quad
\spro{\hpmomi{+}}{q_+} =  \frac{1}{2}\,\kappa_+, \quad
\spro{\hpmomi{+}}{k} =  \frac{1}{2}\,u', \quad
\spro{\hpmomi{+}}{\bk} =  \frac{1}{2}\,t', \nl
\spro{\hQ}{q_+} &=& \frac{1}{2}\,\kappa_-, \quad
\spro{\hQ}{k} = \frac{1}{2}\,t, \quad
\spro{\hQ}{\bk} = \frac{1}{2}\,u, \nl
\spro{q_+}{k} &=& \frac{1}{2}\,\zeta_-, \quad
\spro{q_+}{\bk} = \frac{1}{2}\,\zeta_+, \nl
\spro{k}{\bk} &=& - \frac{1}{2}\,s'.
\eqa
The process takes place at energy $\sqrt{\hy\,\shat}$ and there are only five 
linear-independent invariants, which we choose to be $\hy\,\shat$ and
$\tau,x_1,x_2$ and $z$ defined by
\bq
t = \tau\hy\,\shat, \quad s'-\zeta_- = x_1\hy\,\shat, \quad
s' = x_2\hy\,\shat, \kappa_- = z\hy\,\shat.
\eq
The phase space can be computed in terms of the following object:
\bqa
{{\partial^2\Phi_3}\over {\partial x_1\partial x_2}} &=&
\hy^2\shat^2\int d^4k\,d^4\bk\,d^4q_+\,\delta^+(k^2)\delta^+(\bk^2)
\delta^+(q^2_+)\delta^4\lpar \hpmomi{+} + \hQ - k - \bk - q_+\rpar  \nl
{}&\times& \delta\lpar 2\,\spro{k}{\bk} + x_2\hy\,\shat\rpar\,
\delta\lpar2\,\spro{k}{\bk} + 2\,\spro{q_+}{k} + x_2\hy\,\shat\rpar.
\eqa
\vspace{0.2cm}
\bqas
\ba{ccc}
\vcenter{\hbox{
  \begin{picture}(110,100)(0,0)
  \ArrowLine(50,50)(0,100)
  \Photon(0,0)(50,50){2}{7}
  \ArrowLine(100,100)(50,50)
  \ArrowLine(100,50)(50,50)
  \ArrowLine(50,50)(100,0)
  \GCirc(50,50){15}{0.5}
  \Text(10,100)[lc]{$e^+(\hpmomi{+})$}
  \Text(10,0)[lc]{$\gamma(\hQ=\hQ_-)$}
  \Text(110,100)[lc]{$\barnu_e(q_+)$}
  \Text(110,50)[lc]{$\bard(\bk)$}
  \Text(110,0)[lc]{$u(k)$}
  \end{picture}}}
\ea
\eqas
\vspace{-2mm}
\begin{figure}[h]
\caption[]{The sub-process $e^+\gamma \to \barnu_e u \bard$.}
\label{subp}
\end{figure}
\vskip 5 pt
The integration is most conveniently performed in the system where
\bq
{\hat P} = \hpmomi{+} + \hQ = \lpar 0,0,0,\sqrt{\hy\,\shat}\rpar.
\eq
Moreover, let $\bf k$ be along the positive $z$ axis with $\bf \hQ$ in the
$x-y$ plane (and polar angle denoted by $\theta$) and let $\bf q_+$ be
described by angles $\psi,\phi$. Then we find,
\bqa
{{\partial^2\Phi_3}\over {\partial x_1\partial x_2}} &=&
\hy^2\shat^2 \theta(x_1)\,\int d^4k\,\delta^+(k^2)\,\delta\lpar
2\,E_d\sqrt{\hy\,\shat} - x_1\hy\,\shat\rpar\,J,  \nl
J &=& \theta(1-x_2)\,\frac{1-x_2}{8}\,\int_{-1}^{+1}\,d\cos\psi\,
\int_0^{2\,\pi}\,d\phi  \nl 
{}&\times& \delta\lpar \frac{1}{2}\,x_1(1-x_2)(\cos\psi-1)\hy\,\shat + 
(x_1-x_2)\hy\,\shat \rpar.
\eqa
It is more convenient to introduce
\bqa
{{\partial^3\Phi_3}\over {\partial x_1\partial x_2\partial z}} &=&
\hy^3\shat^3 \theta(x_1)\,\int d^4k\,\delta^+(k^2)\,\delta\lpar
2\,E_d\sqrt{\hy\,\shat} - x_1\hy\,\shat\rpar\,{\bar J},  \nl
{\bar J} &=& \theta(1-x_2)\,\frac{1-x_2}{8}\,\int_{-1}^{+1}\,d\cos\psi\,
\int_0^{2\,\pi}\,d\phi  \nl 
{}&\times& \delta\lpar\lpar
(1-x_2)\sin\theta\sin\psi\cos\phi+(1-x_2)(\cos\theta\cos\psi-1) -2\,z\rpar
\hy\,\shat\rpar  \nl
{}&\times& 
\delta\lpar \frac{1}{2}\,x_1(1-x_2)(\cos\psi-1)\hy\,\shat + (x_1-x_2)\hy\,\shat
\rpar.
\eqa
If we now take into account that 
\bq
\cos\theta = 1 + 2\,\frac{\tau}{x_1},
\eq
the final result follows
\bqa
d\Phi_3 &=& \frac{\pi^2\,\hy\,\shat}{4\,x_1\,R}\,\Theta\,dx_1 dx_2 dz d\tau, \\
\Theta &=& \{\prod_i \theta(x_i)\theta(1-x_i)\}\,\theta(x_1-x_2)\,
\theta(z_+-z)\,\theta(z-z_-)\,\theta(-\tau)\,\theta(\tau+x_1), \nn
\label{phi3}
\eqa
where $\Theta$ gives the boundaries of the phase space and where we have 
introduced
\bqa
R^2 &=& -4\,z^2 + 4\,(1-x_2)(\cos\theta\cos\psi-1)\,z - (1-x_2)^2\,
(\cos\theta - \cos\psi)^2,  \nl
\cos\psi &=& {{2\,x_2 - x_1 - x_1x_2}\over {x_1\,(1-x_2)}},  
\eqa
and where $z_{\pm}$ are the roots of $R^2 = 0$, i.e.
\bq
z_{\pm} = \frac{1-x_2}{2}\,\lpar \cos\theta\cos\psi \pm |\sin\theta\sin\psi| - 
1 \rpar.
\eq
The cross section will be computed with a cut on the invariant mass of the
$u \bard$-pair, i.e. $M^2(u\bard) \ge s_0$ giving
\bq
\frac{s_0}{\hy\,\shat} \le x_2 \le 1, \qquad \hy\,\shat \ge s_0.
\eq
Starting from the four diagrams of \fig{cc20g} we derive
\bqa
{\hat W} &=& - \frac{1}{2}\,{\hat W}_{\mu\mu}\mdls_{\hQs = 0} =
\frac{g^6\stws}{|\Delta_s|^2|\Delta_t|^2 \shat}\,  \nl
{}&\times& \Biggl[ \hy^3\,{\hat W}_{11} +
|\Delta_t|^2\,{\hat W}_{22} + |\Delta_s|^2\,{\hat W}_{33} +
\hy|\Delta_s|^2\,{\hat W}_{44} \nl
{}&+&
\hy^2\,\lpar {\hat W}_{12} + {\hat W}_{13} +
{\hat W}_{14}\rpar +
\hy\,\lpar {\hat W}_{23} + {\hat W}_{24}\rpar +
\hy\,|\Delta_s|^2\,{\hat W}_{34}\Biggr]
\label{wcut}
\eqa
where the propagators are defined, in the fixed width scheme, by
\bq
\Delta_s = -x_2\hy + \mu^2_{_\wb} - i\,\gamma_{_\wb}\mu_{_\wb},  \qquad
\Delta_t = -(1-x_2+z)\hy + \mu^2_{_\wb} - i\,\gamma_{_\wb}\mu_{_\wb},
\label{defred}
\eq
with $\gw = \gamma_{_\wb}\sqrt{s}$ and $\mu_{_\wb} = \mw\sqrt{s}$. The
${\hat W}_{ij}$ represent the $i-j$ interference of diagrams in
\fig{cc20g} and are given in the following list:
\bqa
{\hat W}_{11} &=&
        \tau   ( 1 + z x_1 + 2 z - 2 x_1 x_2 + x_1 + z^2 ) \nl
{}&+&  \tau^2   ( 1 + z - x_2 )
       + z x_1 + z x^2_1 + 2 x_1 x_2 + x_1 z^2 - 2 x_2 x^2_1 - x_2 + x^2_1,
\nl\nl
{\hat W}_{22} &=&  y \tau   ( 1 - x_1 ),
\qquad
{\hat W}_{33} = \frac{4}{9} \frac{y}{\tau} ( 1 -  x_1 ),
\nl\nl
{\hat W}_{44} &=&  - \frac{1}{9} \frac{\tau}{\upsilon}  (  1 + x_1 )
       + \frac{1}{9} \tau   
- - \frac{1}{9} \frac{\tau^2}{\upsilon}  
       - \frac{1}{9} \frac{x_1}{\upsilon}
       + \frac{1}{9} x_1,
\nl\nl
{\hat W}_{12} &=&
        \Reb\Delta_t \Bigl[ \tau   (  - 2 + z x_1 - 2 z + 2 x_1 x_2 ) \nl
{}&+& \tau^2   (  - 1 + x_2 )
         - z x_1 - 2 x_1 x_2 + 2 x_2 x^2_1 + x_2 - x^2_1 \Bigr],
\nl\nl
{\hat W}_{13} &=&  \frac{2}{3} \Reb\Delta_s \Bigl[ \frac{1}{\tau}  
        ( z x_1 - 2 z x^2_1 - 2 x_1 x_2 +  x_1 
+  2 x_2  x^2_1 + x_2 - 2 x^2_1 )  \nl
{}&+& \frac{2}{3} \tau   (  -  x_1 + x_2 )
        - \frac{2}{3} - \frac{2}{3} z x_1 + \frac{4}{3} x_1 x_2 
- \frac{2}{3} x^2_1 \Bigr],
\nl\nl
{\hat W}_{14} &=& \frac{1}{3} \Reb\Delta_s \Bigl[ \frac{\tau}{\upsilon}
           (  - 1 - 2 x_1 x_2 + 3 x_1 - 2 x^2_1 ) 
+ \frac{1}{3} \tau   ( 1 + 2 z ) \nl
{}&+& \frac{1}{3} \frac{\tau^2}{\upsilon}  ( 1 - x_1 - x_2 )
       + \frac{1}{3} \frac{1}{\upsilon}  ( 2 x_1 x_2 - 2 x_2 x^2_1 - x_2 ) 
+ \frac{2}{3} z x_1 + \frac{1}{3} x^2_1 \Bigr],
\nl\nl
{\hat W}_{23} &=&
       - \frac{4}{3} \Reb\Delta_s \Reb\Delta_t \Bigl[ \frac{x_1}{\tau} 
        (  1 - x_1 )^2  
+ \frac{4}{3} x_1 - \frac{4}{3} x^2_1 \Bigr]
       - \frac{4}{3} \frac{\gamma^2_w\mu^2_w}{\tau} x_1 (  1-x_1 )^2  \nl
{}&+& \frac{4}{3} \gamma^2_w\mu^2_w  x_1( 1 - x_1 )
\nl\nl
{\hat W}_{24} &=&
 \frac{1}{3} \Reb\Delta_s \Reb\Delta_t \Bigl[ \frac{\tau}{\upsilon} 
         ( 1 + 2 x_1 x_2 +  x_1 - 2 x^2_1 )  
+ \frac{1}{3} \tau   (  - 2 + x_1 )  
+ \frac{1}{3} \frac{\tau^2}{\upsilon} \nl
{}&\times& ( 1 -  x_1 + x_2 )
       + \frac{1}{3} \frac{1}{\upsilon} (  - 2 x_1 x_2 + 2 x_2 x^2_1 +  x_2 + 
         2 x^2_1 - 2 x^3_1 )  - \frac{1}{3} x_1 \Bigr] \nl
{}&+& \frac{1}{3} \gamma^2_w\mu^2_w\frac{\tau}{\upsilon}
         ( 1 + 2 x_1 x_2 + x_1 - 2 x^2_1 )
       + \frac{1}{3} \gamma^2_w\mu^2_w\tau   (  - 2 + x_1 )  
+ \frac{1}{3} \gamma^2_w\mu^2_w\frac{\tau^2}{\upsilon} \nl
{}&\times& ( 1 - x_1 + x_2 )
       + \frac{1}{3} \frac{\gamma^2_w\mu^2_w}{\upsilon} 
         (  - 2 x_1 x_2 + 2 x_2 x^2_1 +  x_2 + 2 
         x^2_1 - 2 x^3_1 )  - \frac{1}{3} \gamma^2_w\mu^2_w  x_1,
\nl\nl
{\hat W}_{34} &=&
 \frac{2}{9} \frac{1}{\tau\upsilon} (  - 2 x_1 x_2 + 2 x_2 x^2_1 + x_2 )
       + \frac{2}{9} \frac{1}{\tau}   x_1( 1 - 2 x_1 )  \nl
{}&+& \frac{2}{9} \frac{\tau}{\upsilon}   (  - 1 + x_1 + x_2 )
       + \frac{2}{9} \frac{1}{\upsilon}   ( 1 + 2 x_1 x_2 - 3 x_1 + 2 x^2_1 )
       - \frac{2}{9} x_1
\eqa
where we have introduced a new variable $\upsilon$,
\bq
\upsilon= 1 + \tau + z.
\eq
There remains the problem of angular cuts for the outgoing quark(antiquark).
These cuts should be applied with respect to the beam direction and
should be expressed in terms of the variables describing the sub-process.
For CC20$_{\gamma}$ and for $\theta_c \ll 1$ the $e^-$ is lost in a narrow
cone around the beam, so that we can use the approximation 
$\hQ = \hy\hpmomi{-}$. In the laboratory system we have 
\bq
p_{\pm} = \frac{1}{2}\,\sqrt{s}\,\lpar 0,0,\mp 1,1\rpar.
\eq
Let $E_u$ be the energy of the outgoing $u$-quark, so that
\bq
k = E_u\,\lpar \sin\theta_u,0,\cos\theta_u,1\rpar,
\eq
where $\theta_u$ is the scattering angle of the $u$-quark with respect to the
incoming electron. One finds
\bqa
\cos\theta_u &=& {{\spro{p_+}{k}-\spro{p_-}{k}}\over 
{\spro{p_+}{k}+\spro{p_-}{k}}},  \nl
\spro{p_+}{k} &=& \frac{\spro{\hpmomi{+}}{k}}{x_+} = \frac{u'}{2\,x_+},
\qquad
\spro{p_-}{k} = \frac{\spro{\hQ}{k}}{x_-\hy} = \frac{t}{2\,x_-\hy}.
\eqa
The condition $|\cos\theta_u| \le C$ becomes, in terms of invariants,
\bqa
(1+C)\,x_-\hy x_1 + \Bigl[(1+C)\,x_m\hy + (1-C)\,x_+\Bigr]\,\tau \le 0,  \nl
(1-C)\,x_-\hy x_1 + \Bigl[(1-C)\,x_m\hy + (1+C)\,x_+\Bigr]\,\tau \ge 0.
\label{qcuts}
\eqa
With $\bk = E_d\,(\sin\theta_d,0,\cos\theta_d,1)$ we derive two additional 
conditions similar to those of \eqn{qcuts} but with $u' \to t',
t \to u$ and reflecting the cut $|\cos\theta_d| \le C$.

\eqn{wcut}, in conjunction with \eqnsc{sigmag}{defww} and \eqn{phi3}, allows
us to compute the cross section within the WW-approximation.
\section{The fully extrapolated sub-process $e^+ \gamma \to \barnu_e u 
\bard$\label{sect_6}}

Our goal is to compute the CC20 cross section without any kinematical cut, 
apart from imposing that $M(u\bard) \ge \sqrt{s_0}$.
We have seen in sect. \ref{sect_5} that there is a mass singularity in the
total cross section when the quarks are assumed to be massless. Here
we compute again the cross section with finite quark masses. Let us consider
again the process $e^+(\hpmomi{+}) \gamma(\hQ_-) \to \barnu_e(q_+) u(k) 
\bard(\bk)$.
Moreover let $\hQ_+$ be defined as $\hQ_+ = \hpmomi{+}-q_+$ and let 
$\hQ$ be $\hQ_+ + \hQ_-$.
In order to compute ${\hat W}_{\mu\mu}$, as required by \eqn{defww}, we 
neglect for the moment ISR and introduce three master scalar-integrals:
\bq
I_n = \int d^4k\,d^4\bk \delta^+(k^2+\mus)\,\delta^+(\bk^2+\mds)\,
\delta^4(Q-k-\bk)\,\frac{1}{\lpar \spro{Q_-}{k}\rpar^n},   
\label{master}
\eq
with $n=0,\dot,2$. They are easily evaluated in the system where $Q =
(0,0,0,\mu)$ and where the photon four-momentum is $Q_- = X\,(0,0,1,1)$. One 
immediately finds
\bq
X = \frac{\spro{Q_+}{Q_-}}{Q^2}. 
\eq
If we start with $I_2$ then we obtain
\bqa 
I_2 &=& \int d^4k\,\delta^+(k^2+\mus)\,\delta^+(2\,\mu E_u-\mu^2-\mus+\mds)\,
\frac{1}{\lpar\spro{Q_-}{k}\rpar^2}  \nl
{}&=&  \frac{\pi}{4\,\mu X^2}\,\int_0^{\infty}\,k\,\delta(k^2-E^2_u+\mus)\,
\frac{2\,k}{\mus}  \nl
{}&=& \frac{\pi}{4}\,\frac{E}{\mu X^2\mus} \sim - \frac{\pi}{8}\,\frac{Q^2}
{\lpar\spro{Q_+}{Q_-}\rpar^2}\,\frac{1}{\mus} \quad \mbox{for} \quad
\muq \to 0.
\eqa
Similarly we evaluate $I_1$ as follows:
\bqa
I_1 &=&\int d^4k\,\delta^+(k^2+\mus)\,\delta^+(2\,\mu E_u-\mu^2-\mus\mds)\,
\frac{1}{X\lpar k\cos\theta-E_u\rpar}  \nl
{}&=& - \frac{\pi}{2\,\mu X}\,\int_0^{\infty}\,dk\,k\,\delta(k^2+\mus)\,
\ln\frac{E_u-k}{E_u+k}  \nl
{}&=& - \frac{\pi}{2\,\mu X}\,\ln\frac{\muq}{E_u + \sqrt{E^2_u-\mus}} \sim
\frac{\pi}{4\,\spro{Q_+}{Q_-}}\ln\frac{\mus}{-Q^2} \quad \mbox{for} \quad
\muq \to 0.
\label{inti1}
\eqa
Finally, for $I_0$ one gets
\bq
I_0 = \frac{\pi}{2}.
\eq
The complete result for the cross section follows from squaring the matrix 
element,
\bq
W_{\mu\nu} =  \frac{g^6\stws}{64}\,{\cal R}^{\dagger}_{\nu}\,
\lpar - i\,\sla{k} + \muq\rpar\,{\cal R}_{\mu}\,\lpar i\,\sla{\bk} + \md
\rpar
\eq
The function ${\cal R}$ is
\bqa
{\cal R}_{\mu} &=& \gapu{\alpha}\gdp\,{\cal R}^1_{\mu\alpha}\,P(s)P(t) +
\gapu{\alpha}\gdp\,{\cal R}_2{\mu\alpha}\,P(s)P(e)  \nl
{}&+& \gadu{\mu}\,\lpar \sla{\hQ}-\sla{k}-i\,\muq\rpar\gapu{\alpha}\gdp\,
{\cal R}^3_{\alpha}\,P(t)P(f)  \nl
{}&+& \gapu{\alpha}\gdp\,\lpar\sla{Q_-}-\sla{\bk}+i\,\md\rpar\,\gadu{\mu}
{\cal R}^4_{\alpha}\,P(t)P(\barf).
\label{defR}
\eqa
Furthermore we have $\gdp = 1 + \gfd$ and
\bqa
{\cal R}^1_{\mu\alpha} &=& V_{\mu\beta\alpha} \siap(p_+)\gapu{\beta}\gdp
\soap(q_+),
\nl
{\cal R}^2_{\mu\alpha} &=& - \siap(p_+)\gadu{\mu}\,\lpar \sla{p_+}+\sla{Q_-}
\rpar\,\gadu{\alpha}\gdp\soap(q_+),  
\nl
{\cal R}^3_{\mu} &=& -Q_u\,\siap(p_+)\gadu{\mu}\gdp\soap(q_+),
\nl
{\cal R}^4_{\mu} &=& Q_d\,\siap(p_+)\gadu{\mu}\gdp\soap(q_+),
\eqa
where $V_{\mu\alpha\beta}$ is the corresponding triple gauge-boson vertex, 
$Q_u(Q_d)$ is the up-(down-)fermion charge and the propagators appearing in 
\eqn{defR} are
\bqa
P(s) &=& {1\over {\lpar k+\bk\rpar^2 + \mws - i\,\gw\mw}}, \quad
P(t) = {1\over {\lpar p_+ - q_+\rpar^2 + \mws - i\,\gw\mw}},  \nl
P(e) &=& {1\over {\lpar p_+ - Q_-\rpar^2 + \mes}},  \nl
P(f) &=& {1\over {\lpar Q_- - k\rpar^2 + \mus}}, \quad
P(\barf) = {1\over {\lpar Q_- - \bk\rpar^2 + \mds}}.
\eqa
Next, we have to integrate over the phase space. The integration over $k,\bk$
can be performed by introducing additional integrals:
\bqa
I_{n,l}^{\mu_1\dots\mu_n}(u) &=&
\int d^4k\,d^4\bk \delta^+(k^2+\mus)\,\delta^+(\bk^2+\mds)\,
\delta^4(Q-k-\bk) \nl
{}&\times&\frac{k^{\mu_1}\dots k^{\mu_n}}{\lpar \spro{Q_-}{k}\rpar^l},   
\nl\nl
I_{n,l}^{\mu_1\dots\mu_n}(d) &=&
\int d^4k\,d^4\bk \delta^+(k^2+\mus)\,\delta^+(\bk^2+\mds)\,
\delta^4(Q-k-\bk) \nl
{}&\times&\frac{\bk^{\mu_1}\dots \bk^{\mu_n}}{\lpar \spro{Q_-}{\bk}\rpar^l}.   
\label{inti}
\eqa
All these integrals can be reduced to the scalar form factors. Quarks masses
are kept only in front of $I_2$ and the reduction gives
\bqa
I_{3n}^{\mu\nu\alpha}(q) &=&  Q^{\mu} Q^{\nu} Q^{\alpha}I_{3n,31}(q)+
                       Q_-^{\mu} Q_-^{\nu} Q_-^{\alpha}I_{3n,32}(q)+
                       (Q_-^{\mu} Q^{\nu} Q^{\alpha}+
                       Q^{\mu} Q_-^{\nu} Q^{\alpha} \nl
{}&+&                  Q^{\mu} Q^{\nu} Q_-^{\alpha})I_{3n_33}(q)+
                       (Q^{\mu} Q_-^{\nu} Q_-^{\alpha}+Q_-^{\mu} Q^{\nu} 
                       Q_-^{\alpha}+
                       Q_-^{\mu} Q_-^{\nu} Q^{\alpha})I_{3n,34}(q) \nl
{}&+&                 (Q^{\mu} \delta^{\nu\alpha}+Q^{\nu} \delta^{\mu\alpha} +
                       Q^{\alpha} \delta^{\mu\nu})I_{3n,35}(q)  \nl
{}&+&                 (Q_-^{\mu} \delta^{\nu\alpha}+Q_-^{\nu} 
                       \delta^{\mu\alpha}+
                       Q_-^{\alpha} \delta^{\mu\nu})I_{3n,36}(q)
\nl\nl
I_{30}^{\mu\nu\alpha} &=&  \frac{I_0}{4}\,\Bigl[ Q^{\mu} Q^{\nu} Q^{\alpha} -
                       (Q^{\mu} \delta^{\nu\alpha}+Q^{\nu} \delta^{\mu\alpha}+
                       Q^{\alpha} \delta^{\mu\nu}) \frac{Q^2}{6}\Bigr]
\nl\nl
I_{2n}^{\mu\nu}(q) &=&  Q^{\mu} Q^{\nu}I_{2n,21}(q)+
                   Q_-^{\mu} Q_-^{\nu}I_{2n,22}(q)+
                   (Q^{\mu} Q_-^{\nu}+Q^{\nu} Q_-^{\mu})I_{2n,23}(q) \nl
{}&+&              \delta^{\mu\nu}I_{2n,24}(q)
\nl\nl
I_{20}^{\mu\nu} &=&  \frac{I_0}{3}\,\lpar Q^{\mu} Q^{\nu} -
                  \delta^{\mu\nu} \frac{Q^2}{4}\rpar, \quad
I_{10}^{\mu} =  Q^{\mu} \frac{I_0}{2}
\nl\nl
I_{1n}^{\mu}(q) &=&  Q^{\mu}I_{1n,11}(q)+Q_-^{\mu}I_{1n,12}(q),
\eqa
where $q = u,d$ and where all the form factors can be reduced to linear 
combinations of the master scalar integrals of \eqn{master}. After a 
straightforward algebra one obtains
\bqa
 I_{12,11}(q) &=&  0,
\nl
 I_{12,12}(q) &=& 
        I_2(q)   \Bigl[ 1 + \frac{1}{2} \frac{Q^2_+}{\spro{Q_+}{Q_-}} \Bigr],
\nl
 I_{11,11}(q) &=& 
         \frac{I_0}{\spro{Q_+}{Q_-}},
\nl
 I_{11,12}(q) &=& 
  \Bigl[ I_1(q) -2\,\frac{I_0}{\spro{Q_+}{Q_-}}\Bigr]\,
  \Bigl[ 1 + \frac{1}{2} \frac{Q^2_+}{\spro{Q_+}{Q_-}} \Bigr],
\nl
 I_{22,24}(q) &=&  0,
\quad
 I_{22,21}(q) =  0,
\quad
 I_{22,23}(q) =  0,
\nl
 I_{22,22}(q) &=& 
        I_2(q)   \Bigl[ 1 + \frac{1}{2} 
 \frac{Q^2_+}{\spro{Q_+}{Q_-}} \Bigr]^2,
\nl
 I_{21,24}(q) &=& 
    - \frac{1}{2}  I_0 \Bigl[  1 + \frac{1}{2} \frac{Q^2_+}{\spro{Q_+}{Q_-}} 
\Bigr],
\nl
 I_{21,21}(q) &=& 
        \frac{1}{2}  \frac{I_0}{\spro{Q_+}{Q_-}},
\nl
 I_{21,23}(q) &=& 
     \frac{1}{2} \frac{I_0}{\spro{Q_+}{Q_-}}\, \Bigl[ 1 +
 \frac{1}{2} \frac{Q^2_+}{\spro{Q_+}{Q_-}} \Bigr],
\nl
 I_{21,22}(q) &=& 
  \Bigl[   I_1(q) - 3\,\frac{I_0}{\spro{Q_+}{Q_-}}\Bigr]\,  
 \Bigl[ 1 + \frac{1}{2} \frac{Q^2_+}{\spro{Q_+}{Q_-}} \Bigr]^2,
\nl
 I_{32,36}(q) &=&  0,
\quad
 I_{32,35}(q) =  0,
\quad
 I_{32,31}(q) =  0,
\quad
 I_{32,33}(q) =  0,
\quad
 I_{32,34}(q) =  0,
\nl
 I_{32,32}(q) &=& 
        I_2(q)   \Bigl[ 1 + \frac{1}{8} \frac{Q^6_+}{\lpar\spro{Q_+}{Q_-}\rpar^3} + 
       \frac{3}{4} \frac{Q^4_+} {\lpar\spro{Q_+}{Q_-}\rpar^2} + \frac{3}{2} 
         \frac{Q^2_+}{\spro{Q_+}{Q_-}} \Bigr],
\nl
 I_{31,36}(q) &=& 
   - \frac{1}{6}   I_0   \Bigl[  1 + \frac{1}{2} 
        \frac{Q^2_+}{\spro{Q_+}{Q_-}} \Bigr]^2,
\nl
 I_{31,35}(q) &=& 
 - \frac{1}{6}  I_0   \Bigl[  1 + \frac{1}{2} \frac{Q^2_+}{\spro{Q_+}{Q_-}} 
\Bigr],
\nl
 I_{31,31}(q) &=& 
        \frac{1}{3}  \frac{I_0}{\spro{Q_+}{Q_-}},
\nl
 I_{31,33}(q) &=& 
 \frac{1}{6} \frac{I_0}{\spro{Q_+}{Q_-}}\,  \Bigl[ 1 +
 \frac{1}{2} \frac{Q^2_+}{\spro{Q_+}{Q_-}} \Bigr],
\nl
 I_{31,34}(q) &=& 
\frac{1}{3}  \frac{I_0}{\spro{Q_+}{Q_-}} \Bigl[ 1 +
 \frac{1}{2} \frac{Q^2_+}{\spro{Q_+}{Q_-}} \Bigr]^2,
\nl
 I_{31,32}(q) &=& 
        I_1(q)   \Bigl[ 1 + \frac{1}{8} \frac{Q^6_+}{\lpar\spro{Q_+}{Q_-}\rpar^3} +
     \frac{3}{4} \frac{Q^4_+}{\lpar\spro{Q_+}{Q_-}\rpar^2} + \frac{3}{2} 
          \frac{Q^2_+}{\spro{Q_+}{Q_-}} \Bigr] \nl
{}&-& 11 I_0   \Bigl[ \frac{1}{24} \frac{Q^6_+}{\lpar\spro{Q_+}{Q_-}
 \rpar^4} + \frac{1}{4}  \frac{Q^4_+}{\lpar\spro{Q_+}{Q_-}\rpar^3} + 
 \frac{1}{2} \frac{Q^2_+}{\lpar\spro{Q_+}{Q_-}\rpar^2} \nl
{}&+&   \frac{1}{3} \frac{1}{\spro{Q_+}{Q_-}} \Bigr].
\label{intf}
\eqa
ISR is restored by changing in the previous equations $Q,Q_{\pm}$ into
$\hQ,\hQ_{\pm}$.
The kernel cross section for the process, to be convoluted with the
$e^{\pm}$ structure functions, is therefore written as
\bqa
{}&{}& {\hat \sigma} = \frac{g^8\stwf}{\lpar 2\,\pi\rpar^8}\,
\frac{\pi}{64\,\shat}\,\int d\hQ^2_- d\hy_-\,f_{\gamma}\,\int d\Phi_3\,
{\hat W}_{\mu\mu}\mdls_{\hQs = 0},  \nl
{}&{}& \int d\Phi_3\,{\hat W}_{\mu\mu}\mdls_{\hQs = 0} = 
\frac{\pi^2}{4}\,N_c,\int d\hQ^2_+ d\hy_+\,|A|^2,
\label{kcs}
\eqa
where $A$ is the amplitude for the process, function of $\hQ^2_+, \hy_{\pm}$,
with
\bq
\hQ^2_+ = \hpmomi{+} - q_+, \quad \hy_+ = - 2\,\frac{\spro{\hQ_+}{\hQ_-}}
{\hy_-\shat}, \quad
\hy_- = {{\spro{\hpmomi{+}}{\hQ_-}}\over {\spro{\hpmomi{+}}{\hpmomi{-}}}}.
\label{defyhpm}
\eq
Starting from the original $d\Phi_3$ we have been able to perform the
$k,\bk$ integrations, with the help of \eqns{inti}{intf}, arriving at a
twofold, $d\hQs d\hy_+$, integral.
For the purpose of integration it is more useful to change variable from 
$\hQ^2_+$ to $x$, defined by
\bq
\hQ^2_+ = \lpar \hy_+ - x\rpar \hy_-\shat.
\label{defx}
\eq
The limits of integration and the jacobian of the transformation are:
\bq
\frac{s_0}{\hy_-\shat} \le x \le \hy_+, \quad \frac{s_0}{\hy_-\shat} \le
\hy_+ \le 1, \quad d\hQs_+ d\hy_+ = \hy_-\shat d\hy_+ dx.
\label{varia}
\eq
Before giving the complete expression for $|A|^2$ and computing the cross 
sections we have to answer the question of what to
do with the light quark masses. The following section is devoted to a 
clarification of the origin of these additional mass singularities.
\section{QCD corrections\label{sect_7}}

We have already indicated that, for massless quarks, the cross section for 
$e^+e^- \to e^- \barnu_e u \bard$ is dominated by two large logarithms.
One originates in the limit of small scattering angle of the outgoing electron.
The other comes from the propagator of the internal light quark in the
multi-peripheral diagrams.
Another way of looking at it is to reconsider the integral $I_1$ of
\eqn{inti1} and to evaluate it for $\muq = \md = 0$ and arbitrary
$Q^2_{\pm}$. One obtains
\bqa
I^{(0)}_1 &=& \int d^4k\,d^4\bk\,\delta^+(k^2)\,\delta^+(\bk^2)\,
\delta^4(Q-k-\bk)\,\frac{1}{\lpar Q_- - k\rpar^2}  \nl
{}&=& \frac{\pi}{4\,\sqrt{\Delta}}\,\ln{{\spro{Q_+}{Q_-} - \sqrt{\Delta}}
\over {\spro{Q_+}{Q_-} + \sqrt{\Delta}}},
\eqa
with $\Delta$ being a Gram's determinant,
\bq
\Delta = \lpar\spro{Q_+}{Q_-}\rpar^2 - Q^2_+Q^2_-.
\eq
Therefore, for small scattering angles of the outgoing electron, the integral
behaves like
\bq
I^{(0)}_1 \sim \frac{\pi}{4\,\spro{Q_+}{Q_-}}\,\ln{{Q^2_+Q^2_-}\over
{4\,\lpar\spro{Q_+}{Q_-}\rpar^2}}, \quad \mbox{for} \quad
Q^2_- \to 0.
\label{largel}
\eq
Note that $(Q_- - k)^2$ appears in the internal quark propagator of the
multi-peripheral diagrams.
For values of $Q^2_-$ small enough, the lower limit for
$(Q_- - k)^2$ becomes much smaller than $\Lambda_{\scriptstyle QCD}$, well
beyond the limit of applicability of perturbative QCD.
This fact has many similarities to the inelastic $ep$ scattering, see
the work in \cite{kn:eps}.
\vspace{0.2cm}
\bqas
\ba{ccc}
\vcenter{\hbox{
  \SetScale{0.7}
  \begin{picture}(110,100)(0,0)
  \ArrowLine(50,120)(0,140)
  \ArrowLine(100,140)(50,120)
  \Line(50,120)(50,90)
  \ArrowLine(100,90)(50,90)
  \ArrowLine(50,90)(50,45)
  \Gluon(50,45)(100,45){2}{7}
  \ArrowLine(50,45)(50,20)
  \Photon(0,0)(50,20){2}{7}
  \ArrowLine(50,20)(100,20)
  \GCirc(50,20){8}{0.5}
  \end{picture}}}
&{}&
\vcenter{\hbox{
  \SetScale{0.7}
  \begin{picture}(110,100)(0,0)
  \ArrowLine(50,120)(0,140)
  \ArrowLine(100,140)(50,120)
  \Line(50,120)(50,90)
  \ArrowLine(100,90)(50,90)
  \ArrowLine(50,90)(50,45)
  \ArrowLine(50,45)(100,45)
  \Gluon(50,45)(50,20){2}{7}
  \Photon(0,0)(50,20){2}{7}
  \ArrowLine(100,30)(50,20)
  \ArrowLine(50,20)(100,10)
  \GCirc(50,20){8}{0.5}
  \end{picture}}}
\ea
\eqas
\vspace{-2mm}
\begin{figure}[h]
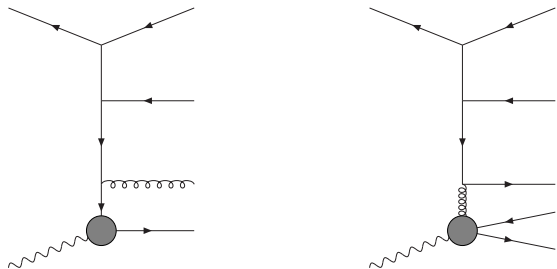

\caption[]{Example of resolved type processes in CC20$_{\gamma}$.}
\label{resol}
\end{figure}
\vskip 5pt
So far, QCD corrections to the CC20 process have been applied within the
context of naive QCD \cite{kn:lep2ww} or with a complete $\ord{\als}$ 
calculation \cite{kn:mpp} which assumes a point-like coupling of the photon 
to quarks. 
However, the large logarithm of \eqn{largel} receives contributions from
any order in $\als$ from multiple gluon radiation. The latter creates a series
of extra quark propagators, each yielding an extra power of the logarithm
compensating the additional power of $\als$. This fact is discussed in
\cite{kn:eps} and in \cite{kn:jap}.
Therefore, logarithmically enhanced terms, of order $\als^n\ln^n(\mqs/Q^2)$
appear at every order in the perturbative expansion and, since the logarithm
is large, the perturbative series does not converge quickly.
Fortunately, this difficulty can be obviated, at least in principle. A 
formalism exists to sum these logarithms to all orders in perturbation theory,
see \cite{kn:sol}.

The point is that a photon has a point-like coupling to the quark-antiquark
pair only for sufficiently high virtuality. On the contrary, for small
electron scattering angle, the photon is quasi-real in our CC20 process and
behaves like a meson. This fact and its consequences are well-known in 
other processes, like the total $\gamma p$ cross section. At low photon
virtualities one also expects contributions from the partonic constituents of
the photon. The two contributions form the so-called resolved photon component,
see \fig{resol}, which is usually added to the direct one, computed to fixed 
order in perturbation theory, where the photon is treated as an elementary 
particle.
The resolved photon component is given in terms of the photon, hadronic, 
structure function. There is a well-known subtlety in adding direct and
resolved components. The direct component, evaluated at some fixed order in 
$\als$ from all corresponding Feynman diagrams, contains singular terms that 
are already re-summed in the resolved component. The correct result 
\cite{kn:eps} is schematically represented in the following equation:
\bq
{\hat\sigma} = \int d\hy_-\,{\cal F}_{\gamma}\,\Bigl[ 
{\hat\sigma}^{\rm res}_{\gamma e} + {\hat\sigma}^{\rm dir,sub}_{\gamma e}
\Bigr],
\label{splits}
\eq
where the superscript {\em sub} indicates that one must subtract the terms
responsible for the large logarithms in the direct photon component and
where ${\hat\sigma}_{\gamma e}$ is the direct or the resolved cross section for
$\gamma e^+ \to \barnu_e u \bard$ at fixed $\hy_-$.
In the resolved cross section the photon interacts via the quark or the gluon
component in its structure function. Therefore one has \cite{kn:jap}
\bq
{\hat\sigma}^{\rm res}_{\gamma e} = \asums{i=q,g}\,
\int d\eta F_{i\gamma}\lpar\eta,M\rpar\,{\hat\sigma}_{i e \to {\rm jet}}.
\eq
here $F_{i\gamma}$ is the PHSF and $M$ is the factorization scale.
As shown in \eqn{splits} the cross section for the CC20 process, 
$e^+e^- \to e^- \barnu_e {\rm jet}$
follows from the cross section for $e^+ \gamma \to \barnu_e {\rm jet}$ 
by applying the equivalent photon or Weizs\"acker-Williams approximation 
which factorizes the flux of quasi-real photons emitted by the $e^-$ from the 
interaction rate between the positron and the photon assumed to be real.

The introduction of a resolved component for the photon is a familiar topic
in $\gamma p$ or $\gamma\gamma$ scattering. Here the situation is slightly
different. The small virtuality of the photon is only needed when CC20
is a background to the high-energy hadronic lineshape or to Higgs boson
searches or for single $\wb$ production. The cross section for 
CC20$_{\gamma}$ is obtained starting from four Feynman diagrams, two being
single $\wb$ resonant and two being multi-peripheral. In single $\wb$
production, where one applies a cut $|\cos\theta(e^-)| \ge 0.997$, the QCD
corrections are important and it appears difficult to obtain a precise
prediction for the total cross section without summing the large logarithms
into the $F_{i\gamma}$ distribution function.

For the hadronic lineshape, on the other end, hadronic events are selected 
based on final state particle multiplicity in the detector. This gives the 
total sample but more interesting is the high-energy $M^2(u\bard) \ge s_0$ 
sample.
The relative dominance of the multi-peripheral diagrams in CC20$_{\gamma}$
is larger in the total sample but not necessarily in the high-energy one.
Therefore the uncertainty associated with the use of the PHSF calculated at 
the zeroth order in $\als$, i.e. in the Born approximation, is less
relevant if we apply a strong invariant mass cut. 

Our strategy, for the moment, will be to use the parton model result, i.e.
zeroth order in $\als$, and to cure the ill-defined massless limit by
replacing the quark masses with a factorization scale $M$, in our
case $\muq = \md \to M$. The total cross section for the CC20 process will,
therefore, depend on the scale $M$.
The amplitude squared, to be inserted in \eqn{kcs}, becomes
\bqa
|A|^2 &=& \frac{4}{3} \frac{\hy_-}{|\Delta_s|^2} v^2 \gamma^2_{_\wb} 
         \mu^2_{_\wb} \frac{x}{Y}  
         \frac{1-\hy_+}{\hy_+^4} 
\Bigl[ \hy_+^2 + x ( - 3 \hy_+  + 2 x ) \Bigr]
\nl\nl
{}&+& \frac{1}{3} \frac{\hy_-}{|\Delta_s|^2} v^2 \frac{x}{\hy_+^2} \Bigl[ 2 - 
         4 \hy_+ - \hy_+^3 + \hy_+^2 + 2 x (1-\hy_+) ( - \hy_+ + x) \Bigr]
\nl\nl
{}&+& \frac{2}{3} \frac{\hy^2_-}{|\Delta_s|^2} v^3 \mu^2_{_\wb} \frac{x}{Y} 
      \frac{1-\hy_+}{\hy_+^3}  \Bigl[ \hy_+^2 + x ( - 3 \hy_+ + 2 x ) \Bigr]
\nl\nl
{}&+& \frac{2}{3} \frac{\hy^3_-}{|\Delta_s|^2} v^4 \frac{x^2}{Y} 
      \frac{1-\hy_+}{\hy_+^3} \Bigl[ - \hy_+^2 + x (3 \hy_+ - 2 x ) \Bigr]
\nl\nl
{}&+& \frac{1}{9} \frac{\hy_-}{|\Delta_t|^2} L \frac{v^2}{\hy_+^4} 
         \Bigl[ - 5 \hy_+^3 - \hy_+^5 + 2 \hy_+^4 
 + x ( 15 \hy_+^2 - 6 \hy_+^3 + 3 - 20 x \hy_+ + 8 x \hy_+^2 \nl
{}&-& 4 x \hy_+^3 + 10 x^2 - 4 x^2 \hy_+ + 2 x^2 \hy_+^2 ) \Bigr]
\nl\nl
{}&+& \frac{4}{3} \frac{\hy_-}{|\Delta_t|^2} v^2 \gamma^2_{_\wb} 
         \mu^2_{_\wb} \frac{x}{Y} 
         \frac{1-\hy_+}{\hy_+^4} 
\Bigl[ \hy_+^2 + x ( - 3 \hy_+ + 2 x ) \Bigr]
\nl\nl
{}&+& \frac{1}{9} \frac{\hy_-}{|\Delta_t|^2} \frac{v^2}{\hy_+^4} \Bigl[ - 
         11 \hy_+^3 -  \frac{5}{2} \hy_+^5 + 5 \hy_+^4 
 + x ( 31 \hy_+^2 - 
         19 \hy_+^3 + \frac{11}{2} \hy_+^4 - 50 x \hy_+ \nl
{}&+& 32 x \hy_+^2 - 
         8 x \hy_+^3 + 30 x^2 - 18 x^2 \hy_+ + 5 x^2 \hy_+^2 ) \Bigr]
\nl\nl
{}&+& \frac{2}{3} \frac{\hy^2_-}{|\Delta_t|^2} v^3 \mu^2_{_\wb} \frac{x}{Y}  
      \frac{1-\hy_+}{\hy_+^3} \Bigl[ - \hy_+^2 + x (3 \hy_+ - 2 x ) \Bigr]
\nl\nl
{}&+& \frac{2}{3} \frac{\hy^3_-}{|\Delta_t|^2} v^4 \frac{x}{Y}  
         \frac{1-\hy_+}{\hy_+^3} \Bigl[ - \hy_+^3 + x ( 4 \hy_+^2 - 5 x \hy_+ + 
         2 x^2 ) \Bigr].
\eqa
The propagators, in the fixed width scheme, become
\bq
\Delta_s = {1\over {- x\hy_- v + \mu^2_{_\wb} - i\,\gamma_{_\wb}\mu_{_\wb}}},
\quad
\Delta_t = {1\over {- \lpar x - \hy_+\rpar \hy_- v + \mu^2_{_\wb} - 
i\,\gamma_{_\wb}\mu_{_\wb}}},
\eq
where 
\bq
v = x_+x_-, \quad Y = {1\over {\hy^2_+\hy^2_- + 4\,\gamma^2_{_\wb}
\mu^2_{_\wb}}}, \quad L = \ln \frac{M^2}{x \hy_- \shat},
\eq
and $\hy_{\pm}, x$ are given in \eqnsc{defyhpm}{defx}, $\mu_{_\wb},
\gamma_{_\wb}$ after \eqn{defred}.
With $|A|^2$ at our disposal we can use \eqn{kcs} and \eqn{varia}, apply the
convolution with QED structure functions, and derive the total cross section.

\section{Numerical results and conclusions\label{sect_8}}

In this section we present all relevant numerical results for the CC20 
processes as computed by the FORTRAN program WTO version 2.0 \cite{kn:wto}. 
The chosen setup is specified by the following list:
\bq
\sqrt{s} = 186\,\GeV, \quad \mw = 80.39\,\GeV, \quad \mz= 91.1867\,\GeV
\eq
Naive QCD is not introduced which implies, in particular, that the $\wb$
width is included without QCD corrections. For our setup this results into
$\gw = 2.0459\,$GeV.
The QED radiation is included by means of the structure function approach 
(in the so-called $\beta$-scheme~\cite{kn:sf1,kn:sf2,kn:sf3}).
First we consider a cut on the scattering angle of the outgoing quarks
with respect to the beam axis, $10^{\circ} \le \theta_q \le 170^{\circ}$.
We also fix a lower cut on the invariant mass of the $u \bard$ system,
$M^2(u\bard) \ge 0.01\,s$.
According to the procedure described in \eqn{split} we introduce a
separating angle $\theta_c$ and compute the following cross sections:
\begin{itemize}

\item[$\sigma_{<}$] or $|\mbox{CC20}^{<}_{\gamma}(\me)|^2$ for $\theta 
\le \theta_c$,

\item[$\sigma_{>}$] or $|\mbox{CC20}^{>}(0)|^2$ for $\theta_c \le \theta 
\le \pi$,

\item[$\sigma_{<\rm int}$] or $2\,\Bigl[ \mbox{CC20}^{<}_{\gamma}(0)
\Bigr]^{\dagger}\,\mbox{CC20}^{<}_{\rm R}(0) + |\mbox{CC20}^{<}_{\rm R}(0)|^2$
for for $\theta \le \theta_c$.

\end{itemize}

Our reference values will be $\theta_c = 0.3^{\circ},0.4^{\circ}$ and
$0.5^{\circ}$. We have verified that $\sigma_{<\rm int}$ is completely 
negligible for
our choice of the separator $\theta_c$ so that the total is safely
given by the sum $\sigma_{<} + \sigma_{>}$. Indeed we find 
$\sigma_{<\rm int} = 6\div 5 \div 3 \times 10^{-5}\,$
pb for $\theta_c = 0.5^{\circ} \div 0.4^{\circ} \div 0.3^{\circ}$.
Always for $10^{\circ} \le \theta_q \le 170^{\circ}$ we find for $\sigma_{<}/
\sigma_{>}$ the results shown in \tabn{tab1}.
 \begin{table}[htbp]\centering
 \begin{tabular}{|c|c|c|c|}
 \hline
 $\theta_c\,$[Deg] &  $\sigma_{<}$  & $\sigma_{>}$  & $\sigma_{<}+\sigma_{>}$ \\
 \hline
   $0.3^{\circ}$   &  0.0527   & 0.6316(9) & 0.6843(9)     \\
 \hline
   $0.4^{\circ}$   &  0.0554   & 0.6289(7) & 0.6843(7)     \\
 \hline
   $0.5^{\circ}$   &  0.0575   & 0.6269(6) & 0.6844(6)     \\
 \hline
 \end{tabular}
\vspace*{3mm}
 \caption[
 ]{\it
Cross section in pb for the process $e^+e^- \to e^- \barnu_e u \bard$, for 
$10^{\circ} \le \theta_q \le 170^{\circ}, q=u,d$ as a function of $\theta_c$.}
 \label{tab1}
 \end{table}
 \normalsize

\begin{figure}[t]
\begin{minipage}[t]{14cm}
{\begin{center}
\vspace*{-2.0cm}
\hspace*{-1.6cm}
\mbox{\epsfysize=14cm\epsfxsize=15cm\epsffile{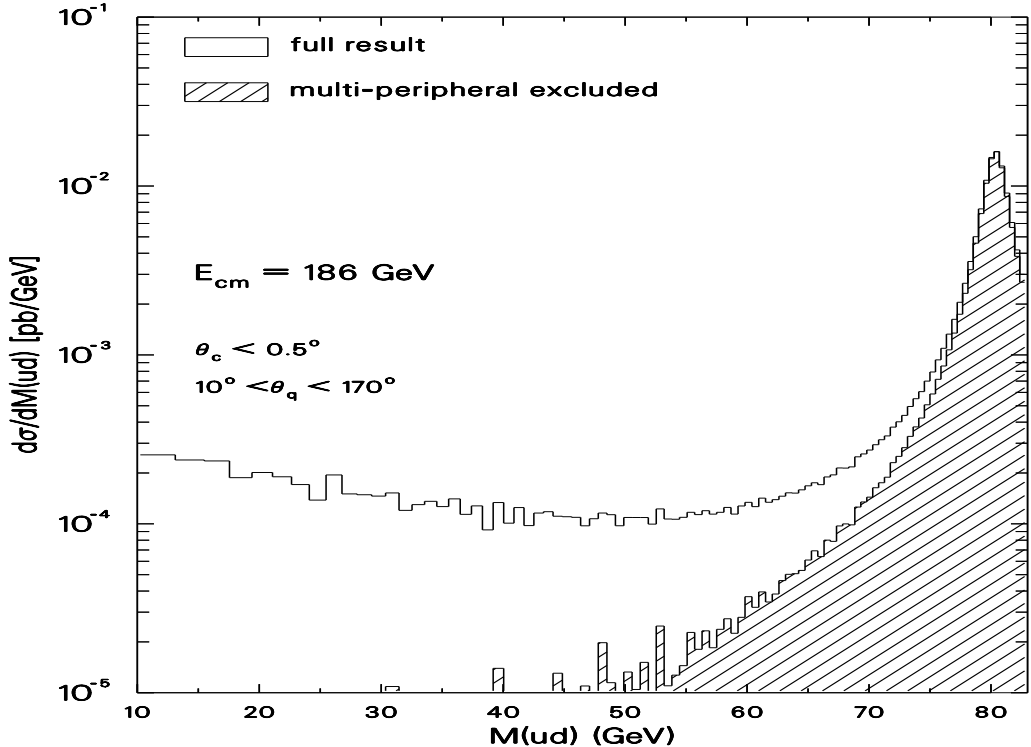}}
\vspace*{-2.7cm}
\end{center}}
\end{minipage}
\vspace*{0.5cm}
\caption[]{Invariant mass distribution for $e^+e^- \to \barnu_e u \bard$
showing the effect of the multi-peripheral component.}
\label{massdist}
\end{figure}

\tabn{tab1} clearly shows that there is a smooth matching
of the two components, $<(\me)$ and $>(0)$ at $\theta = \theta_c$. This result 
justifies the application of the WW-approximation in the narrow cone around 
the electron axis and we conclude by quoting the following result:
\bq
\sigma_{\rm CC20}\lpar 186\,\GeV,\,10^{\circ} \le \theta_q 
\le 170^{\circ},\, M^2(u\bard) \ge 0.01\,s\rpar = 0.6843(4)\,\pb.
\eq
We have also varied the angular cut on the outgoing quarks, keeping
$M^2(u\bard) \ge 0.01\,s$ and $\theta_c = 0.5^{\circ}$. The latter is fully 
justified by the tiny dependence of the cross section on the separating
angle $\theta_c$. The results are illustrated in \tabn{tab2}.
 \begin{table}[htbp]\centering
 \begin{tabular}{|c|c|c|c|}
 \hline
 $\theta_q\,$[Deg] &  $\sigma_{<}$  & $\sigma_{>}$ & $\sigma_{<}+\sigma_{>}$ \\
 \hline
   $5^{\circ}$   & 0.0607    & 0.6404(6) & 0.7011(6)     \\
 \hline
   $6^{\circ}$   & 0.0601    & 0.6384(6) & 0.6985(6)     \\
 \hline
   $8^{\circ}$   & 0.0588    & 0.6333(6) & 0.6921(6)     \\
 \hline
   $10^{\circ}$  & 0.0575    & 0.6269(6) & 0.6844(6)     \\
 \hline
 \end{tabular}
\vspace*{3mm}
 \caption[
 ]{\it
Cross section in pb for the process $e^+e^- \to e^- \barnu_e u \bard$, for 
$\theta_c = 0.5^{\circ}$ as a function of $\theta_q$.}
 \label{tab2}
 \end{table}
 \normalsize
Next, we consider the total CC20 cross section, without angular cuts on the
outgoing quarks and with massless quarks. As explained in Sect. \ref{sect_7},
where we have discussed QCD corrections, the resulting cross section depends 
on a factorization scale $M$. For $M = 1\,$GeV we find the results of 
\tabn{tab3}.
 \begin{table}[htbp]\centering
 \begin{tabular}{|c|c|c|c|}
 \hline
 $\theta_c\,$[Deg] &  $\sigma_{<}$ & $\sigma_{>}$  & $\sigma_{<}+\sigma_{>}$ \\
 \hline
   $0.3^{\circ}$   &  0.0850   & 0.6503(8) & 0.7356(8)     \\
 \hline
   $0.4^{\circ}$   &  0.0881   & 0.6473(7) & 0.7354(7)     \\
 \hline
   $0.5^{\circ}$   &  0.0905   & 0.6451(6) & 0.7354(6)     \\
 \hline
 \end{tabular}
\vspace*{3mm}
 \caption[
 ]{\it
Cross section in pb for the process $e^+e^- \to e^- \barnu_e u \bard$, for the
factorization scale $M = 1\,$GeV, as a function of $\theta_c$.}
 \label{tab3}
 \end{table}
 \normalsize
From \tabn{tab3} we derive the CC20 cross section in a fully extrapolated
setup.
\bq
\sigma_{\rm CC20}\lpar 186\,\GeV,\,M = 1\,GeV\rpar = 
0.7354(4)\,\pb.
\eq
We have also investigated the dependence of the cross section on the 
factorization scale $M$. With $\theta_c = 0.5^{\circ}$ and $M^2(u\bard) \ge 
0.01 s$ the results are presented in \tabn{tab4}, showing a mild dependence
on $M$ of the total.
 \begin{table}[htbp]\centering
 \begin{tabular}{|c|c|c|c|}
 \hline
 $M\,$[GeV]    &  $\sigma_{<}$ & $\sigma_{>}$  & $\sigma_{<}+\sigma_{>}$ \\
 \hline
   $0.1$   &  0.0908   & 0.6451(6) & 0.7359(6)     \\
 \hline
   $1$     &  0.0905   & 0.6451(6) & 0.7356(6)     \\
 \hline
   $10$    &  0.0902   & 0.6451(6) & 0.7353(6)     \\
 \hline
   $100$   &  0.0900   & 0.6451(6) & 0.7351(6)     \\
 \hline
 \end{tabular}
\vspace*{3mm}
 \caption[
 ]{\it
Cross section in pb for the process $e^+e^- \to e^- \barnu_e u \bard$, for 
$\theta_c = 0.5^{\circ}$ as a function of the factorization scale $M$.}
 \label{tab4}
 \end{table}
 \normalsize
Finally we have analyzed single $\wb$ production with $\theta(e^-) < 
0.5^{\circ}$ in WW-approximation. In \fig{massdist} we have reported the
$M(u\bard)$ distribution for $10^{\circ} \le \theta_q \le 170^{\circ}$.
In order to understand the role of the different components we have plotted
the distribution with and without the multi-peripheral component. It follows
that this component dominates at low invariant masses, while above
$M(u\bard) \approx 70\,$GeV it is practically without influence.
A final comment is devoted to the validity of the WW-approximation.
Note that we only use this approximation in a narrow cone around the electron
axis, typically $\theta \le 0.5^{\circ} = 8.73\,$mrad, and the complete 
calculation outside the cone. Corrections to \eqn{defww} of $\ord{Q^2}$
have been discussed in \cite{kn:wwa} where it has been shown that, after
integration, the cross section receives additional contributions proportional
to powers of $Q^2_c/(2\,\spro{Q}{p_+})$. Since
\bq
\frac{Q^2_c}{2\,\spro{Q}{p_+}} < \frac{s}{4\,s_0}\,\theta^2_c,
\eq
and, in our case, $\sqrt{s} = 186\,$GeV, $s_0 = 0.01\,s$ we find
$Q^2_c/(2\,\spro{Q}{p_+}) < 1.9 \times 10^{-3}$. Therefore non-factorizable
corrections are formally negligible.

The mild dependence of the total cross section on the factorization scale $M$ 
can be understood from \tabn{tab5}. Here, for $\theta \le 0.5^{\circ}$, we 
have reported: 1) the total cross section , i.e the sum of the single-resonant
and of the multi-peripheral, non-resonant, components; 2) the multi-peripheral 
component alone. Therefore the non-resonant, $M$-dependent, terms are strongly 
suppressed and the $M$-dependence has little influence on the total,
justifying our approximation of working at zeroth order in $\als$.
 \begin{table}[htbp]\centering
 \begin{tabular}{|c|c|c|}
 \hline
               &                         &                         \\
 $M\,$[GeV]    &  $\sigma_{<}^{\rm TOT}$ & $\sigma_{<}^{\rm MP}$   \\
               &                         &                         \\
 \hline
   $0.1$   &  0.0908   &  0.0009    \\
 \hline
   $1$     &  0.0905   &  0.0007    \\
 \hline
   $10$    &  0.0902   &  0.0005    \\
 \hline
   $100$   &  0.0900   &  0.0003    \\
 \hline
 \end{tabular}
\vspace*{3mm}
 \caption[
 ]{\it
Total Cross section and Multi-Peripheral component in pb for 
$e^+e^- \to e^- \barnu_e u \bard$, for $\theta_c = 0.5^{\circ}$ as a function 
of the factorization scale $M$.}
 \label{tab5}
 \end{table}
 \normalsize

The experimental Collaborations at LEP are now recording and analyzing a 
sizeable fraction of events with four fermions in the final state.
Outgoing electrons represent a notorious problem because of the presence
of $t$-channel photons interacting with $\wb$ bosons or coupling to 
quark-antiquark pairs.

The collinear limit forbids a calculation where the massless limit for fermions
is assumed from the beginning and, in turn, this may induce numerical
instabilities in computing the total cross section, even for a fully
massive MonteCarlo. It should be mentioned also that the majority of the
MonteCarlos used in the analysis are built in the massless approximation and
because of that the total cross section is not available.

We have suggested a simple but realistic solution based on the use of the
Weizs\"acker-Williams approximation, to abe applied in a narrow cone
around the beam axis. The large logarithms, $\ln(\mes/s)$, are correctly 
described by our numerical solution. Furthermore the improved 
WW-approximation that we are using is valid beyond the leading logarithmic 
approximation, as explained in \eqn{defww}, and correctly integrates also the
$\mes/Q^4$ terms present in the photon flux.

The correct treatment of the kinematics, accounting for the introduction of 
QED initial state radiation, is also emphasized. We have derived a version of 
the flux-function which describes quasi-real photons emitted by the electron 
after QED radiation.
A second logarithmic enhancement in the cross section, arising from internal
fermion propagators, is also described and a link is established with the
familiar examples of $\gamma p$ or $\gamma\gamma$ scattering. Finally, several
numerical results are shown, proving the goodness of the adopted solution.
\section{Acknowledgements}

I would like to express special thanks to Martin Gr\"unewald for several
important discussions on the experimental aspects of the problem, to
Michelangelo Mangano for mentioning, long ago, the possible relevance of the 
Weizs\"acker-Williams approximation in this context, to Alessandro Ballestrero
for a discussion on the CC20 phase space, to Roberto Pittau for information 
on the status of EXCALIBUR.
%
%

\end{document}